\documentclass[10pt,aps,pra,a4paper,twocolumn,superscriptaddress,floatfix,longbibliography]{revtex4-1}
\usepackage[table, svgnames, dvipsnames]{xcolor}
\usepackage{makecell, cellspace}

\usepackage{bbold}
\usepackage{amsmath, dsfont}
\usepackage{bm}
\usepackage{amssymb}
\usepackage[normalem]{ulem}
\usepackage{braket}
\usepackage{hyperref}
\usepackage{tikz}
\usepackage{ifthen}
\usepackage{stmaryrd}
\usetikzlibrary{tikzmark}

\def\ri{i}
\def\re{\mathrm e}
\def\rd{\mathrm d}
\def\BF{\mathcal F}
\def\hc{{\rm H.c.}}

\def\Up{\uparrow}
\def\Dn{\downarrow}
\def\MH{\mathcal{H}}

\def\DF{\theta}
\def\MF{\varphi}

\newcommand{\ave}[1]{\left\langle #1 \right\rangle}

\newcommand\bwt{\begin{widetext}}
\newcommand\ewt{\end{widetext}}

\begin{document}

\title{Exploring helical phases of matter in bosonic ladders}

\author{Andreas Haller}
\affiliation{Institute of Physics, Johannes Gutenberg University, D-55099 Mainz, Germany}
\author{Apollonas S. Matsoukas-Roubeas}
\affiliation{Niels Bohr International Academy and Center for Quantum Devices,
University of Copenhagen, Universitetsparken 5, 2100 Copenhagen, Denmark}
\author{Yueting Pan}
\affiliation{Center for Advanced Quantum Studies, Department of Physics, Beijing Normal University, 100875 Beijing, China}
\author{Matteo Rizzi}
\affiliation{Forschungszentrum Jülich, Institute of Quantum Control, Peter Grünberg Institut (PGI-8), 52425 Jülich, Germany}
\affiliation{Institute for Theoretical Physics, University of Cologne, D-50937 K\"oln, Germany}
\author{Michele Burrello}
\affiliation{Niels Bohr International Academy and Center for Quantum Devices,
University of Copenhagen, Universitetsparken 5, 2100 Copenhagen, Denmark}

\begin{abstract}

Ladder models of ultracold atoms offer a versatile platform for the experimental and theoretical study of different phenomena and phases of matter linked to the interplay between artificial gauge fields and interactions. Strongly correlated helical states are known to appear for specific ratios of the particle and magnetic flux densities and they can often be interpreted as a one-dimensional limit of fractional quantum Hall states, thus being called pretopological. Their signatures, however, are typically hard to observe due to the small gaps characterizing these states. Here we investigate bosonic ladder models at filling factor $\nu=1$. Based on bosonization, renormalization group and matrix product state simulations we pinpoint two strongly correlated helical phases appearing at this resonance. We show that one of them can be accessed in systems with two-species hardcore bosons and on-site repulsions only, thus amenable for optical lattice experiments. Its signatures are sizable and stable over a broad range of parameters for realistic system sizes.

\end{abstract}

\maketitle

{\em Introduction.--}
The experimental investigation of ultracold atomic gases has developed in the last years with remarkable strides.
Ultracold atom simulators have been adopted to examine several many-body quantum problems, with the opportunity of tuning the ratio between their interactions and kinetic energies and to trap atomic gases in different geometries. In this context, the study of ladder geometries generated through optical lattices offer the possibility of testing the behavior of quantum systems at the border between one- and two-dimensional systems \cite{atala2014}. Their kinetic energy is firmly one-dimensional, but the presence of plaquettes around which the atoms move allows for the introduction of artificial gauge potentials with non-trivial effects.

In the last few years, the effects of these artificial magnetic fluxes have been tested for ladder geometries with a transverse direction defined in either real or ``synthetic'' dimension \cite{Celi2012}. In the latter case, an inner degree of freedom of the atoms is used to represent the transverse coordinate in the ladder. Synthetic dimensions offer, in this way, the possibility of defining both sharp edges and artificial gauge fluxes with suitable Raman couplings \cite{Celi2014}.

The introduction of artificial gauge potentials in general breaks time-reversal symmetry and allows for the onset of helical many-body states, characterized by the appearance of a net chiral current running in counterpropagating directions at the edges of the ladder. Such phenomena have been investigated for both fermionic  \cite{mancini2015,fallani2016,Kang2018,Han2019} and bosonic \cite{spielman2015,Genkina2019} non-interacting gases.

Ultracold gases trapped in ladder systems are currently at the focus of great attention, due to their non-trivial behavior reminiscent of both superconducting \cite{Orignac2001} and quantum Hall systems \cite{Cornfeld2015,Hugel2014,Wauters2018,Giamarchi2019}, and the possibility of engineering interacting topological phases of matter \cite{Burrello2015,Bermudez2017,Nehra2018,Barbarino2018,Tirrito2019,Barbarino2019}. In particular, these systems are characterized by several commensuration effects. The first kind is related to the physics of Mott insulators and appears for commensurate ratios between the number of particles and sites in the ladder \cite{Carr2006,Crepin2011,Keles2015,Piraud2015,Strinati2018,Chen2019}. The second kind concerns the ratio between the total artificial magnetic flux enclosed by the ladder and the number of plaquettes in the geometry: several non-trivial phases corresponding to crystals of magnetic vortices alternate when increasing the flux per plaquette at fixed particle density \cite{Greschner2015,Greschner2016,Greschner2017,Buser2019}. A third, and more elusive kind of commensuration has been at the center of an intensive study in the last years and it is related to the ratio between the number of particles and magnetic fluxes. For several fractional values of this ratio, suitable repulsive interactions among the atoms cause the onset of partially gapped states, characterized by a helical current. For both fermions and bosons, it has been shown that many of these helical many-body states are indeed the one-dimensional ladder limit of fractional quantum Hall states \cite{Petrescu2015,Mazza2017,Taddia2017,Petrescu2017,Haller2018,CalvaneseStrinati2019,Rosson2019,Santos2019,Yang2020}. They can be considered as pretopological one-dimensional chiral states.

The physics of these many-body states can be understood in terms of specific resonances between the modulation length of the density of particles and the modulation of the hopping phases determining the external fluxes. In particular, bosonization techniques allow us to understand through semiclassical approximations the fermionic states appearing at filling factor $1/3$ and the bosonic states at filling factor $1/2$ as Laughlin-like pretopological states.

In this work, we focus on the case of 2-leg ladders of bosonic atoms at filling factor $\nu=1$. Similarly to their fermionic counterpart at $\nu=1/2$ \cite{Haller2018}, these states cannot be discussed in terms of a simple semiclassical approximation due to the competition of several operators becoming resonant.
The physics at this resonance is indeed dictated by competing interactions that concur in forming two  different phases. The one dominating for hardcore bosons with rung repulsions is originated by the same effective interaction responsible for the formation of paradigmatic examples of intrinsically gapless symmetry-protected one-dimensional phases of matter \cite{vishwanath2020}.
In this respect, resonant ladder models offer a mechanism to form partially gapped phases from interactions that are usually irrelevant, thus providing a platform to study the features of novel strongly correlated helical systems that are intrinsically gapless.

Two-dimensional spinless bosonic systems in Hofstadter-Hubbard models in ladder geometries at $\nu=1$ have already been numerically studied \cite{Hugel2017,Pollet2018}, and it was pointed out that for this specific ratio and certain ranges of the hopping parameters, signatures of a second incommensurability appear in the correlation functions \cite{Chiofalo2015,Orignac2016,Orignac2017,Citro2018}. A systematic analysis of the appearance and characterization of the energy gaps in this system, however, is still missing.

In this work we aim to fill this gap, and we discuss in detail the physics of this system through bosonization and matrix-product-state simulation. We show, in particular, that helical states appear in a ladder of hardcore bosons. With suitable choices of the interleg interactions, the signatures of these states are considerably stronger than their fermionic counterpart at filling $\nu=1/2$. This opens the path for an experimental study of these strongly correlated and helical phases of matter: our simulations and analysis confirm indeed that one of the two helical phases under investigation can be accessed in bosonic ladders with contact interactions only, which is the most realistic scenario for atoms trapped in optical lattices.

In the following, we will discuss the possible phases of matter characterizing this system in the limit of weak tunneling along the rungs, and we will investigate in detail its main observables and correlation functions. Sec. \ref{sec:model} defines the lattice model and its interactions, which are described through an effective low-energy field-theory in Sec. \ref{sec:bosonization}. The renormalization group analysis of the system at the $\nu=1$ resonance is presented in Sec. \ref{sec:RG} and the main features of the model are analyzed in Sec. \ref{sec:MPS} through matrix product state simulations. Finally, our conclusions are summarized in Sec. \ref{sec:concl}. The appendixes present the detail of our bosonization and renormalization group analysis.

\section{The model} \label{sec:model}
Ladder models of ultracold bosons trapped in optical lattices have been experimentally realized with rubidium gases both in spacial ladders \cite{atala2014}, by using optical superlattices to isolate single two-leg ladders, and in synthetic dimensions \cite{spielman2015,Genkina2019}. In both cases, suitable pairs of Raman lasers allow us to engineer an effective long-time dynamics of the system described by a low-energy tight-binding Hamiltonian characterized by position-dependent phases in the hopping amplitudes of the atoms (see, for example, Refs. \cite{goldman2014,goldman2015,burrello2017,aidelsburger2018}). The resulting artificial gauge flux $\chi$ in each plaquette depends on the recoil momentum of the Raman lasers and can be varied by tuning the angle between them (see, for example, the experiments \cite{Aidelsburger2013,Miyake2013,Aidelsburger2015}).

In this work, we focus on the case of ultracold bosons trapped in a two-leg ladder geometry (see Fig. \ref{fig:model}a). Their long-time dynamics is described by an interacting Hamiltonian of the form:
\begin{equation} \label{hamtot}
    \MH = \MH_0 + \MH_U + \MH_\perp\,,
\end{equation}
where $\MH_0$ represents the kinetic energy of the atoms, whereas $\MH_U$ and $\MH_\perp$ define respectively the intraleg and interleg repulsive interactions. We consider a ladder geometry with $L$ rungs and a gas of $N$ atoms.

\subsection{The single--particle physics}

\begin{figure}[ht]
    \includegraphics{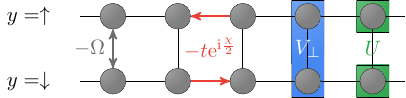}\llap{\parbox[b]{12.7cm}{(a)\\\rule{0ex}{0.7cm}}}
    \includegraphics{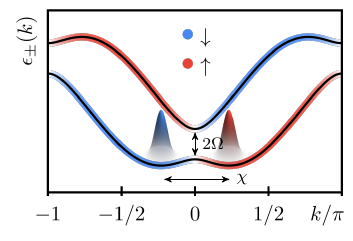}\llap{\parbox[b]{11.8cm}{(b)\\\rule{0ex}{2.4cm}}}
    \caption{(a) Graphical representation of the model Hamiltonian. Arrows represent the tight binding terms in Eq.~\eqref{eq:kinetics} and density-density interactions of Eq.~\eqref{eq:interactions} are depicted by colored boxes.
    (b) Band structure with polarization (in colors, red corresponds to $\uparrow$ and blue to $\downarrow$) for $\Omega/t=0.25$. The partial gap induced by inter-wire transitions is of size $2\Omega$.}
    \label{fig:model}
\end{figure}

We begin our analysis from the single--particle features of the model. The kinetic energy $\MH_0$ is given by two hopping terms
\begin{align}
    \MH_0 = \MH_t + \MH_\Omega
    \,,
\end{align}
describing a tight-binding model of spinless bosonic particles hopping in the ladder geometry. Our model is characterized by an intraleg tunneling amplitude $t$ and an interleg hopping $\Omega$ (see Fig.~\ref{fig:model}):
\begin{align}\label{eq:kinetics}
    \MH_t &= -t\sum_{x=1}^{L-1} \sum_{y=\pm 1} b^\dag_{x,y}b^{}_{x+1,y}\re^{\ri\frac{\chi y}{2}} + \hc
    \,,
    \nonumber\\
    \MH_\Omega &= -\Omega \sum_{x=1}^{L-1} b^\dag_{x,\Up}b^{}_{x,\Dn} + \hc
    \,.
\end{align}
Hereafter we set the lattice spacing $a=1$ for simplicity; the pseudospin index $y=\pm 1$, or, equivalently, $y=\Up,\Dn$, labels the transverse direction distinguishing two legs of the ladder, and the bosonic annihilation/creation operators $b_{x,y}/b^\dag_{x,y}$ satisfy the bosonic commutation algebra.
We have chosen a gauge which preserves the translational invariance along the $x$ direction,
 and the hopping phases along this direction define an artificial magnetic flux per plaquette $\chi = \oint \vec A\rd \vec{l} $ defined by the counter-clockwise Aharonov-Bohm phase acquired by an atom moving along any of the ladder plaquettes.

In a translational invariant case, the single-particle Hamiltonian $\MH_0$ can be rewritten in momentum space:
\begin{equation} \label{hamk}
\MH_0 = -\sum_k {\bm b}^\dag_k
\begin{pmatrix}
2t\cos\left(k+\chi/2\right) & \Omega \\
\Omega & 2t\cos\left(k-\chi/2\right)
 \end{pmatrix} {\bm b}_k
\end{equation}
where $\bm b_k= \left(b_{k,\Up}, b_{k,\Dn}\right)^\intercal$ is a two-component spinor.


The spectrum of the Hamiltonian~\eqref{hamk} is depicted in Fig. \ref{fig:model} (b) for a small value of $\Omega/t$; it matches the typical spectrum of one-dimensional systems characterized by a spin-orbit coupling (whose role is played by $\chi$ in $\MH_t$) and the Zeeman splitting $\Omega$. $\MH_t$ defines indeed two cosine dispersions for pseudospin $y=\,\Up$ and $y=\,\Dn$,  displaced in momentum space by the flux $\chi$.
In the case of bosons, the particles condense around the two minima of these dispersions and their momentum, which can be measured through time-of-flight experiments.
As such, the momentum is locked to their spins and therefore with the leg degree of freedom (see, for example, Refs. \cite{spielman2015,Genkina2019}).

The role of $\MH_\Omega$ is to open a gap $2\Omega$ at $k=0$. This gap can be interpreted as the bulk gap of a quantum Hall system in the one-dimensional ladder limit \cite{Cornfeld2015}. When $\Omega$ exceeds a threshold $\Omega_c (\chi)$, the minimum of the energy dispersion becomes non-degenerate at $k=0$ and the system becomes a superfluid whose atoms are aligned in the $x$ direction of the pseudospin. For non-interacting bosonic systems, this corresponds to the Meissner phase characterized by a chiral current proportional to the flux $\chi$ \cite{atala2014}.

The non-interacting system is indeed characterized by two main phases: the Meissner phase appearing at small fluxes and sufficiently large interleg tunneling $\Omega$, and a vortex phase appearing at larger fluxes or smaller values of $\Omega$ \cite{Orignac2001,atala2014}.

The introduction of repulsive interactions strengthen the Meissner phase and allows for the appearance of several partially gapped and strongly correlated states in the vortex phase.

\subsection{Introducing the interactions}

In ultracold atom experiments, repulsive contact interactions among atoms are common. For a realization of the ladder model in synthetic dimension, atoms whose position differs only by the transverse $y$ coordinate are actually located on the same physical position. Therefore contact repulsions are represented by the following density-density interactions
\begin{align}\label{eq:interactions}
    \MH_U &= \frac U2 \sum_{x} n_{x,y}(n_{x,y}-1)\,,
    \nonumber\\
    \MH_\perp &= V_\perp \sum_{x,y} n_{x,y}n_{x,y+1}\,,
\end{align}
with $n_{x,y}=b^\dag_{x,y}b^{}_{x,y}$.
Such interactions have the potential to drive the system from the vortex to different helical phases, if the total density $n$ is commensurate to the flux $\chi$~\cite{Cornfeld2015,Petrescu2017,CalvaneseStrinati2019}.

For ultracold atoms trapped in optical lattices, the ratio between interaction and kinetic energy can be varied by changing the amplitude of the trapping lasers which define the depth of the trapping lattice.
Concerning the ratio between $U$ and $V_\perp$, tunable intraspecies and interspecies interactions can be achieved through the introduction of suitable magnetic fields via Fano-Feshbach resonances \cite{Chin2010}.
While rubidium gases might be difficult to tune in a regime with considerable differences between intraspecies and interspecies interactions, a more promising platform might be offered by potassium~\cite{Fattori1,Fattori2,Tarruell1,Tarruell2}.

For the remainder of this work, and, especially, in our numerical simulations, we will mostly consider the ladder hard--core bosonic (HCB) scenario, in which $U/t\rightarrow\infty$ is the dominating interaction which excludes any double occupation of a single site $(x,y)$, i.e. $n_{x,y}\leq1$. We will consider instead a tunable interspecies interaction $V_{\perp}$.

\section{Effective low--energy description of the model} \label{sec:bosonization}
To study the physics of the interacting gas at zero temperature we set up a low-energy description of the system in the continuum limit based on bosonization \cite{giamarchibook,gogolinbook}.
In order to examine the chiral states appearing at commensurate values of the filling factor $\nu = \pi N/ L\chi$, we adopt the following bosonization identity for bosons \cite{Cazalilla2004}, accounting for higher harmonics of the density modulations:
\begin{align}
    b^\dag_{x,y} \rightarrow \Psi^\dag_{y}(x) \sim \sum_{p\in2\mathbb Z}\beta_p \re^{\ri [(pk_0+y\chi)x - p\DF_y(x) - \MF_y(x)]}\,;
    \label{eq:bosonization_identity}
\end{align}
here $k_0=\pi N/(2L)$ is the wavevector associated to the density modulations and we introduced two pairs of dual fields that satisfy the commutation algebra $\left[\DF_{y'}(x'),\MF_{y\vphantom'}(x)\right] = \ri\pi\delta_{yy'}\Theta(x'-x)$. The fields $\theta_y$ and $\varphi_y$ are associated to the charge and current fluctuations of the particles in the two legs and the density operator can be approximated with $n(x,y) \approx (k_0 - \partial_x \theta_y(x))/\pi$, whereas the current density $j_y$ is proportional to $\partial_x\varphi_y$. Finally, $\beta_p$ is a set of non-universal parameters, and, for our purposes, we will set $\beta_0=1$, $\beta_{\pm 2}=1/2$, whereas all the other $\beta$'s are set to 0. This choice is suitable to approximate hardcore bosons in the limit $\Omega,V_\perp \to 0$ (see Appendix \ref{app:bosonization} for more details).

The general effective interacting Hamiltonian is mapped into a two-species Luttinger liquid, which can be described in terms of two decoupled charge $c$ and spin $s$ sectors. In the following we adopt the standard notation $\varphi_{c/s}=(\varphi_\Up \pm \varphi_\Dn)/\sqrt{2}$ and $\theta_{c/s}=(\theta_\Up \pm \theta_\Dn)/\sqrt{2}$. We also introduce the parameters $u_q$ and $K_q$ (with $q\in\{c,s\}$) which represent the velocities and Luttinger parameters of the bosonic modes.
The latter encode the effects of the intraleg interactions in the system.
The interleg tunneling $\Omega$ and repulsion $V_\perp$, instead, determine the presence of additional interactions, which can be cast into a generalized sine-Gordon form. The effective bosonized Hamiltonian is derived in Appendix \ref{app:bosonization} and assumes the form:
\begin{multline}
    \MH
    =
    \sum_{q\in\{c,s\}}\int\frac{\rd x}{2\pi}\left[u_q K_{q}(\partial_x\varphi_{q})^2 + \frac{u_q}{K_q}(\partial_x\theta_{q})^2\right]
    \\
    +\int{\rd x}\left[h \mathcal{O}_h
    +g\left(\mathcal \re^{\ri(2k_0-\chi)x}O_g + {\rm H.c.}\right)\right]\,,
    \label{eq:LL_model}
\end{multline}
with:
\begin{align}
\mathcal O_h(x) &= \cos{2\sqrt{2}\theta_s}\,, \label{oh}\\
\mathcal O_g(x) &= \re^{\ri \sqrt2(\theta_c+\theta_s-\varphi_s)}+\re^{\ri \sqrt2(\theta_c-\theta_s-\varphi_s)} \,. \label{og}
\end{align}
See also Refs. \cite{Orignac2016,Citro2018} for an equivalent derivation. The operator $\mathcal{O}_g$ is, in general, fast-oscillating due to the phase $(2k_0-\chi)x$; therefore, it averages to a non-zero value only if its phase becomes position independent, thus the system approaches the resonance $2k_0 = \chi$, corresponding to the commensurate filling factor $\nu=1$.

The operator $\mathcal{O}_h$ matches the interaction adopted in Ref. \cite{vishwanath2020} to design intrinsically gapless symmetry-protected topological phases of matter. Analogously to these topological states of matter, $h>0$ for the repulsive bosonic model, as well as for its fermionic counterpart studied in Ref. \cite{Haller2018}.

In the Hamiltonian~\eqref{eq:LL_model} we neglected other fast-oscillating terms that are responsible for the onset of different pretopological states (see, for example, Refs. \cite{Cornfeld2015,Petrescu2015,Mazza2017}) but are irrelevant in proximity of $\nu=1$. We also considered systems with densities far from the Mott commensurate states \cite{Carr2006,Piraud2015}, therefore we also neglected further oscillating sine-Gordon interactions stemming from the onsite repulsion $U$. The main role of the onsite repulsion in our model is to affect the Luttinger parameters $K_q$.
Additional terms, less relevant in the renormalization group (RG) sense, have been neglected as well.

Finally, the values of the (bare) coupling constants $h$ and $g$ depend on $V_\perp$ and $\Omega$ respectively and can be approximated as:
\begin{equation}  \label{barehg}
h= \frac{k_0^2 V_\perp\beta_2^2}{2\pi^2}\,,\quad g = -\frac{\Omega\beta_2k_0}{2\pi^2}\,;
\end{equation}
see Appendix \ref{app:bosonization} for more details.

Differently from the pretopological Laughlin-like states \cite{Cornfeld2015,Petrescu2015,Mazza2017}, the physics of bosons at $\nu=1$ is dictated by the interplay of the non-commuting operators appearing in $\mathcal{O}_g$ and $\mathcal{O}_h$. The $g$-term, in particular, requires special attention: both operators in Eq.~\eqref{og} have the same amplitudes and scaling dimension, such that a simple semiclassical approach may not suffice for a clear understanding of the behavior of this system in the thermodynamic limit. To overcome this difficulty, we resort to a second-order RG analysis, in analogy with the fermionic ladder models at filling $\nu=1/2$ \cite{Haller2018}, and numerical simulations based on matrix product states (MPS).

\section{Renormalization Group analysis} \label{sec:RG}

A first insight of the possible thermodynamical phases and properties of the model is given by a simple scaling analysis of the interactions in the effective Hamiltonian~\eqref{eq:LL_model}.
The linear combination of fields appearing in the two terms of the $\mathcal{O}_g$ operator~\eqref{og} and in $\mathcal{O}_h$~\eqref{oh} do not commute. Therefore, in the ground states of the model, each of these three terms favors the minimization of different combinations of densities and currents which are not compatible with each other. Both the contributions in $\mathcal{O}_g$ are characterized by a scaling dimension $D_g = \left(K_c+K_s+K_s^{-1}\right)/2$ and, having the same coupling constant, none of the two can dominate over the other. The scaling dimension of $\mathcal{O}_h$ is instead $D_h = 2K_s$, such that this operator would be always irrelevant, in the RG sense, for repulsive contact interaction for which $K_s>1$.
The form of the Hamiltonian~\eqref{eq:LL_model} suggests the potential existence of three phases:

The first is a Luttinger phase in which both interactions are irrelevant. It corresponds to the vortex phase of the ladder, in which both the charge and spin sectors are gapless and all the two-point correlation functions decay algebraically.

The second phase is the phase dominated by the interaction $\mathcal{O}_h$. Hereafter, we will call it the $h$-phase. In this phase the spin sector is gapped and the charge sector is gapless. The $h$-phase, semiclassically, corresponds to the situation in which the field $\theta_s$ is pinned to one of the minima of $\mathcal{O}_h$. As a consequence, the pseudospin fluctuations and rung current on the ladder are suppressed in the bulk of the system. However, it is important to remark that, for finite systems with open boundary conditions, a rung current appears at the edges of the system (as depicted in Fig. \ref{fig:correlations} d below).  We emphasize that this phase is qualitatively different from the Meissner phase appearing at small fluxes since, in the Meissner phase, it is the rung current being ordered rather than the pseudospin density, corresponding to the field $\varphi_s$ being pinned to a semiclassical minimum.

The third phase is the one in which the operator $\mathcal{O}_g$ dominates, and we will refer to it as the $g$-phase. The study of this phase is less straightforward and its characteristics can be intuitively understood through a mean-field analysis based on the mapping into a Wess-Zumino-Witten (WZW) model (see, for example Ref. \cite{gogolinbook}) proposed in Refs. \cite{Chiofalo2015,Orignac2016,Citro2018}.
We sketch in the following the key elements to gain insight into this phase. At the mean-field level, we can separate the  $\mathcal{O}_g$ operator into the operators $o_c=e^{i\sqrt{2}\theta_c}$, acting on the charge sector, and $o_{s,\pm}=e^{i\sqrt{2}\left(\varphi_s\pm \theta_s\right)}$, acting on the spin sector. Assuming that the operators $o_{s,\pm}$ acquire a non-zero expectation value $\left\langle o_{s,\pm} \right\rangle$, the operator $o_c$ opens a gap in the charge sector and suppresses the fluctuations of the charge density. The situation is different in the spin sector: the operators $o_{s,\pm}$ can be mapped into the chiral current operators $J_R^y$ and $J_L^y$ of a WZW model describing the spin sector. With an additional mapping from this WZW model into a rotated Luttinger liquid Hamiltonian \cite{Orignac2016,Citro2018}, it is possible to show that these operators do not open a gap in the spin sector for small values of their coupling constants. This can be understood by considering that the current operator $J_R^y + J_L^y$ constitutes a perturbation proportional to $\partial_x\tilde{\varphi}_s$ in the rotated Luttinger liquid: such perturbation shifts the expectation value $\left\langle \partial_x\tilde{\varphi}_s\right\rangle$, thus it introduces incommensurability without opening a spin gap \cite{Orignac2016}.

In conclusion, mean-field arguments suggest that the $g$-phase is characterized by a gapped charge sector and a gapless spin sector. Such mean-field analysis neglects the interactions mixing the spin and charge sectors, and may provide only an approximate description of the $g$-phase.  This study was performed in Refs. \cite{Orignac2016,Citro2018} to investigate the appearance of a second incommensurability effect in the correlation functions of the system for values of $\Omega$ comparable with $t$, with ladders typically close or within the Meissner phase. We believe, however, that  a similar analysis can be extended also in our regime for $\Omega \ll t$ and systems with well-separated Meissner and $\nu=1$ resonances. We mention, that for contact interactions only we do not observe the onset of the $g$-phase in this regime.

\begin{figure}[t!]
    \centering
    \includegraphics[width=\columnwidth]{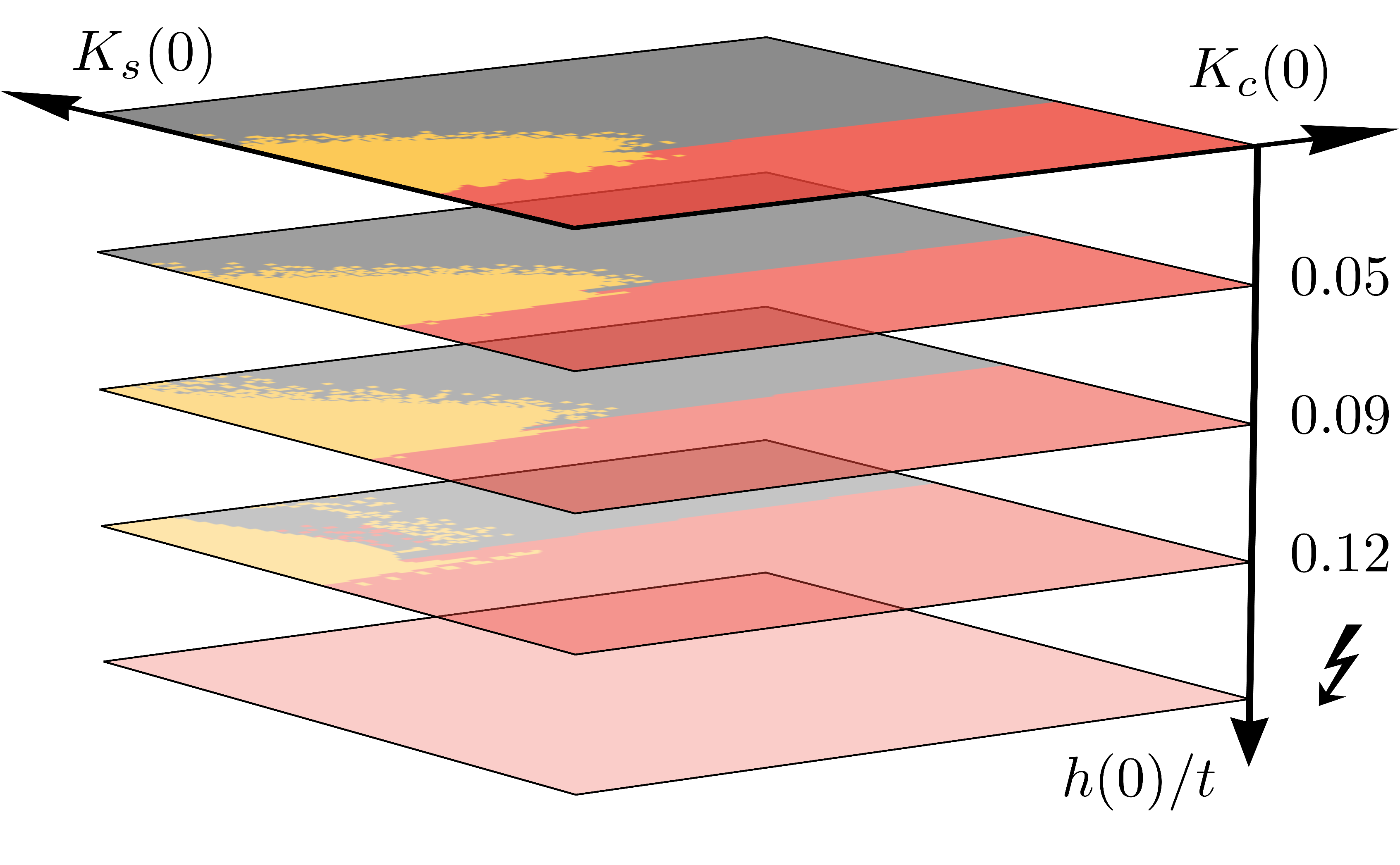}\llap{\parbox[b]{15cm}{\color{black}(a)\\\rule{0ex}{4cm}}}
    \includegraphics[width=\columnwidth]{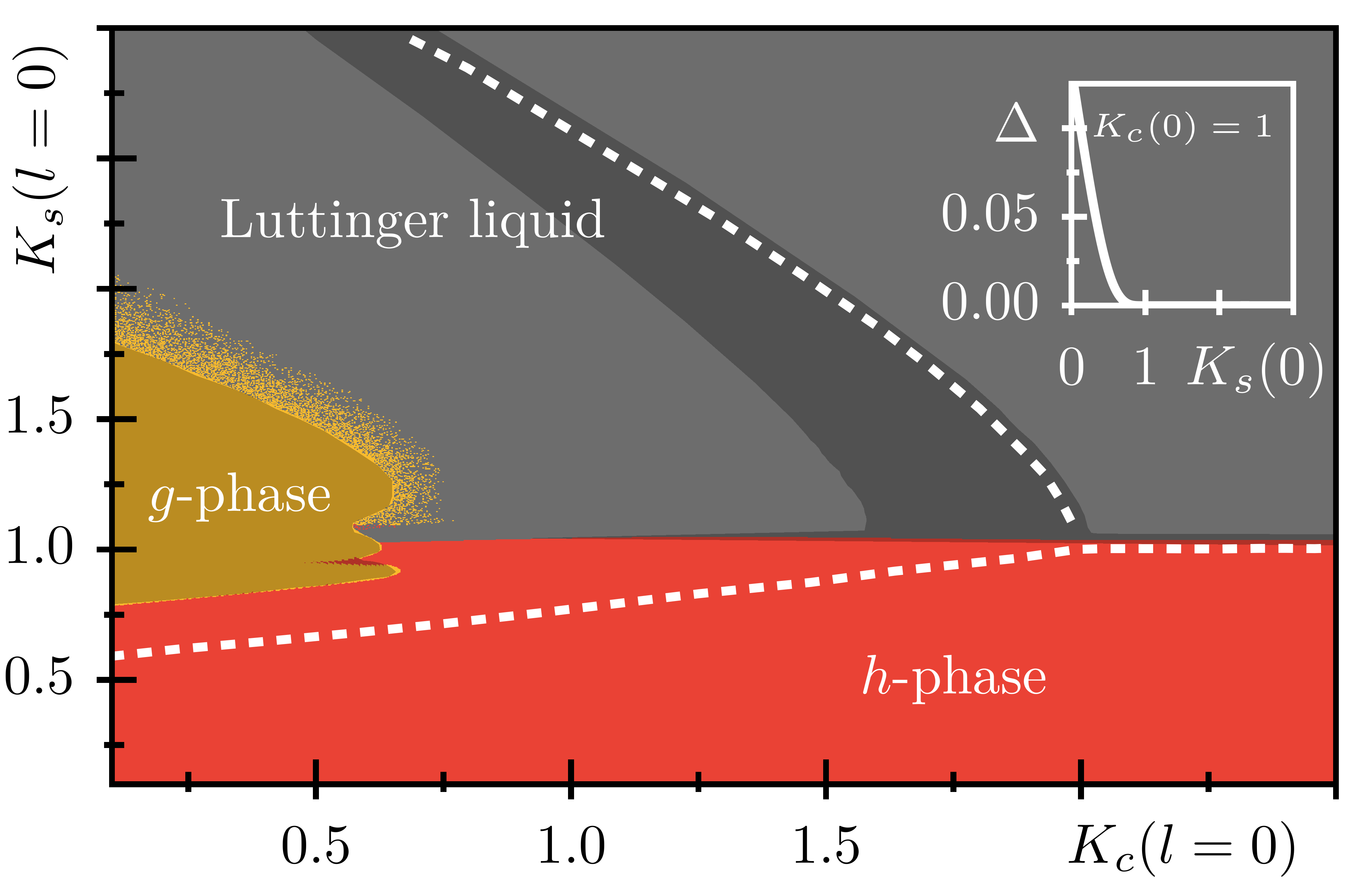}\llap{\parbox[b]{15cm}{\color{white}(b)\\\rule{0ex}{5cm}}}
    \caption{(a) Expected phases at $\chi/\pi=n$ for different values of the initial values $K_s$ and $K_c$ and $h/t$.
    Up to interactions $h(0)/t\approx0.12$ [$V_\perp\approx 7t$], the phase diagram maintains similar features, and we resort to a detailed discussion of the case $V_\perp=t$ presented in panel (b).
    The colored regions correspond to the numerical outcomes of the second order RG flow (see main text): the gray region depicts the Luttinger/vortex phase, the yellow region corresponds to the $g$-phase and the red region to the $h$-phase (shaded regions mark stiff or divergent solutions of the RG equations).
    The dotted lines signal the boundaries of the phases, obtained from a first-order scaling analysis: $D_{g/h}>2$ (Luttinger), $D_g<2,D_h$ ($g$-dominated), $D_h<2,D_g$ ($h$-dominated).
    From first to second order, the size of the $g$-phase shrinks significantly.
    The inset presents estimates of the gap $\Delta \propto  t^* e^{-l^*}$ of a vertical slice of the phase diagram at $K_c(0)=1$.
    }
    \label{fig:RG_flow}
\end{figure}

The scaling dimensions $D_h$ and $D_g$ allow us to obtain a simplified phase diagram as a function of the Luttinger parameters [dashed lines in Fig. \ref{fig:RG_flow} (b)]: the system is gapless for large values of $K_c$ and $K_s$ (Luttinger/vortex phase); the $h$-phase appears for small values of $K_s$, such that $\mathcal{O}_h$ is more and more relevant; the $g$-phase occupies instead a region for small values of $K_c$ and intermediate values of $K_s$.
This suggests the possibility of accessing such a peculiar phase within our setup with suitable interactions.

Given the complexity of the $g$-interactions, we apply a Wilsonian RG study of the Hamiltonian~\eqref{eq:LL_model} at second order in the interaction parameters $g$ and $h$ to obtain a more accurate phase diagram. In Appendix \ref{app:RG} we derive the following RG differential equations:
\begin{align}\label{eq:RG_flow}
    \frac{\rd h}{\rd l} &= h(2-D_h)+2g^2\left(\frac{K_c}{u_c}-\frac{K_s}{u_s}+\frac{1}{K_su_s}\right)\,,\nonumber\\
    \frac{\rd g}{\rd l} &= g(2-D_g)-hg\frac{K_s}{u_s}\,,\nonumber\\
    \frac{\rd K_c}{\rd l} &= 4\pi g^2\left[\frac{K_c}{u_c^2}+\left(K_s+K_s^{-1}\right)\frac{u_s^2+u_c^2}{2u_cu_s^3}\right]K_c^2 \,, \\
    \frac{\rd K_s}{\rd l} &= -4\pi\left[\frac{2h^2K_s^3}{u_s^2}\right. \nonumber \\
		&\qquad\left.+ g^2  \left(1-K_s^2\right)\left(\frac{u_c^2+u_s^2}{2u_su_c^3}K_c+\frac{K_s+K_s^{-1}}{u_s^2}\right)\right].\nonumber
\end{align}
Note that in the above we only consider the flow of couplings and Luttinger parameters, neglecting the flow of velocities. However, we checked that the velocities flow very slowly and thus yield only minor corrections to the phase diagram, justifying our assumption of constant $u_q$'s.

Before analyzing in more detail the solution of these four equations, let us observe that some limiting case can be easily extracted. Setting $g=0$ corresponds with the system away from the $\nu=1$ resonance; in this case only $h$ and $K_s$ flow and they follow the standard (second-order) behavior of the sine-Gordon model \cite{giamarchibook,gogolinbook} dictated by the Thouless equations. The operator $\mathcal{O}_h$ is relevant for $K_s < 1$, such that the system at $g=0$ flows in the $h$-dominated phase for any bare $K_s<1$; for bare $K_s>1$, the system will flow to the gapped $h$-phase for sufficiently large bare values of $h$.

This general behavior describes also what happens to the system for large values of the $K_c$ Luttinger parameter. For $K_c$, thus $D_g$ sufficiently large, the $g$ interaction is highly irrelevant and we expect to find the Luttinger phase for values of $K_s\gtrsim 1$ (if $h/t \ll 1$) and the gapped $h$-phase for $K_s<1$ [see the right side of Fig. \ref{fig:RG_flow} (b)].

Let us address in the following the main features of the phase diagram, that, in its full complexity, depends on the four bare parameters of $g,h,K_c$ and $K_s$.

In Fig. \ref{fig:RG_flow} (a) we present stacks of a restricted two-parameter phase diagram obtained from the numerical solution of the RG equations, as a function of the bare initial values of $K_c$ and $K_s$, and considering fixed values of the bare coupling constants in Eq.~\eqref{barehg} obtained by setting the non-universal parameter $\beta_2=1/2$ (see Appendixes \ref{app:bosonization} and \ref{App:nrg}).
The numerical results are obtained by setting $\Omega = 0.05t$ and varying values of $V_{\perp}/t=\{1,3,5,7,9\}$.
We observe that larger values of $\Omega$ would in principle enhance the extension of the $g$-phase. However, they would also considerably increase the extension of the Meissner phase, such that it is convenient to consider $\Omega/t \ll 1$ in order not to overshadow the properties of the system around $\nu=1$.
This is also consistent with the validity of the perturbation theory approach to the renormalization group analysis which relies on the bare parameters in Eq.~\eqref{barehg} being smaller than $t$.
Concerning the non-universal constant $\beta_2$, its value is expected to weakly depend on $\Omega$ and $V_\perp$ as in analogous systems \cite{Giamarchi2011,Imambekov2012}. The phase diagram displayed in Fig. \ref{fig:RG_flow} is qualitatively stable under small variations of this parameter. In general, a smaller value of $\beta_2$ increases the ratio of the bare coupling constants $g/h$ and favors the $g$-phase over the $h$-phase.

Our strategy to determine the phase diagram is to set an upper threshold $t^* \approx t$ and a lower threshold $\tilde \tau \approx  t/1000$ to the moduli of $g$ and $h$. If, during the flow, any of the two coupling constants increases beyond the threshold, we consider that the system reached a strong-coupling (partially gapped) phase at a value of the flow parameter $l^*\in\mathds R$
such that $g(l^*) \text{ or }h(l^*) =t^*\in\mathds R$. Depending on which coupling constant is dominating, we characterize the gapped phase as being of the $g$ or $h$ kind. In these cases, the gap of the system can be estimated through $\Delta \sim t^*\re^{-l^*}$. If instead, both the coupling constants drop below the lower threshold, we conclude that the system flows into the Luttinger/vortex phase. An example of the RG estimate of the gap is depicted in the inset of Fig. \ref{fig:RG_flow} (b) for the bare value $K_c=1$ and $K_s$ that varies between the gapped $h$-phase ($K_s<1$) and the gapless Luttinger phase ($K_s>1$), separated by the hard-core boson model with $V_\perp=0$, thus $K_c=K_s=1$.

This strategy, however, must take into account possible divergences of the Luttinger parameters: for certain ranges of the bare Luttinger parameters, $K_s$ diverges for a finite value of the flow parameter $l$ (see Appendix \ref{App:nrg} for further details). This happens, in particular, in regions where $g$ dominates over $h$, whereas $h$ tends to flow below $\tilde \tau$. When such a divergence occurs, we label the resulting phase based on the dominant coupling in proximity of the divergence. This corresponds to lowering the upper threshold $t^*$ to a value around $0.2 t$ for most of the $g$-phase and we represented this difficulty in Fig. \ref{fig:RG_flow} (b) by shading the regions where either divergences appears or the numerical solution converges slowly. The divergences of the Luttinger parameters suggest that the second order RG equations may not be sufficient to rigorously determine the phase diagram and the third order contributions should be considered as well.

Therefore, the resulting phase diagram in Fig. \ref{fig:RG_flow} (b) must be considered as a qualitative phase diagram valid for small values of the bare parameters in Eq.~\eqref{barehg}: we checked that its features are essentially stable for $V_\perp <5t$ (thus for a bare value \eqref{barehg} $h < 0.09t$). Above this threshold, non-physical features appear for large values of $K_s$  indicating that higher orders of the perturbation theory are not negligible in this regime.

The second-order results show that the $g$-phase is considerably smaller than what expected by the simple first-order scaling analysis, whereas the $h$ and Luttinger phases increase their extension. The behavior of the system close to the border between the $h$ and $g$ phase cannot be precisely determined: we cannot distinguish whether there is a direct phase transition between these two gapped phases or rather an extended Luttinger region that separates them.

The phase diagram in Fig. \ref{fig:RG_flow} (b) presents an overview of the possible phases at moderate values of the bare coupling constants $g$ and $h$. The point $K_s=K_c=1$ represents the ladder of hardcore bosons with $V_{\perp}=0$, which lies in proximity of the phase transition between the Luttinger and the $h$-phase. In such a scenario, $\mathcal{O}_h$ is indeed marginal and the role of the resonant $\mathcal{O}_g$ interaction becomes crucial in determining the thermodynamic behavior of the system. This is due to the second order term, proportional to $g^2$ which enhances the value of $h$ during the RG flow. Away from the resonance $\chi=2k_0$, $g$ averages to 0, and $h$ is suppressed for repulsive interactions, which yield $K_s>1$ (see Eq.~\eqref{eq:initial_conditions} below). At resonance, instead, the coupling $g$ not only can give rise to the $g$-phase, but it also strengthen the $h$-phase.

To gain a more realistic insight of the system we restrict our discussion to the case of hard-core bosons on the ladder geometry. Based on the mapping of the single leg subsystems into fermions (see Appendix \ref{app:bosonization}) we can estimate the bare values of the Luttinger parameters as:
\begin{equation}\label{eq:initial_conditions}
    K_{c/s} (0) = \left( 1 \pm \frac{V_\perp}{2\pi t \sin(k_0 a)}  \right)^{-1/2}  \,.
 \end{equation}
It is important to stress that these relations provide an estimate of the bare parameters in the perturbative limit $V_\perp \ll 2\pi t$ and $\Omega\ll t$, only. In this respect, we emphasize that there is an important distinction between $K_c$ and $K_s$. The spin sector is considerably influenced by the interaction $\mathcal{O}_h$, which tends to reduce the value of $K_s$ during the RG flow. This implies that, even in the vortex phase, when $\mathcal{O}_h$ is irrelevant, the physical value of $K_s$ displays major deviations from Eq. \eqref{eq:initial_conditions} which considerably overestimates it.
The discrepancy between the bare value of $K_s$ in \eqref{eq:initial_conditions} and the corresponding renormalized parameter is even more severe at the resonances. In the Meissner resonance, for example, the tunneling $\Omega$ gaps the spin sector and $K_s$ is supposed to diverge during the RG flow. At $\nu=1$, instead, the coupling constant $h$ can be further increased by the $g^2$ contribution and the flow of $K_s$ is modified by $g$ in a non-trivial way. For these reasons, Eq. \eqref{eq:initial_conditions} fails in predicting the physical value of $K_s$ for $\nu=1$ beyond the perturbative regime $V_\perp \ll 2\pi t$.

The charge sector, instead, is not affected by $\mathcal{O}_h$ and, considering the system outside the resonance or away from the $g$-phase, $K_c$ does not considerably change during the RG flow. Hence, differently from the spin sector, Eq. \eqref{eq:initial_conditions} provides a reasonable estimate for the physical value of $K_c$ and, indeed, it qualitatively agrees with the $K_c$ parameter extracted from the numerical tensor network simulation (see Fig. \ref{fig:LL_parameters} and the next section). Nonetheless, we expect major deviations for very strong interactions: in the limit $\Omega \to 0$ the model can be mapped into a spinless fermionic Hubbard model \cite{Cornfeld2015} and $K_c \to 1/2$ for $V_\perp \to \infty$. The lower limit of $K_c$ for large $V_\perp$, however, depends on $\Omega$ and it is not captured by our simple approximation, as showed by the comparison with the numerical results in Fig. \ref{fig:LL_parameters} (d).

When we consider a system with fixed $\Omega$, thus a fixed bare value of $g$, the phase diagram depends on the remaining three initial values of the RG flow. However, based on Eqs. \eqref{barehg} and \eqref{eq:initial_conditions} for the model of hard-core bosons, the bare parameters $K_c$, $K_s$, and $h$ are determined from the value of $V_\perp$ only.
This implies that when we vary $V_\perp$, the system defines a trajectory in this three-parameter phase diagram.
With respect to Fig. \ref{fig:RG_flow} (b), we observe that for larger values of $V_\perp$ the $g$-phase is shifted towards larger values of $K_s$.


The results from Eqs. (\ref{barehg},\ref{eq:RG_flow}) and~\eqref{eq:initial_conditions} predict an evolution of the system from the gapless Luttinger phase towards the gapped phases with increasing $V_\perp$. We emphasize, however, that this perturbative RG analysis provides accurate results only for $h < t$, thus for sufficiently small $V_\perp k_0^2$.
It is thus unclear whether the physical trajectory for non-perturbative interactions evolves inside the $g$, or inside the $h$-dominated phase.
Therefore, in the next sections we complement the previous RG predictions with the MPS results bridging the gap into the non-perturbative regime.

\section{Numerical Results} \label{sec:MPS}

To obtain more quantitative predictions about the behavior of the model, we simulated the ground state of Hamiltonian~\eqref{hamtot} for hard-core bosons and system sizes on the order of typical experiments, i.e. $L=64$.
The simulations were performed through density matrix renormalization group, adopting a matrix product state (MPS) ansatz with open boundary conditions and bond dimensions varying up to $512$ different states being kept in the approximation of the reduced density matrix. This leads to a discarded probability of $\Delta\rho<10^{-8}$, enough to assume converged values roughly up to the eighth digit.

In the following we present the main features of the system, including the estimates of the chiral current its main fluctuations, correlation functions and an overview of some dynamical properties.

\subsection{Chiral current}

\begin{figure}[th]
    \center
    \includegraphics{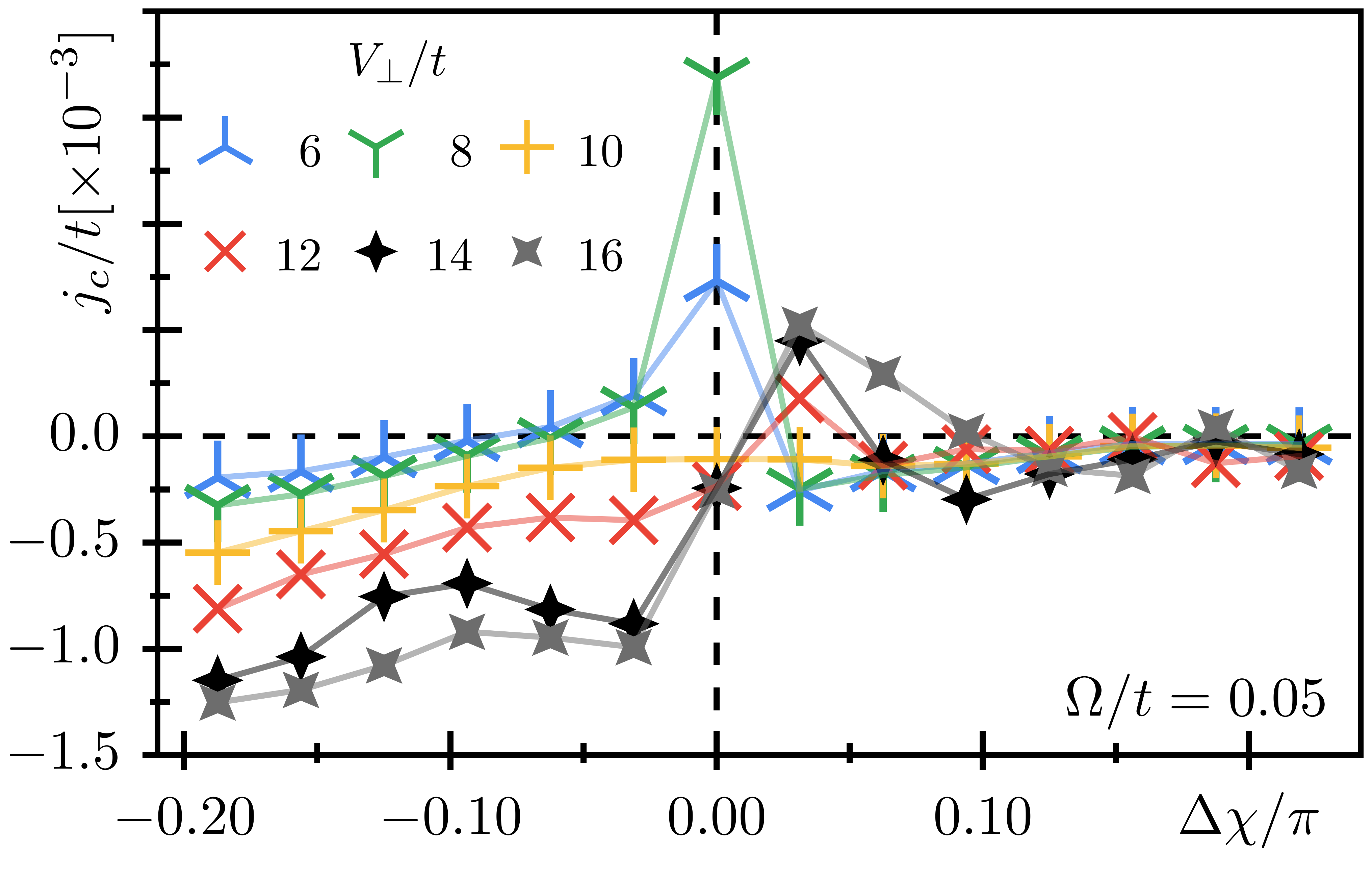}
    \caption{Chiral current $j_c$ as a function of the magnetic flux $\Delta\chi/\pi=\chi/\pi-n$ for different interaction strengths $V_\perp/t$ for $N=48$, $L=64$ and $\Omega=0.05t$. For large interactions, the spin sector is gapped at integer filling factors imposing the condition $\chi/\pi=n$. As a consequence, we observe a double cusp-signature of the chiral current.}
    \label{fig:current}
\end{figure}

\begin{figure}[th]
    \center
    \includegraphics{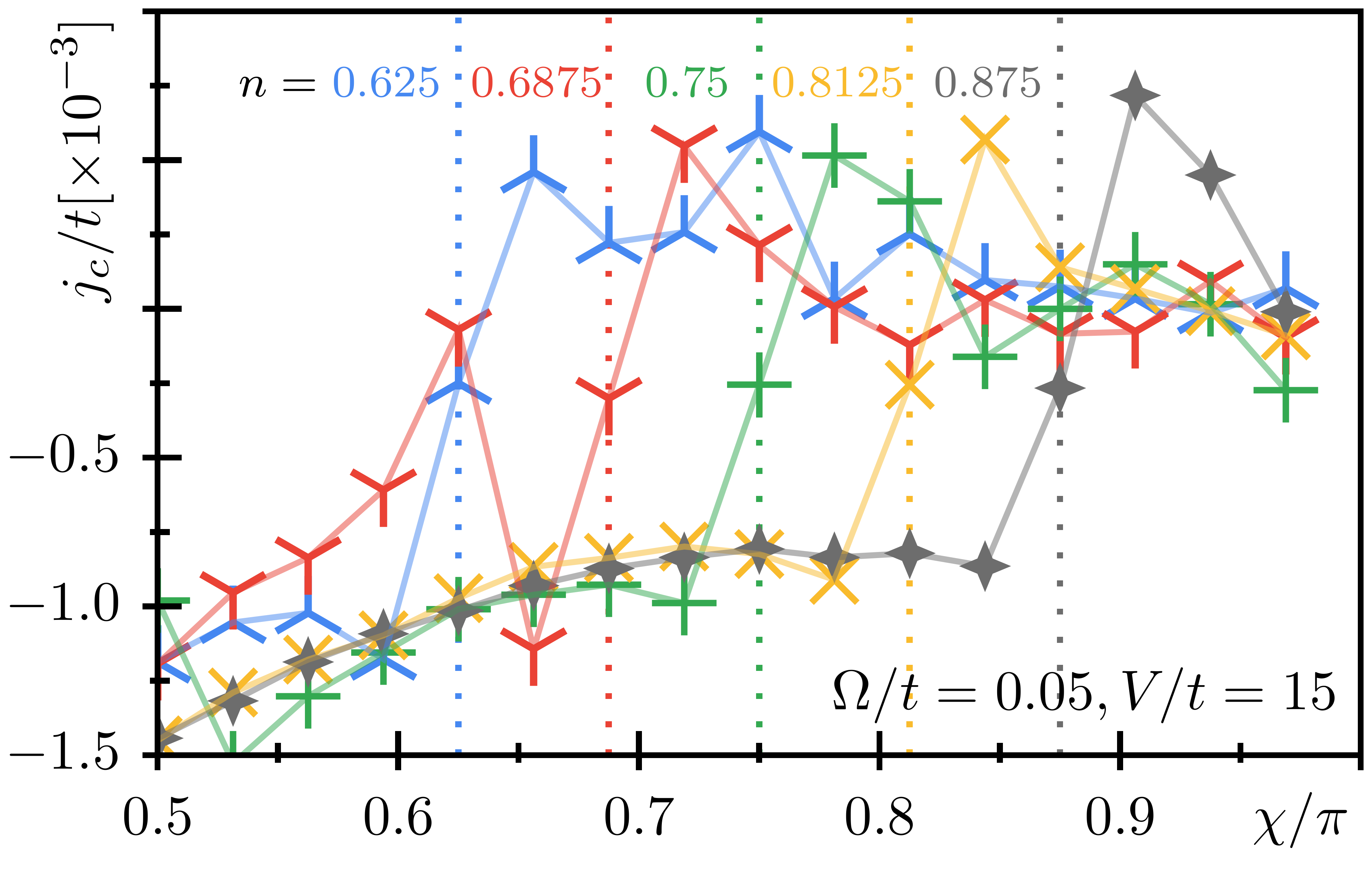}
    \caption{Chiral current $j_c$ as a function of the magnetic flux $\chi/\pi$ for different average densities $n=N/L$ (different colors and symbols) of the system at $V_\perp/t=15$ and $\Omega/t=0.05$. The position of integer filling factor imposes the commensuration $\chi/\pi=n$ and thus shifts the double-cusp signature in the $\chi$-axis.}
    \label{fig:current2}
\end{figure}

We focus on values of the density of hard-core bosons and flux per plaquette in proximity to the $\nu=1$ (for example $N=48$ and $\chi=3\pi/4$ for $L=64$) and we vary the interactions $V_\perp$ and the flux $\chi$. In particular, we vary the flux in units of $2\pi/L$ because all the observables in the open system display strong and regular oscillations when continuously varying $\chi$. The chosen discretization corresponds to the standard unit quantum of momentum for translationally invariant chains. We observe however that other choices are equally well justified: similar results are obtained by slightly larger/smaller flux variations $2\pi/(L\pm1)$ which respectively correspond to the quantum of momentum and the flux quantum in ladders with Dirichlet boundary conditions.

The chiral current $j_c$ is a crucial observable to detect the onset of chiral phases of the ladder.
In the case of non-interacting bosonic systems, it shows the typical linear dependence on the flux in the one-dimensional analog of the Meissner phase followed by a shark-fin transition to a vortex phase \cite{atala2014,Orignac2001}, of which we still spot the decaying tail on the left sides of Figs.~\ref{fig:current} and \ref{fig:current2}.
The corresponding operator reads:
\begin{equation}
j_c = i\frac{t}{2L} \sum_x \left(e^{i\frac{\chi}{2}}b^\dag_{x,\Up}b_{x+1,\Up} - e^{-i\frac{\chi}{2}}b^\dag_{x,\Dn}b_{x+1,\Dn}\right) + \hc\,.
\end{equation}
Here the chiral current is defined based on the gauge choice in Eq.~\eqref{eq:kinetics} and we consider its average over the whole chain length.

Fig. \ref{fig:current} depicts $j_c$ for $N=48$ as a function of the flux displacement around the resonance at filling factor $\nu=1$.
For values of $V_\perp < 6t$ (not shown) the chiral current does not display any discontinuity with respect to the vortex phase.
This is compatible with the system being in a Luttinger liquid phase, consistently with the second-order RG results close to $K_c=K_s=1$ (see Fig. \ref{fig:RG_flow}).
When increasing $V_\perp$ in the approximate range $(6t,8t)$ the system develops a high peak in the chiral current exactly at the resonance with filling factor $\nu=1$ (blue and green curves).
We interpret this peak as a feature due to the system evolving towards the critical region at the edge between the $h$ and $g$ phases.
By increasing further the interaction, at $V_\perp=10t$, the current signature flattens again.
Finally, for $10t < V_\perp < 18t$ the chiral current develops a double cusp pattern, which is a hall mark of the two commensurate-incommensurate phase transitions \cite{giamarchibook,Nersesyan1978,Pokrovsky1979} that separate a (partially) gapped commensurate and chiral phase at the resonance (the $h$ phase) from the truly gapless vortex phase that dominates when the flux $\chi$ is sufficiently displaced from the resonant point.

We observe that, for the system sizes we analyzed ($L=64$), the chiral current cusps are considerably more evident than the analogous cusps appearing for the pretopological $\nu=1/2$ Laughlin-like states at $\chi=4k_0$ and $\chi=2\pi - 4k_0$ (the latter being the particle-hole symmetric state of the Laughlin-like state).
This is in agreement with previous works \cite{Cornfeld2015,Mazza2017,Petrescu2017}, in which the fully resolved Laughlin-like gaps are presented for both larger interactions and larger system sizes.
Contrary to the weak and ambiguous signal in the proximity of $\nu=1/2$, we consistently observe three aligned $j_c$ points across the resonance (see Figures \ref{fig:current} and \ref{fig:current2}), which indicate a sizable gap.
This allows us to conclude that the main signatures of the $\nu=1$ states are more feasible from an experimental viewpoint.
\\

The behavior of the chiral current in the $h$ phase can be deduced by the relation:
\begin{equation}
    j_\chi = -\frac1L\frac{\partial E_{\rm GS}(\chi)}{\partial\chi}
\end{equation}
in which $E_{\rm GS}$ denotes the ground state energy.
If the interaction associated with amplitude $h$ is the dominating term, a gap in the spin sector is formed and the chiral current follows a linear behavior with respect the flux $\chi$ between the two cusps and is predicted to cross $0$ exactly at the resonance.
This is true provided there are no background effects such as the long decaying tail from the strong shark-fin signal at small fluxes.
In reality, the resonance at $\nu=1$ is thus established on top of a slowly decaying background current which is a remnant of the vortex phase.

We verified that, for different particle numbers, the position of the two cusps of the chiral current shift accordingly to the resonance following the relation $\chi=2k_0$ (see Fig. \ref{fig:current2}): the signature of a chiral phase at integer resonance appears indeed for a broad range of particle numbers and its features are visible as long as the $\nu=1$ resonance is sufficiently separated from the Meissner phase of the system.

We point out that, in our simulations, the typical double cusp pattern of the resonant state is mostly evident for rung interactions below $V_\perp=18t$ when $\Omega=0.05t$.
For larger values of $V_\perp$ the Meissner and melted-vortex states dominate more than half of the phase diagram accessible through tuning the flux, ultimately resulting in a poor resolution of the $\nu=1$ signal.

\subsection{Fluctuations and estimate of the Luttinger parameters}

To better compare the numerical results with the RG predictions, we must locate the state of the system as a function of $V_\perp$ in the phase diagram obtained by the RG equations (see the colored dots in Fig. \ref{fig:RG_flow}). To this purpose we must estimate the Luttinger parameters $K_c$ and $K_s$ from the DMRG results.
The Luttinger parameters determine most of the features of the ground state of the system, including its spin and charge fluctuations, the decay of its two-point correlation functions and even the dynamics of its excitations.

The first estimates we derive for the Luttinger parameters are obtained by fitting the bipartite charge and spin fluctuations~\cite{Song2012}.
In particular, we define
\begin{equation}
    \BF^{c/s}(\ell) = \langle [N^{c/s}(\ell)]^2\rangle_{\rm conn.}
\end{equation}
where we introduced the total number of particles / the total magnetization $N^{c/s}$ in a bipartition of size $\ell$.
In bosonization, the leading order of these fluctuations read
\begin{equation} \label{fluct1}
    \BF^{c/s}(\ell) = \frac2{\pi^2}\langle\left[\DF_{c/s}(\ell)-\DF_{c/s}(0)\right]^2\rangle
\end{equation}
and depend thus on the correlations of the density fields $\DF_{c/s}$.
If the corresponding sector is gapless, bipartite fluctuations follow a logarithmic dependence (see Appendix \ref{App:corr}):
\begin{equation}\label{fluct2}
    \BF^{q}(\ell) \propto \frac{K_{q}}{\pi^2}\ln[d(\ell|L)] + f_q \,.
\end{equation}
In the above, $d$ denotes the chord distance of open boundary systems which is defined as
\begin{equation}
    d(\ell|L) = \frac{L}\pi|\sin(\pi \ell/L)|.
\end{equation}
The function $f_{c/s}$ accounts for the total charge/spin fluctuation of the system, which is related to the $U(1)$ gauge transformations in the corresponding sectors: if the total charge/magnetization is conserved, $f_{c/s}=\varepsilon$ (small constant); otherwise, $f_{c/s}$ can be approximated by a linear contribution in the bipartition length, $f_{c/s}=\gamma + f_{0}\ell$, which equally distributes the total fluctuations on each site.

\begin{figure}[t]
    \center
    \includegraphics[width=0.4944\columnwidth]{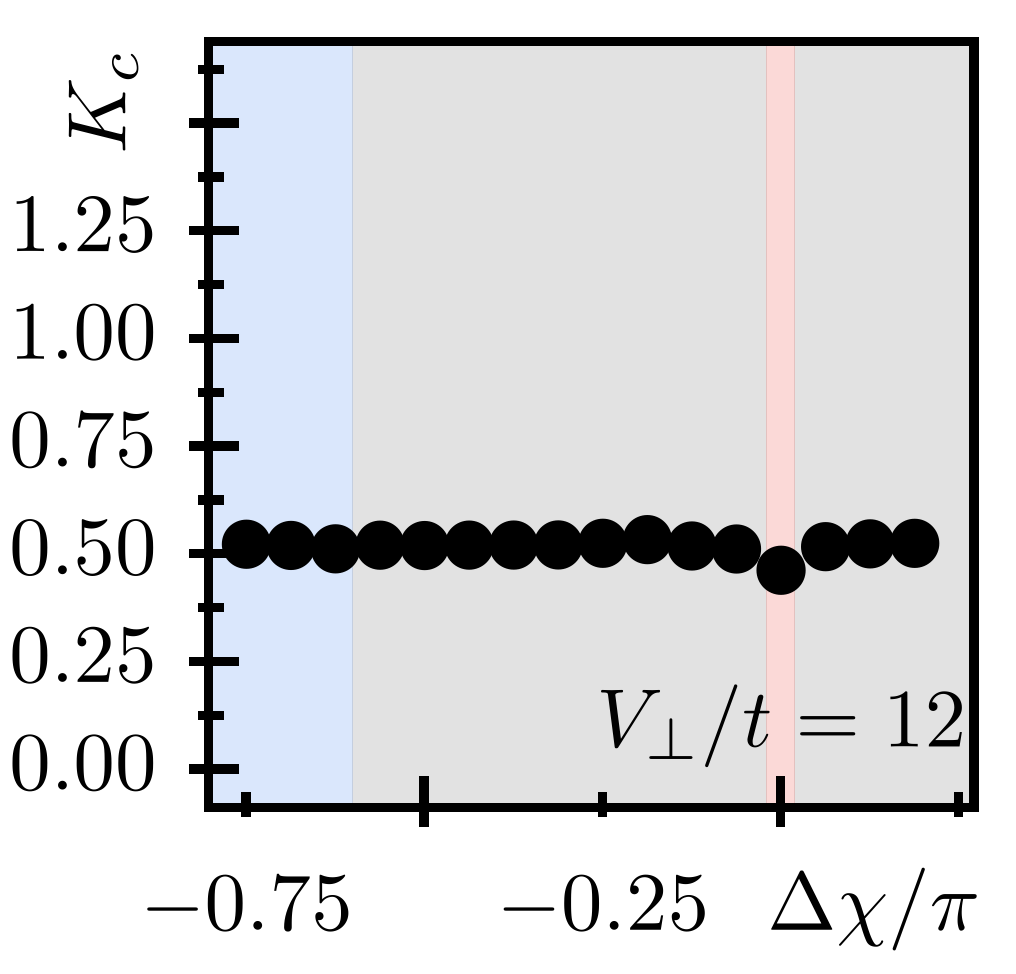}\llap{\parbox[b]{6cm}{\color{black}(a)\\\rule{0ex}{3.32cm}}}
    \hfill
    \includegraphics[width=0.4944\columnwidth]{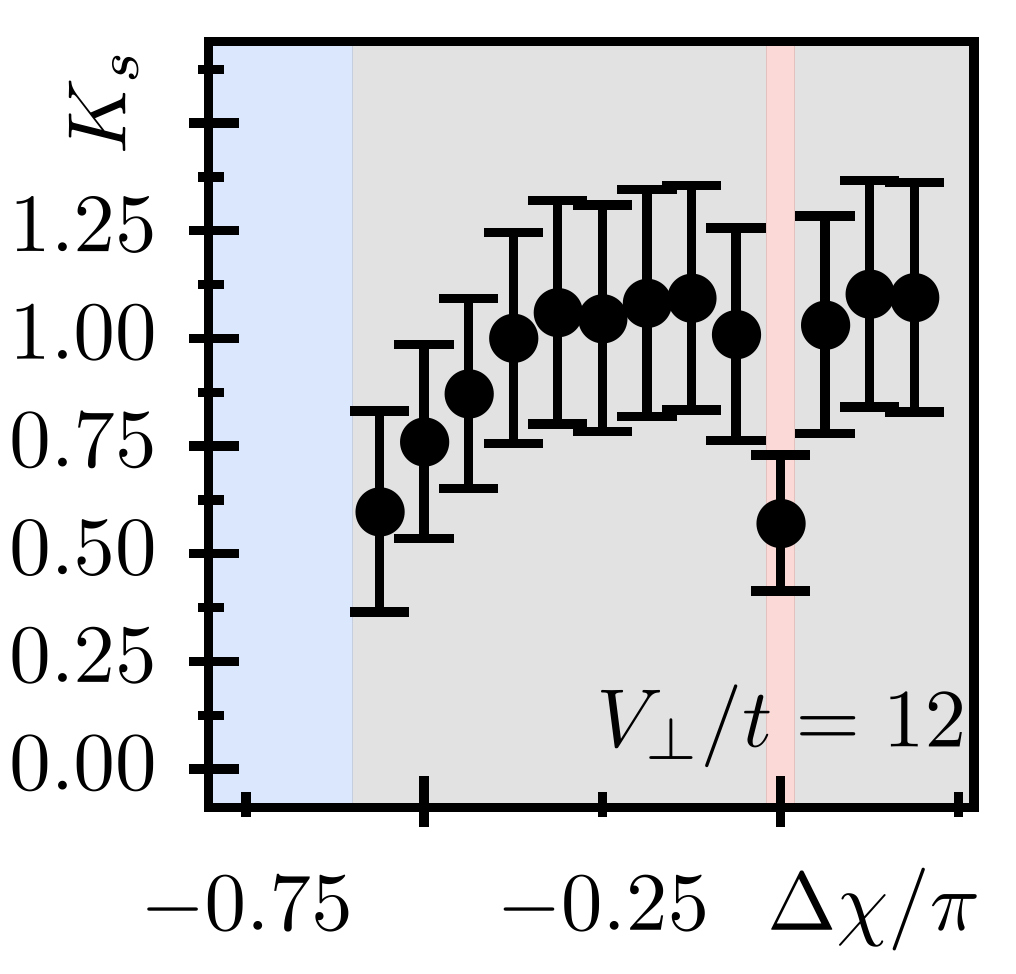}\llap{\parbox[b]{6cm}{\color{black}(b)\\\rule{0ex}{3.32cm}}}
    \includegraphics[width=0.4944\columnwidth]{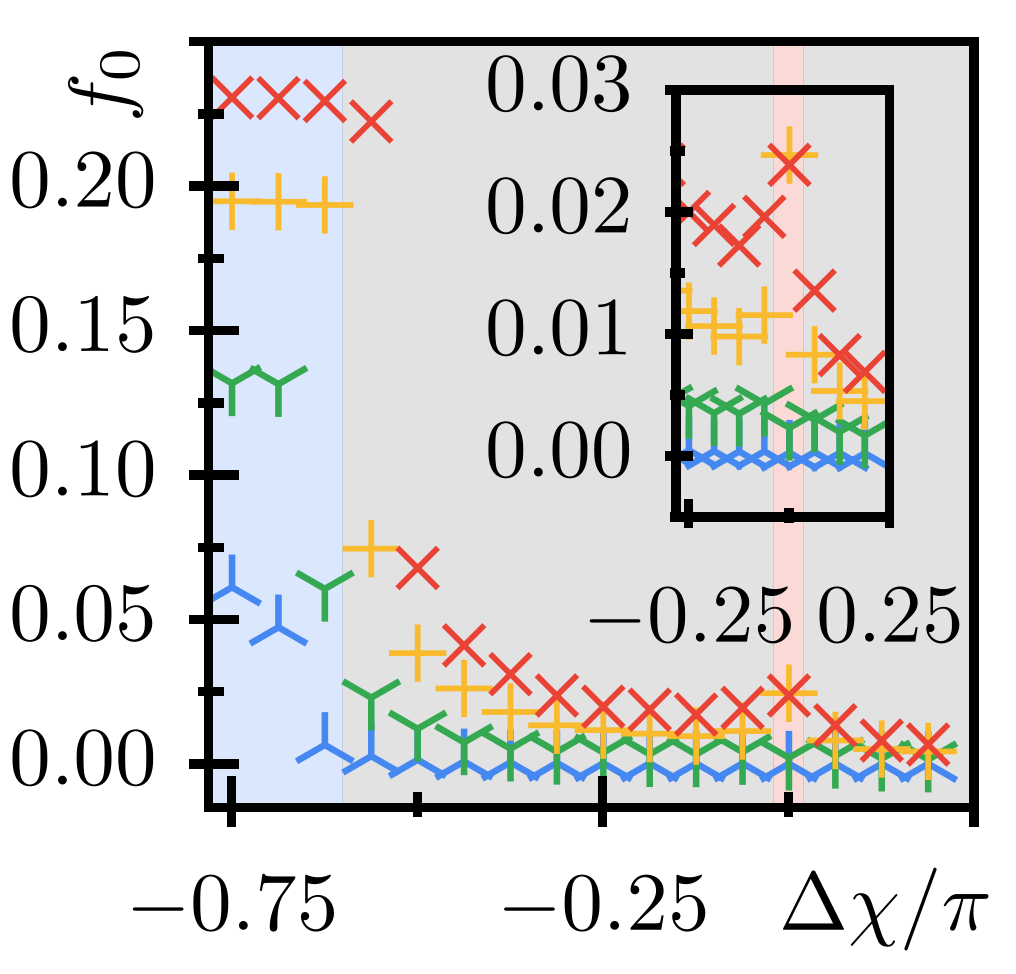}\llap{\parbox[b]{6cm}{\color{black}(c)\\\rule{0ex}{0.8cm}}}
    \hfill
    \includegraphics[width=0.4944\columnwidth]{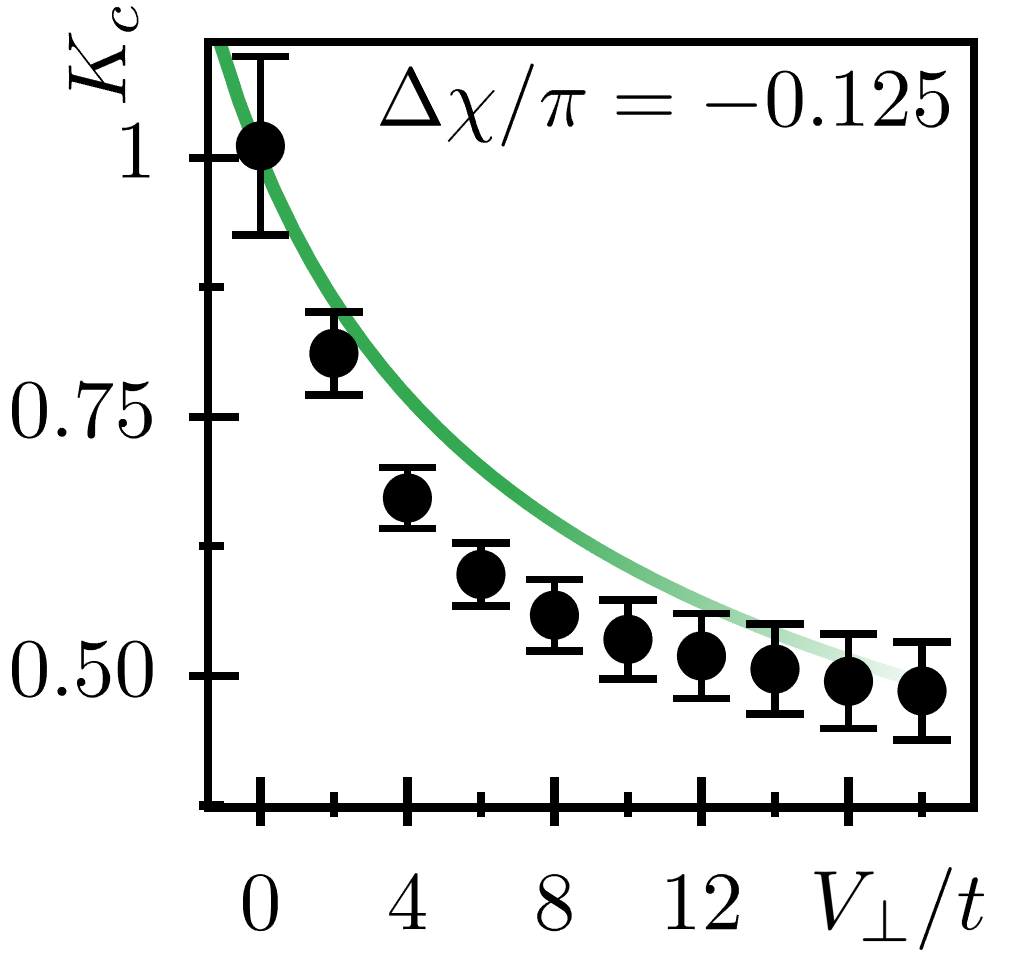}\llap{\parbox[b]{6cm}{\color{black}(d)\\\rule{0ex}{0.8cm}}}
    \includegraphics[width=0.4944\columnwidth]{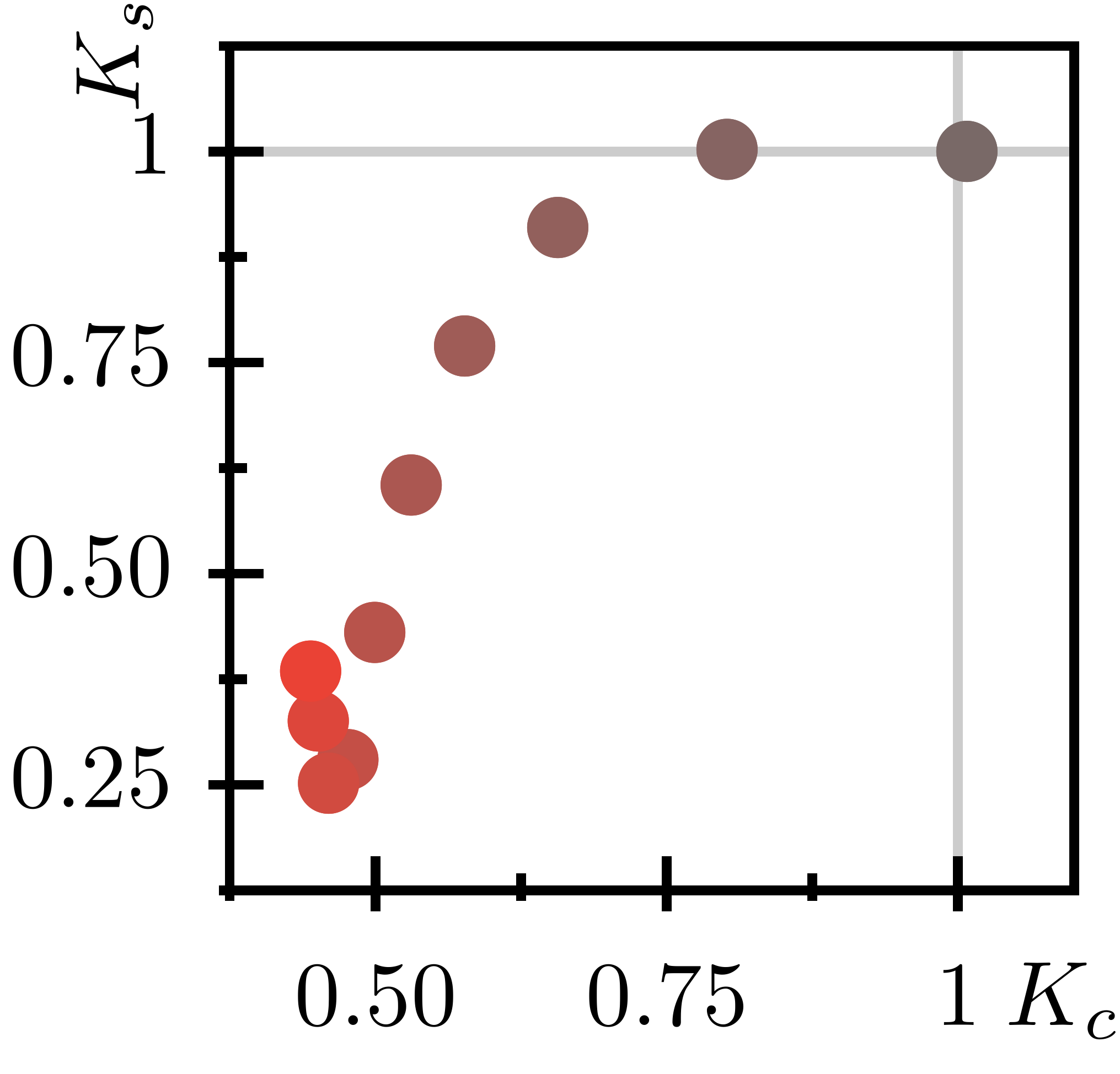}\llap{\parbox[b]{6cm}{\color{black}(e)\\\rule{0ex}{3.32cm}}}
    \hfill
    \includegraphics[width=0.4944\columnwidth]{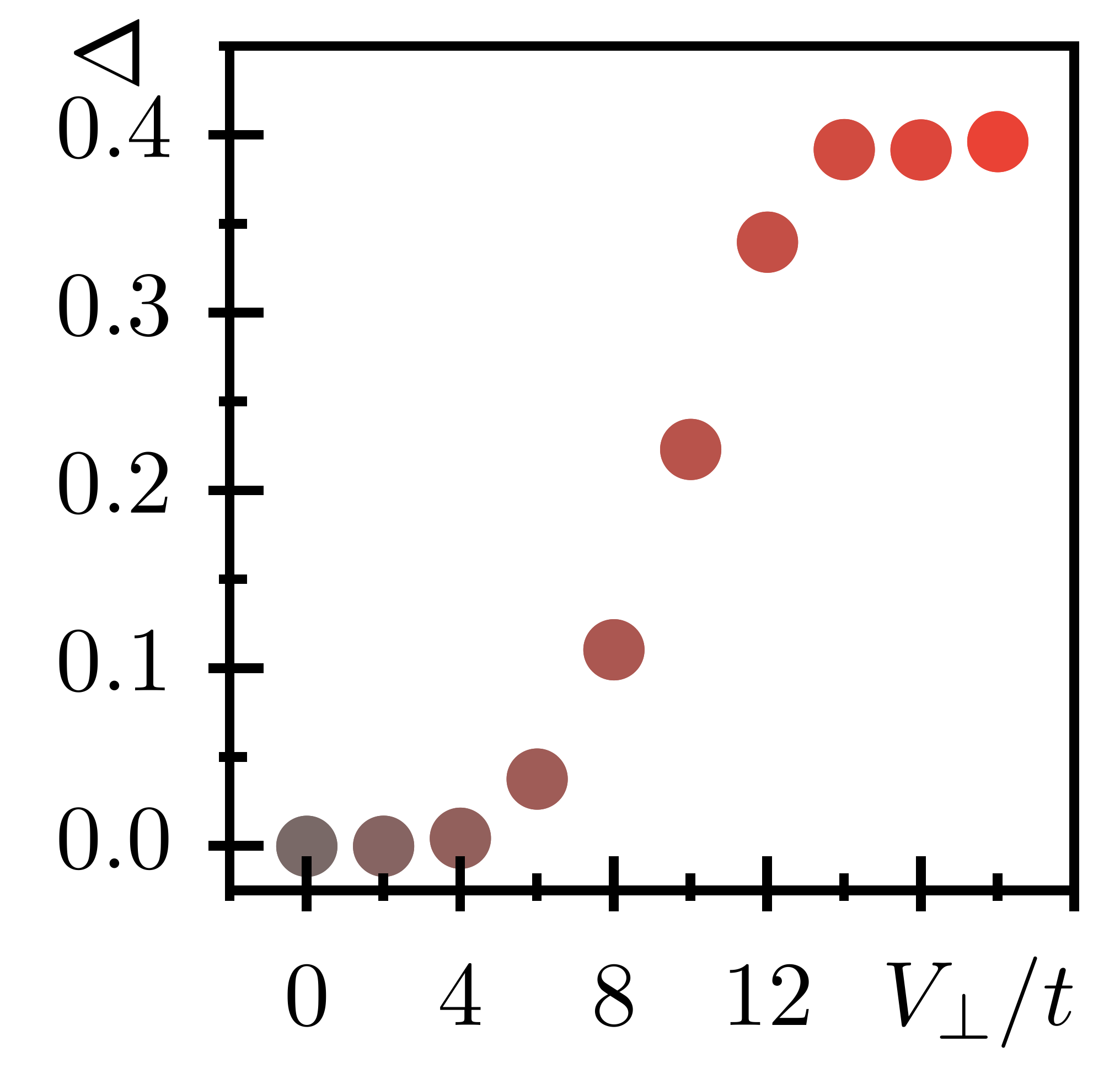}\llap{\parbox[b]{6cm}{\color{black}(f)\\\rule{0ex}{3.32cm}}}
    \caption{Luttinger liquid parameters for hard core bosons, with $\Omega/t=0.05$ and $n=N/L=48/64$. (a/b) $K_{c/s}$ for $V_\perp/t=12$, fitted from the fluctuations of the total density / magnetization $N_{c/s}(\ell)$. We clearly see the pinning of the spin fields in the Meissner phase (shaded in blue), and in the resonance at $\chi/\pi=n$ (shaded in red).
    The pinning of the spin field is further confirmed by fits of the constant $f_0$ (c).
    Different colors represent different interactions $V_\perp/t\in\{0\ \text{(blue), 6\ \text{(green)}, 12\ \text{(yellow)}, 16\ \text{(red)}}\}$. (d) Typical $K_c$ for bosons in the vortex phase as a function of the interaction $V_\perp$. The green curve depicts the approximation in Eq. \eqref{eq:initial_conditions}.
(e) The physical trajectory explored by the MPS simulations. As the color changes from gray to red, the interaction $V_\perp$ is increased and the system evolves from the Luttinger liquid to the $h$-phase. (f) RG-estimates of the energy gap calculated by using the extracted values of $K_s$ and $K_c$ of our MPS simulations, presented in panel (e). The so-obtained estimates predict a transition from Luttinger liquid to $h$-phase with sizable energy gap in the thermodynamic limit.
    }
    \label{fig:LL_parameters}
\end{figure}

From the fits of the DMRG results we derive that the value of $K_c$ remains essentially constant as a function of the flux $\chi$ [see Fig. \ref{fig:LL_parameters} (a)]. This is the expected behavior across the Meissner-vortex phase, since the Meissner phase presents a gap in the spin sector only. When considering the integer resonance at $\chi=2k_0$, the independence of $K_c$ from the flux $\chi$ suggests that also in this case the physics at the resonance is not determined by the charge sector.
The situation is the opposite for the spin fluctuations. In the Meissner phase the spin fluctuations are maximized, the linear term given by $f_s$ in Eq. \eqref{fluct2} is completely dominating over the logarithmic contribution and $K_s$ ceases to be meaningful in this gapped phase (its estimate based on Eq. \eqref{fluct2} is systematically wrong). For large values of the fluxes in the vortex phase, $K_s\gtrsim 1$, consistently with having repulsive rung interaction. Finally, exactly at the resonance, we can observe a dip in the fitted value of $K_s$ [see Fig. \ref{fig:LL_parameters} (b)], which clearly shows that the spin fluctuations are suppressed at the resonance, compatible with the formation of a new gap in the spin sector. This is the first indication that, at $\nu=1$ and sufficiently large $V_\perp$, the system enters the $h$-phase, whereas the $g$-phase, expected to have a gapped charge sector, is not reached.

Interestingly, we find also a signature of the pinned spin sector by fitting the constant $f_{0}$ in the fluctuations of the total magnetization: it is large in the Meissner and small, but significantly increasing with respect to the vortex environment [see the inset in Fig. \ref{fig:LL_parameters} (c)], inside the resonant state at filling $n=\chi/\pi$, thus indicating relevant corrections to the logarithmic law.

Considering the system exactly at the resonance, we can therefore fit the values of the Luttinger parameters as a function of $V_\perp$ based on Eq.~\eqref{fluct2}. The results are used to define the dots of the physical trajectory presented in Fig. \ref{fig:LL_parameters} (e). The values of $V_\perp$ and of the fitted Luttinger parameters can then be adopted as boundary conditions of the RG flow determined by the equations~\eqref{eq:RG_flow}. The numerical solutions of these flow equations allows for an estimate of the gap of the system based on the final value of the flow parameter: $\Delta \approx t^* e^{-l^*}$ (see the previous section). The so-obtained gap is represented in Fig. \ref{fig:LL_parameters} (f) as a function of $V_\perp$ and it is in striking agreement with the previous analysis of the measured chiral current: the gap is negligible for $V_\perp < 6t$ and it opens in correspondence of the first peak of the chiral current at $V_\perp \sim 6t,8t$.

The transition from a gapless to a gapped region around $V_\perp \sim 6t$ is also consistent with the solution of the RG flow equations.
By comparing the fitted values of the Luttinger parameters to the phase diagram in Fig. \ref{fig:RG_flow}, we observe that the system at $V_\perp =0$ lies in proximity of the edge between the  Luttinger liquid and the $h$-phase and it evolves close to the boundary between Luttinger liquid, $h$ and $g$ phases for intermediate values of $V_\perp$. Beyond $V_\perp\gtrsim 6t$ it enters the $h-$phase. For stronger interactions, the predicted gap $\Delta$ remains sizable, consistently with the system entering deep within the $h$-phase.
The observed plateau of the energy gap after $V_\perp/t\sim12$ in Fig. \ref{fig:LL_parameters} (f) is likely a consequence of the (partial) breakdown of the second order RG calculation. Nonetheless, all the signatures observed in the numerical simulations are consistent with the onset of a gap in the spin sector.

We emphasize that our numerical results and the estimate of the spin correlation functions (see the next subsection) suggest a tiny physical gap, such that the related correlation length $\xi_s$ is comparable with a large portion of the system size ($L=64$). This implies a small non-universal constant which should enter as a multiplicative factor in the estimate of the gap presented in the inset of Fig. \ref{fig:LL_parameters} (f). As such, the inset describes the general trend of the gap in arbitrary units.

\begin{figure}[t!]
    \center
    \includegraphics{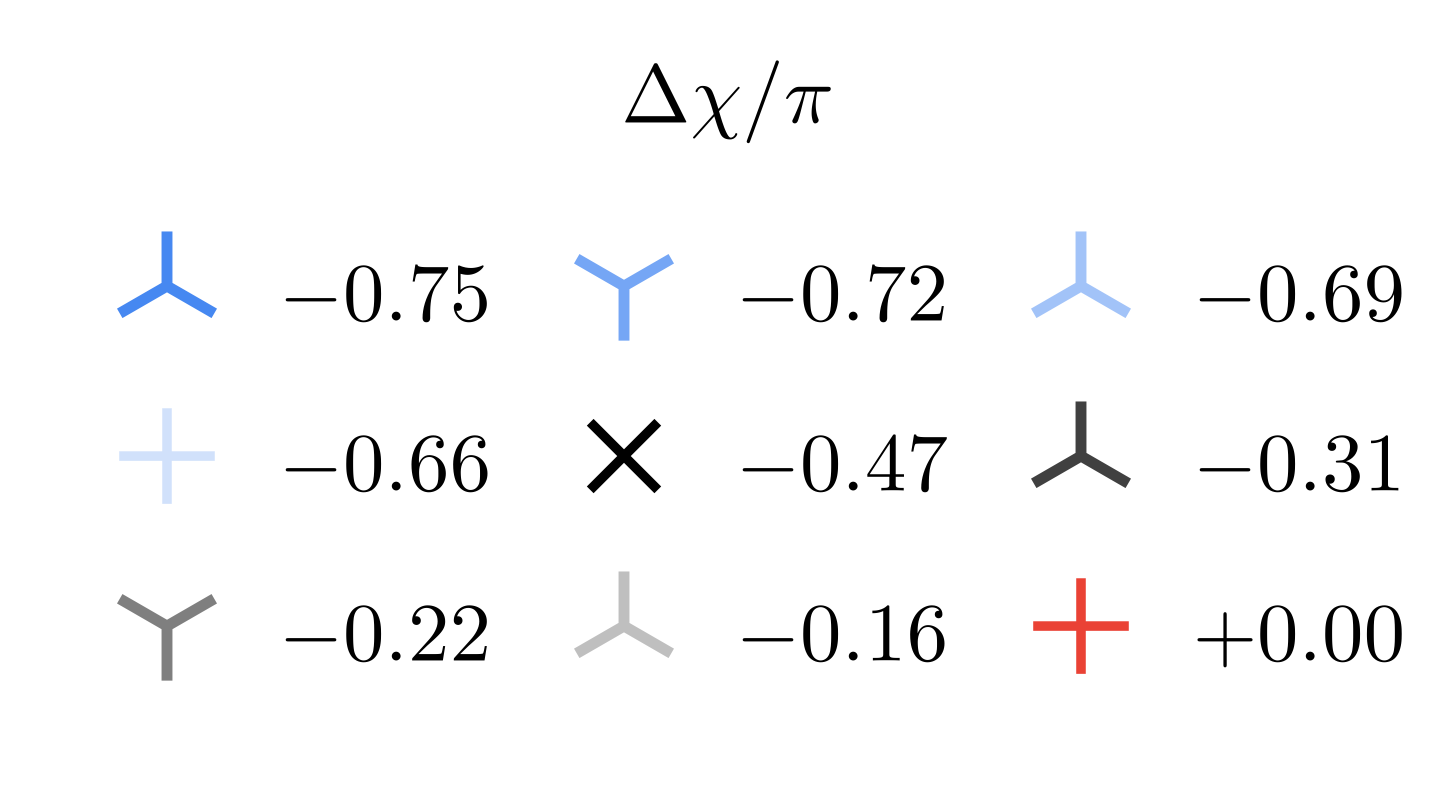}
    \includegraphics[width=0.4944\columnwidth]{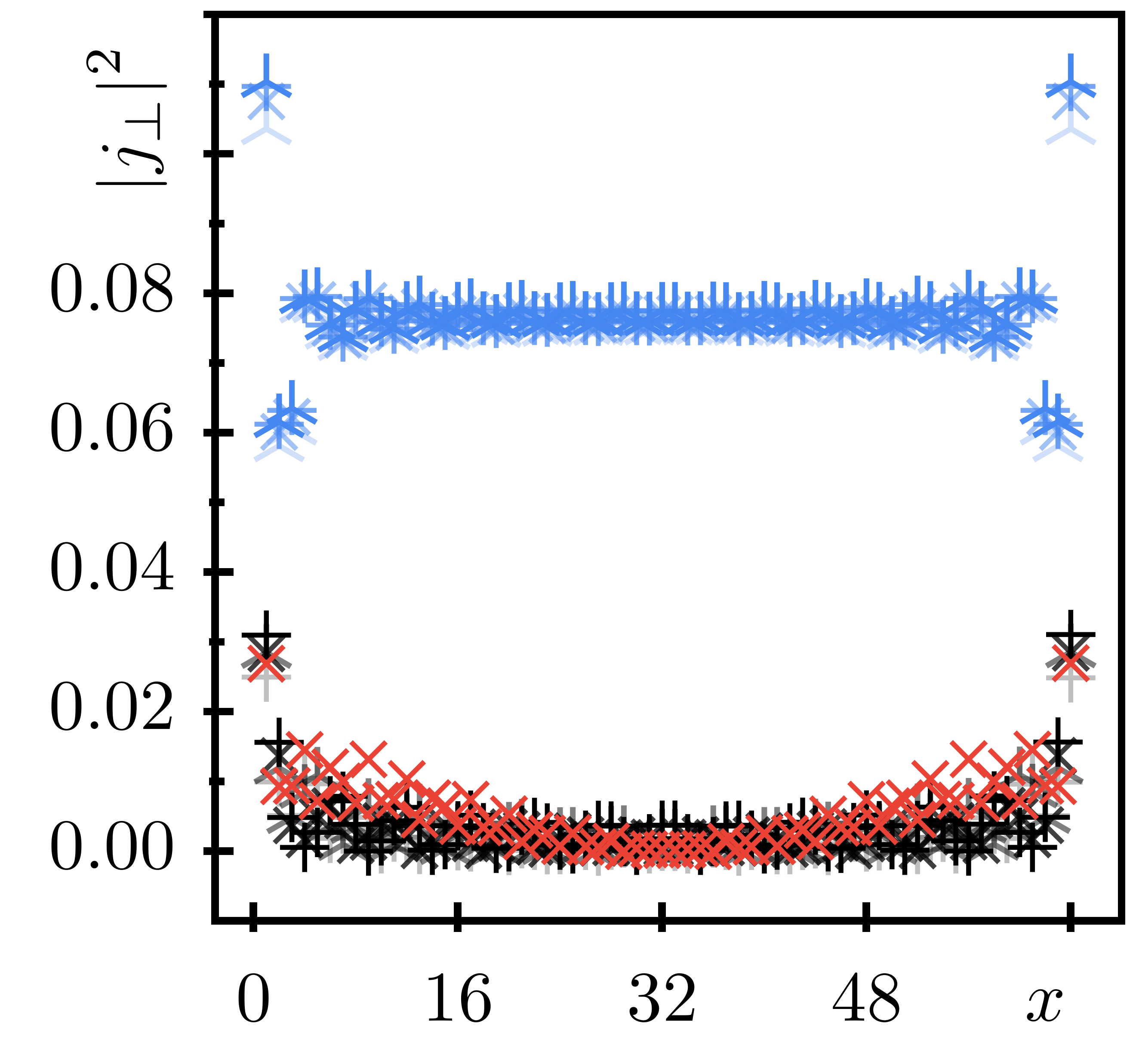}\llap{\parbox[b]{5.5cm}{\color{black}(a)\\\rule{0ex}{1.32cm}}}
    \includegraphics[width=0.4944\columnwidth]{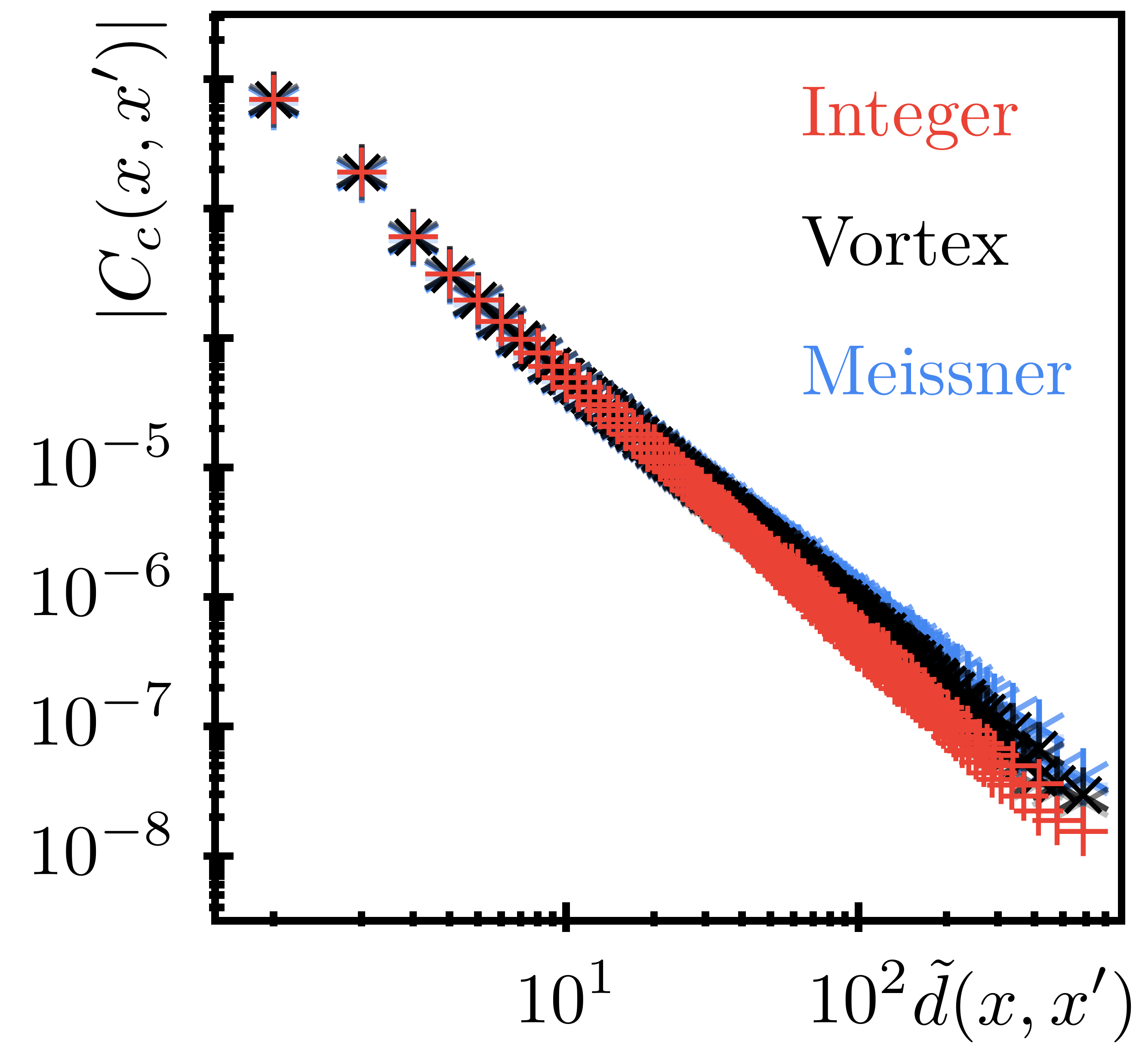}\llap{\parbox[b]{5.5cm}{\color{black}(b)\\\rule{0ex}{1.32cm}}}
    \includegraphics[width=0.4944\columnwidth]{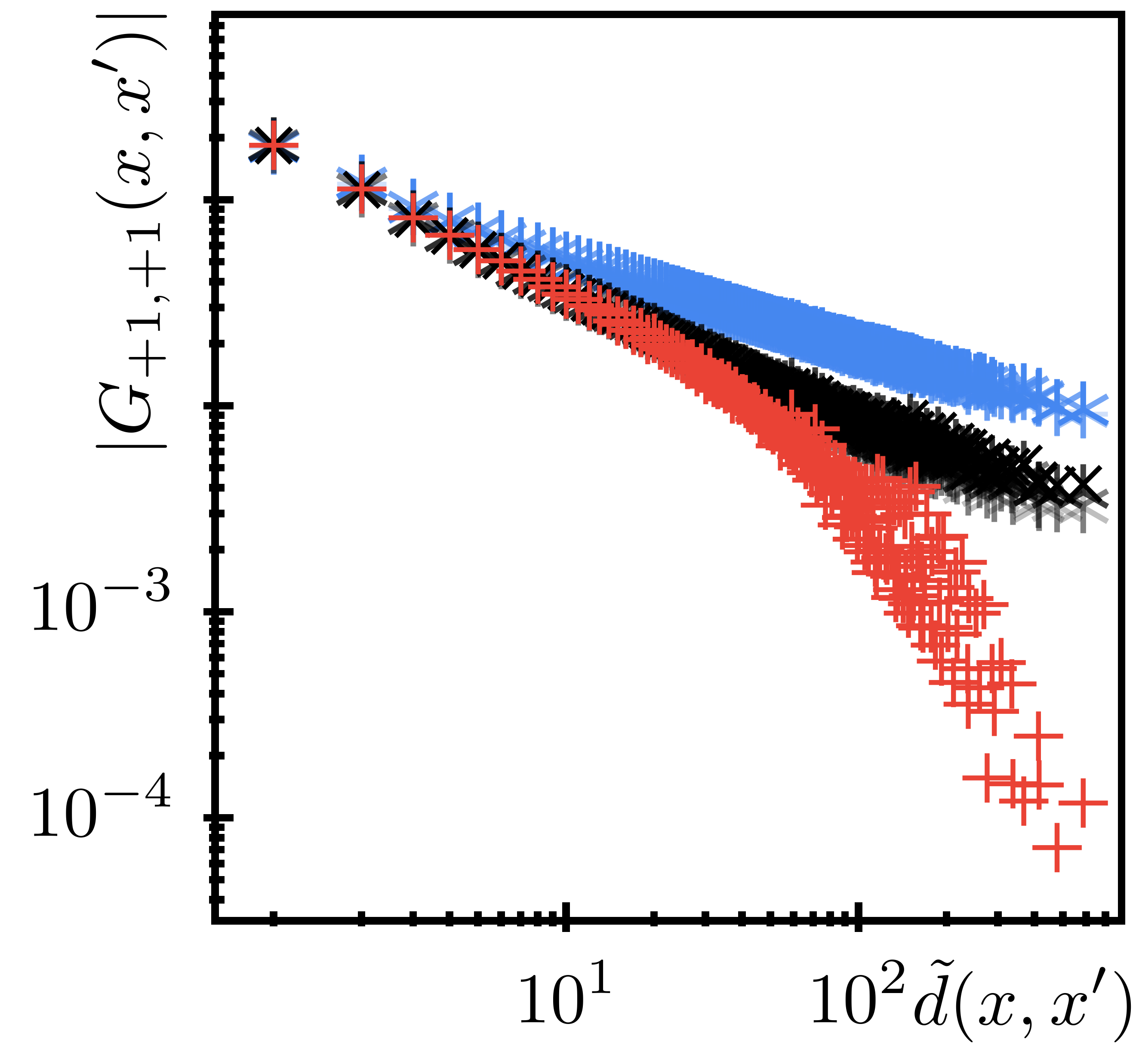}\llap{\parbox[b]{5.5cm}{\color{black}(c)\\\rule{0ex}{1.32cm}}}
    \includegraphics[width=0.4944\columnwidth]{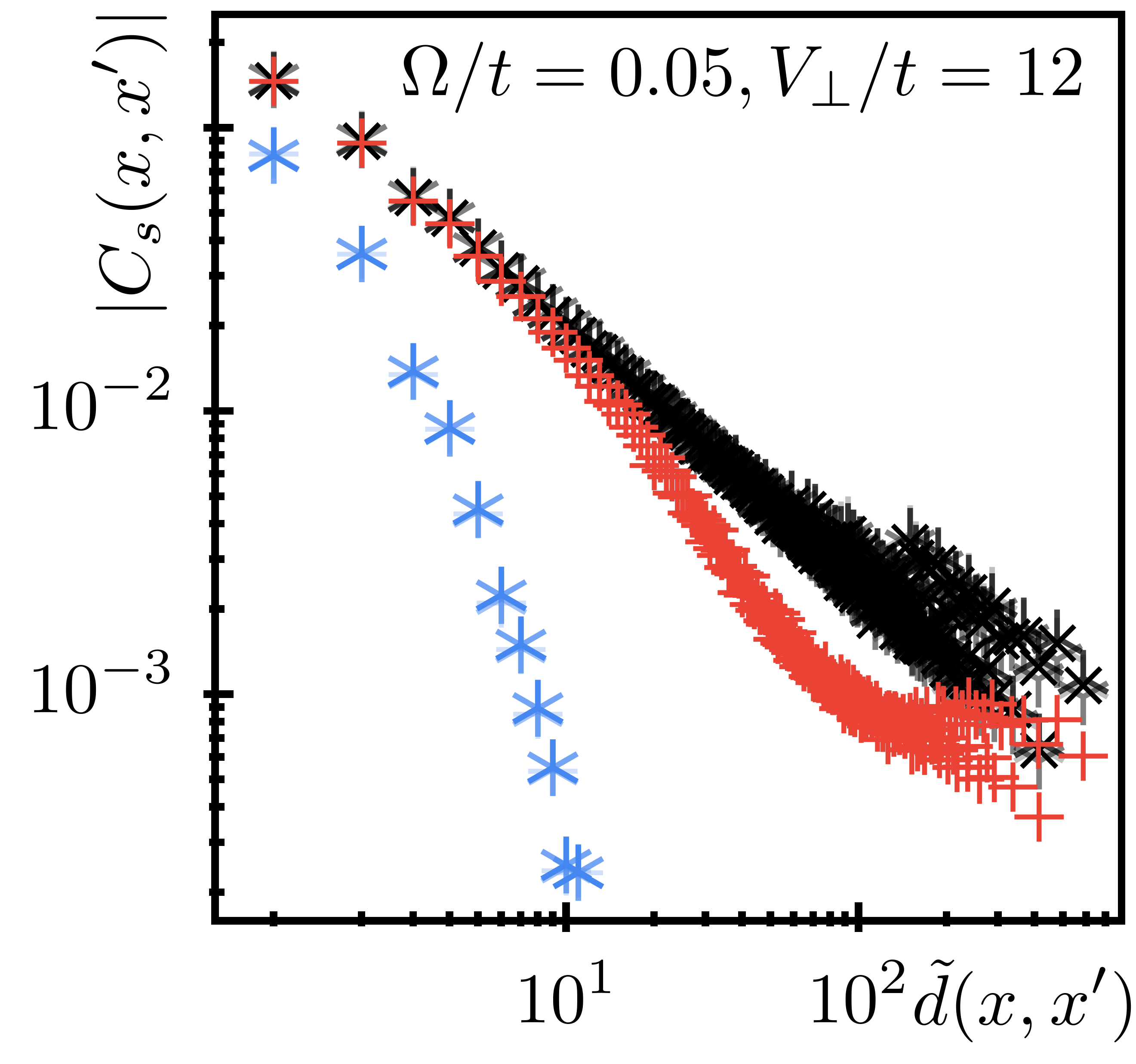}\llap{\parbox[b]{5.5cm}{\color{black}(d)\\\rule{0ex}{1.32cm}}}
    \caption{The local expectation values of the rung current is depicted in panel (a). (b)-(d) Correlation functions $C_{c/s}$ and $G_{+1,+1}$ given by Eq.~\eqref{eq:correlations} versus renormalized chord distance $\tilde d$ given in Eq.~\eqref{eq:renormalized_chord}. We see a clear pinning of $\varphi_s$ in the Meissner phase and at $\chi/\pi=n$, as visible by the exponential behavior in $C_s$ at small and intermediate distances.}
    \label{fig:correlations}
\end{figure}

We conclude our analysis of the suppression of the spin fluctuations by considering its implication on the rung current. When the $\theta_s$ field becomes semiclassically pinned due to the opening of the gap in the $h$-phase, the rung current must be suppressed as well. For systems with open boundary conditions, however, the rung current has a non-zero value at the edges and it is expected to decay in its bulk. This is what we observe in our results [see Fig. \ref{fig:correlations} (a)] where the rung current is slightly suppressed in the bulk, compared to the vortex phase.

\subsection{Correlations}

In order to better characterize the gapped phase at the resonance, we study the decay of several two-point correlation functions.
Indeed, the correlation functions allow us to easily distinguish the Meissner and vortex phases, and provide further indications that the system develops a gap in the spin sector at the integer resonance.
In particular, the following set of observables is studied in Fig.~\ref{fig:correlations}:
\begin{align}
    C_s(i,j) &= \langle b^\dag_{i,\Up}b^{\vphantom\dag}_{i,\Dn}b^{\dag}_{j,\Dn}b^{\vphantom\dag}_{j,\Up}\rangle - \langle b^\dag_{i,\Up}b^{\vphantom\dag}_{i,\Dn}\rangle\langle b^{\dag}_{j,\Dn}b^{\vphantom\dag}_{j,\Up}\rangle,\nonumber
    \\
    C_c(i,j) &= \langle b^\dag_{i,\Up}b^{\dag}_{i,\Dn}b^{\vphantom\dag}_{j,\Dn}b^{\vphantom\dag}_{j,\Up}\rangle,
    \\
    G_{y,y'}(i,j) &= \langle b^\dag_{i,y}b^{\vphantom\dag}_{j,y'}\rangle.
\end{align}

We rewrite them in bosonized form and find
\begin{align}
    C_s(x,x') &\propto\re^{-\langle[\MF_s(x)-\MF_s(x')]^2\rangle} f_s(x,x'),
    \\
    C_c(x,x') &\propto\re^{-\langle[\MF_c(x)-\MF_c(x')]^2\rangle}(1+f_c(x,x')),
    \\
    G_{y,y'}(x,x') &\propto
    \re^{-\frac12\langle[\varphi_{y\vphantom'}(x)-\varphi_{y'}(x')]^2\rangle}(1+g_{y,y'}(x,x')),
    \label{eq:correlations}
\end{align}
in which $f_{s/c}$ and $g_{y,y'}$ denote the corrections to the leading order.
In particular, the leading order is obtained by separating the $p=0$ contribution from higher harmonics $p\neq0$ in Eq.~\eqref{eq:bosonization_identity}.



It is most convenient to plot the above correlation functions as a function of the renormalized chord distance
\begin{equation}
    \tilde d(x,x')=\frac{d(x-x'|2L)d(x+x'|2L)}{\sqrt{d(2x|2L)d(2x'|2L)}},
    \label{eq:renormalized_chord}
\end{equation}
which highlights the leading order decay caused by the momentum field expectation values.


In particular, the connected part of $C_s$ shows the expected behavior: it follows an algebraic decay ($C_s\propto\tilde{d}^{-1/K_s}$) in the vortex phase; it decays exponentially in the Meissner phase, in which the related unconnected correlation rapidly saturates to a constant due to the pinning of $\varphi_s$ (the exponential correction depicted in blue in Fig. \ref{fig:correlations} (c) is due to the subleading terms $f_s$); finally, it displays an exponential tail in the $h$-phase at filling $\nu=1$ due to the gap in the spin sector and the pinning of $\DF_s$ [see the bending in the red curve in Fig. \ref{fig:correlations} (c)].

In contrast to $C_s$, $C_c$ decays algebraically everywhere ($C_c\propto\tilde{d}^{-1/K_c}$), with no signature of any exponential tail [Fig. \ref{fig:correlations} (a)]. This is expected as the charge sector remains always gapless.

Finally, the intra-wire Green's function $G_{y,y}$ shows a clear exponential tail in the $h$-phase [Fig. \ref{fig:correlations} (b)], in agreement to our statements above.
In the Meissner phase instead, we find $G_{y,y}\approx G_{y,-y}\propto\tilde{d}^{-1/(4K_c)}$ which is predicted by the leading order of Eq.\eqref{eq:correlations}.

\begin{figure}[t!]
    \centering
    \includegraphics[width=0.4944\columnwidth]{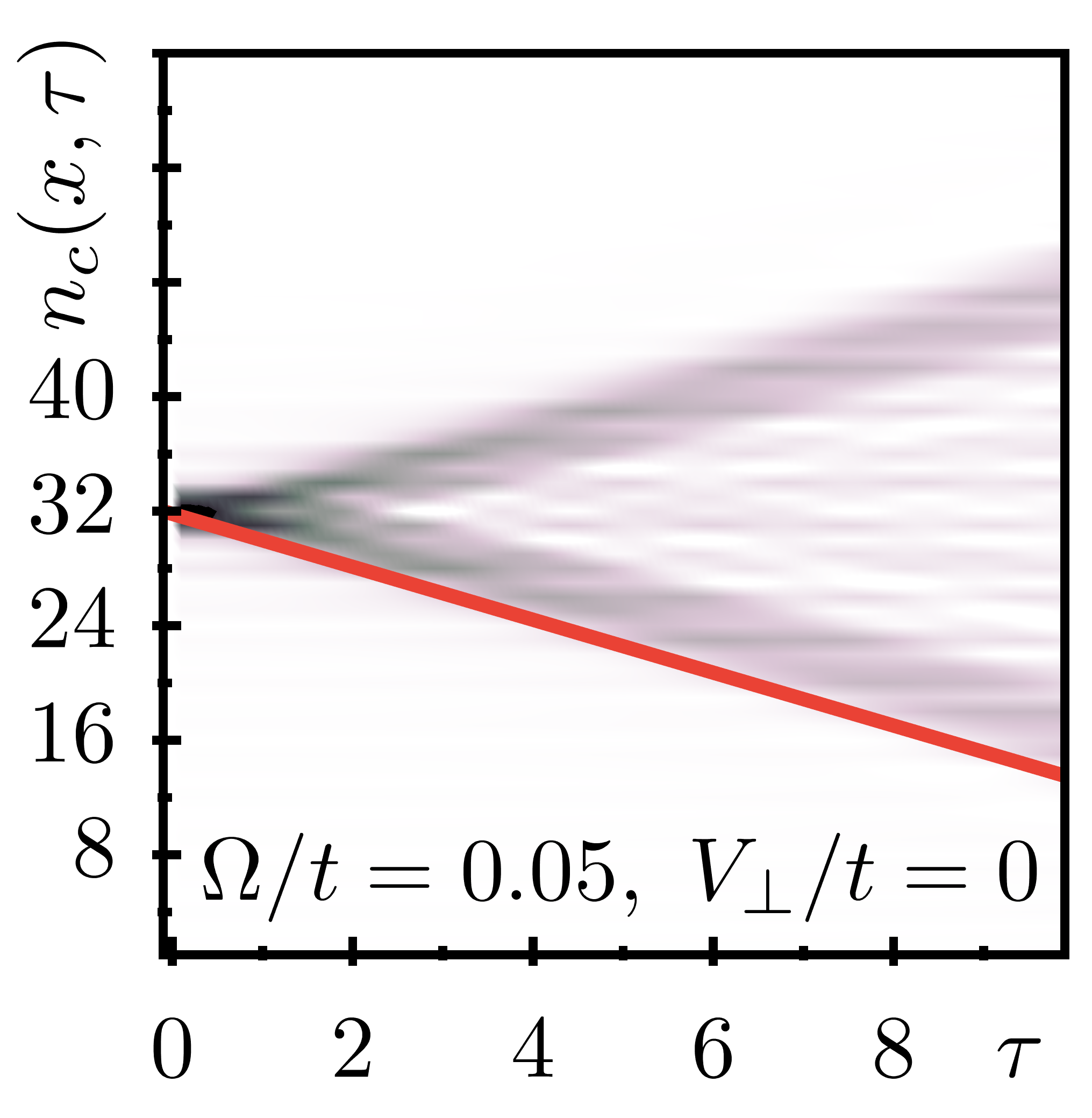}\llap{\parbox[b]{6cm}{\color{black}(a)\\\rule{0ex}{3.6cm}}}
    \includegraphics[width=0.4944\columnwidth]{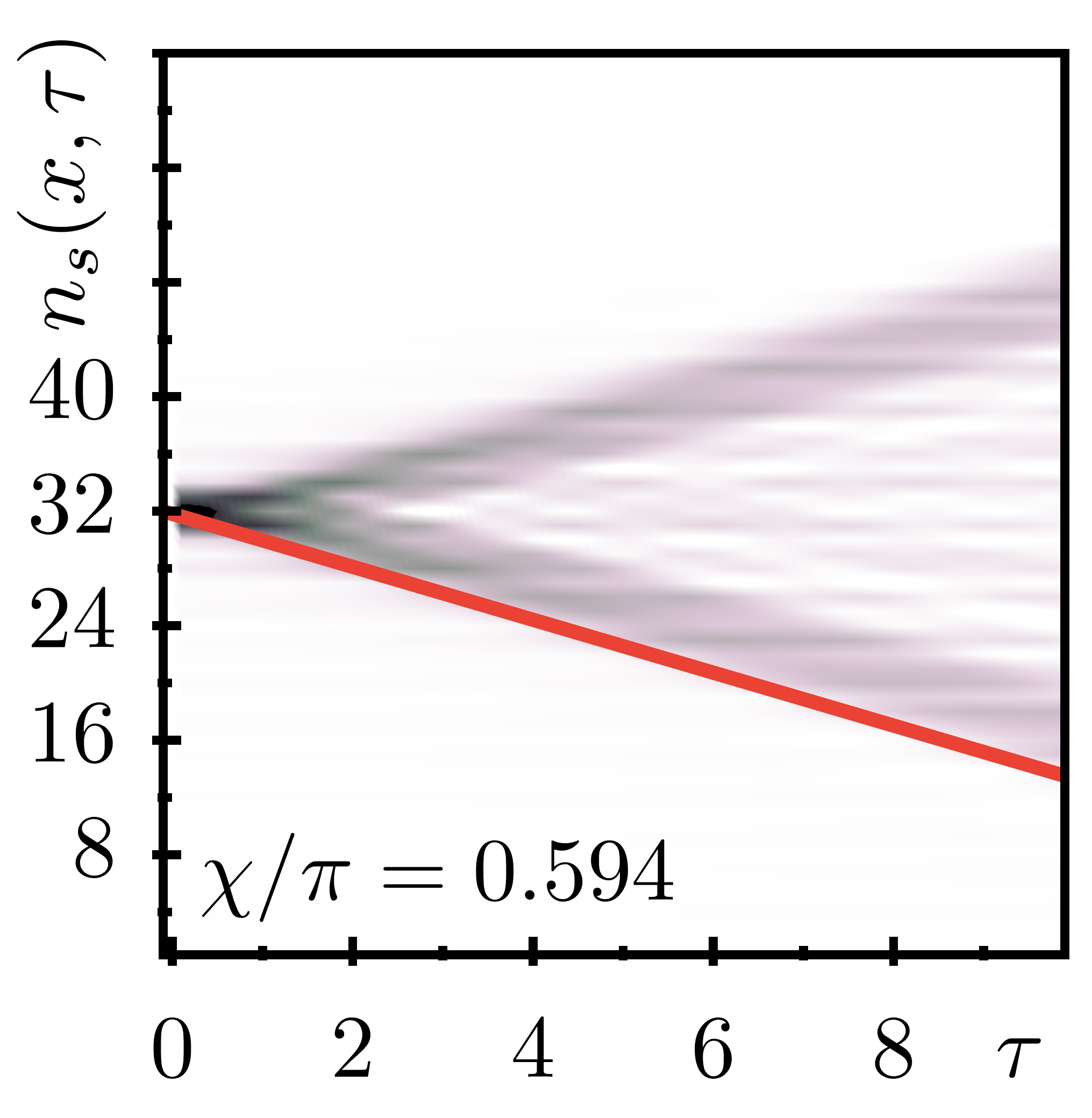}\llap{\parbox[b]{6cm}{\color{black}(b)\\\rule{0ex}{3.6cm}}}
    \includegraphics[width=0.4944\columnwidth]{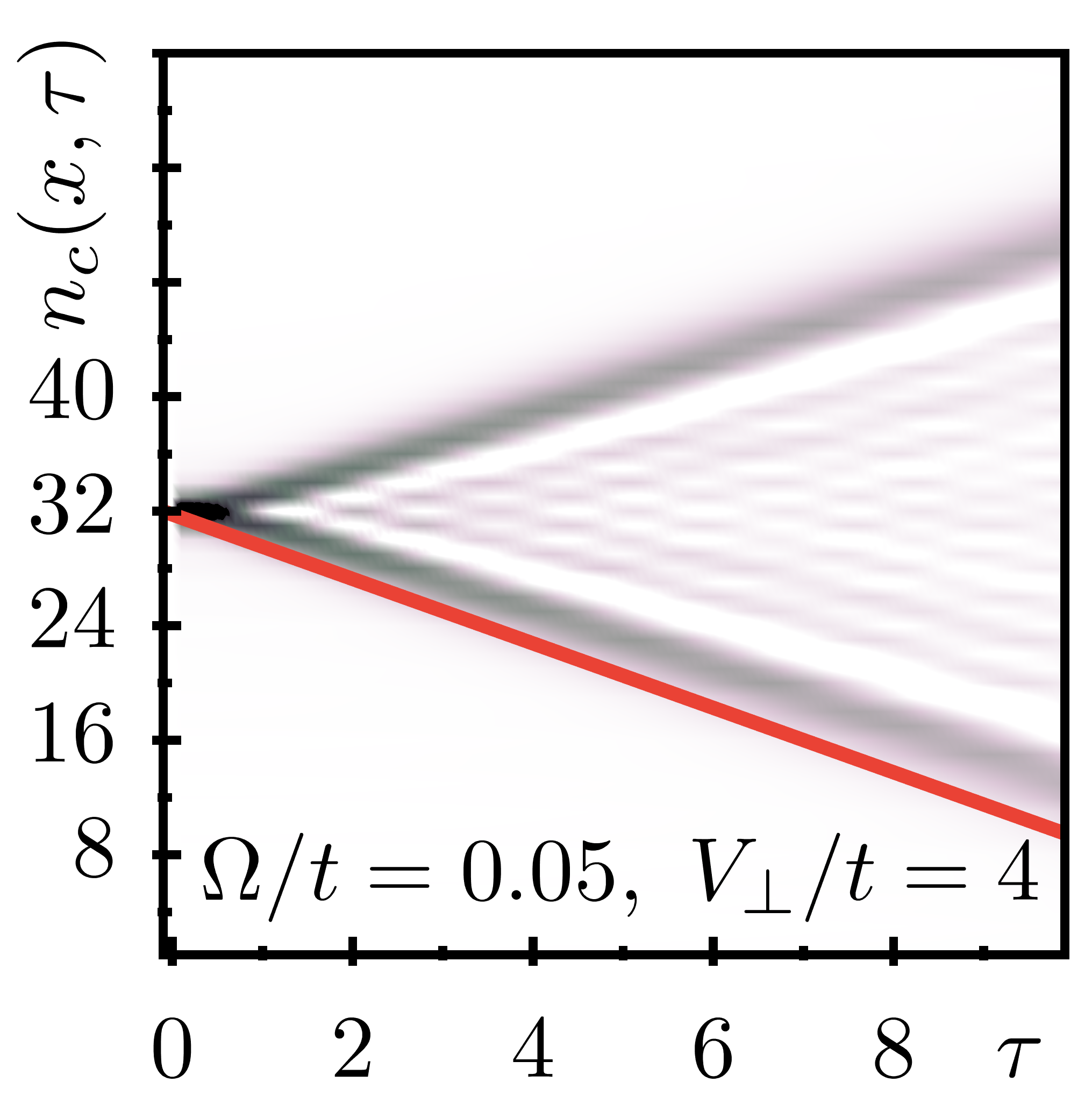}\llap{\parbox[b]{6cm}{\color{black}(c)\\\rule{0ex}{3.6cm}}}
    \includegraphics[width=0.4944\columnwidth]{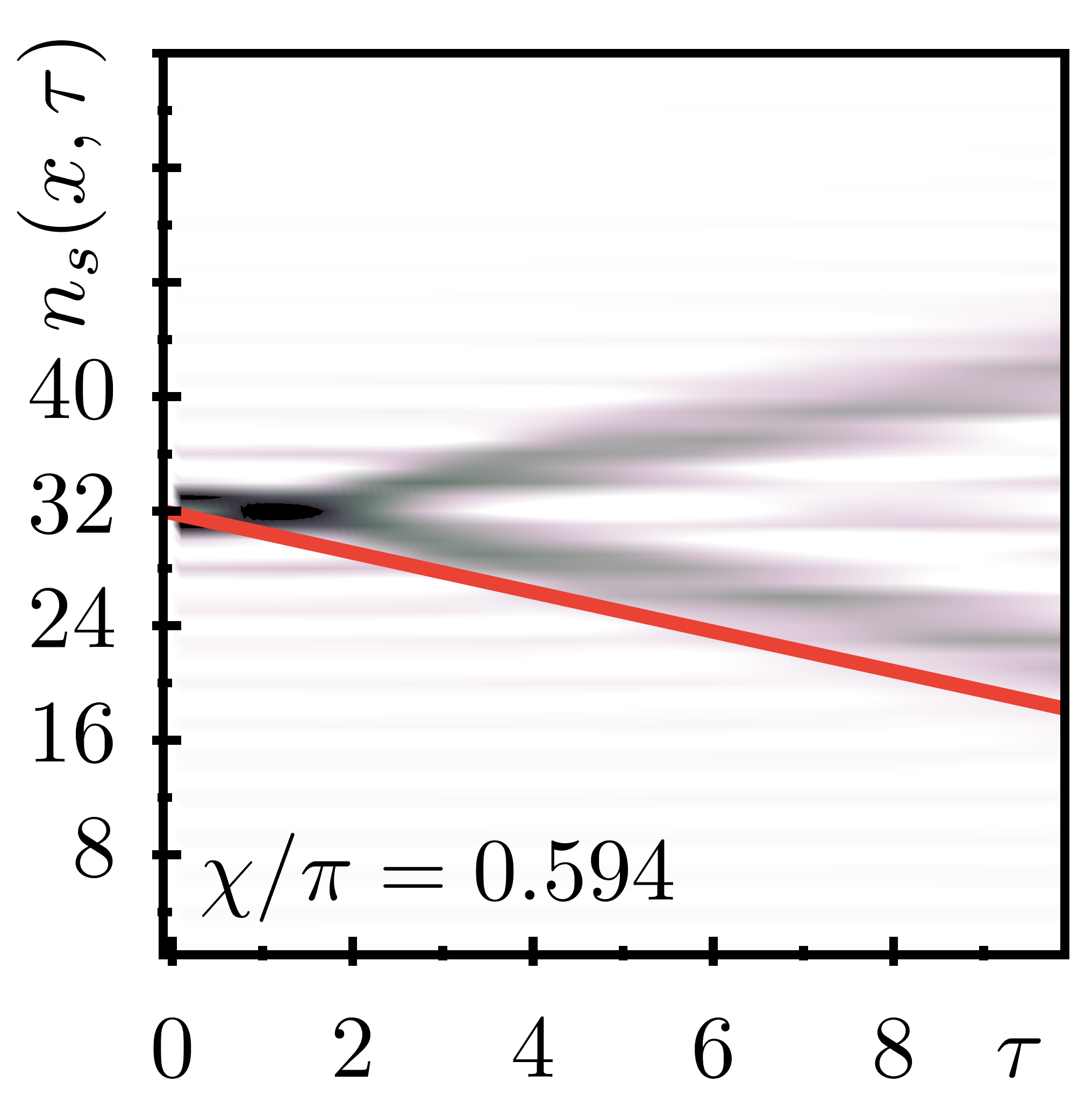}\llap{\parbox[b]{6cm}{\color{black}(d)\\\rule{0ex}{3.6cm}}}
    \caption{Density / spin propagation pattern of the excitation $|\psi(t)\rangle=\re^{-\ri Ht}b_{L/2,\uparrow}|GS\rangle$ for fixed flux and interactions linked to the vortex phase (top row at interaction $V_\perp=0$, bottom row at $V_\perp=4t$, all of them at flux $\chi=0.594\pi$ and $\Omega=0.05t$). We estimate the velocity by fitting the outmost part of the expanding wavefront, indicated by the red line.}
    \label{fig:dynamics_density_propagation}
\end{figure}

\subsection{Dynamics and velocities}

To further probe the validity of our RG predictions, we extract the velocities of the Luttinger liquid through the time evolution of spin and charge excitations, which we can then pair with the outcomes of the Luttinger parameters extracted from the bipartite fluctuations.

The velocities can be approximated by tracking the propagation of a local excitation in time.
For convenience, we choose to generate a hole at the central site of the lattice:
at time $t=0$ we apply $b_{L/2,\uparrow}$ to the (static) ground state $|GS\rangle$ and evolve it in time, i.e. we simulate
\begin{align}
    |\psi(t)\rangle = \re^{-\ri H t}b_{L/2,\uparrow}|GS\rangle
\end{align}
and measure the local density / magnetization at all times $n_{c/s}(x,t)=\langle\psi(t)|\hat n_{c/s}(x,t=0)|\psi(t)\rangle$.

\begin{figure}[t!]
    \centering
    \includegraphics[width=.4944\columnwidth]{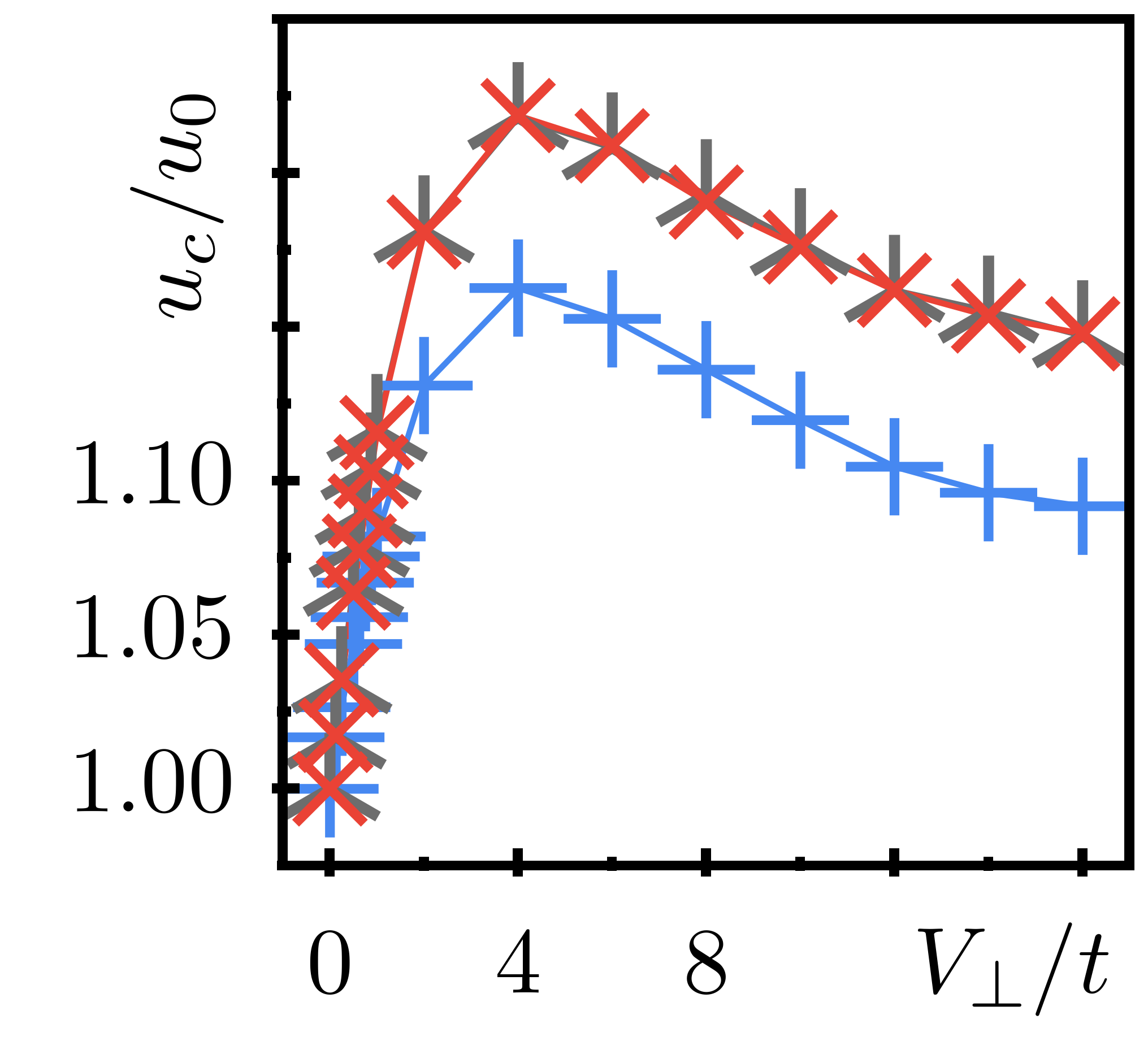}\llap{\parbox[b]{5cm}{\color{black}(a)\\\rule{0ex}{0.9cm}}}
    \includegraphics[width=.4944\columnwidth]{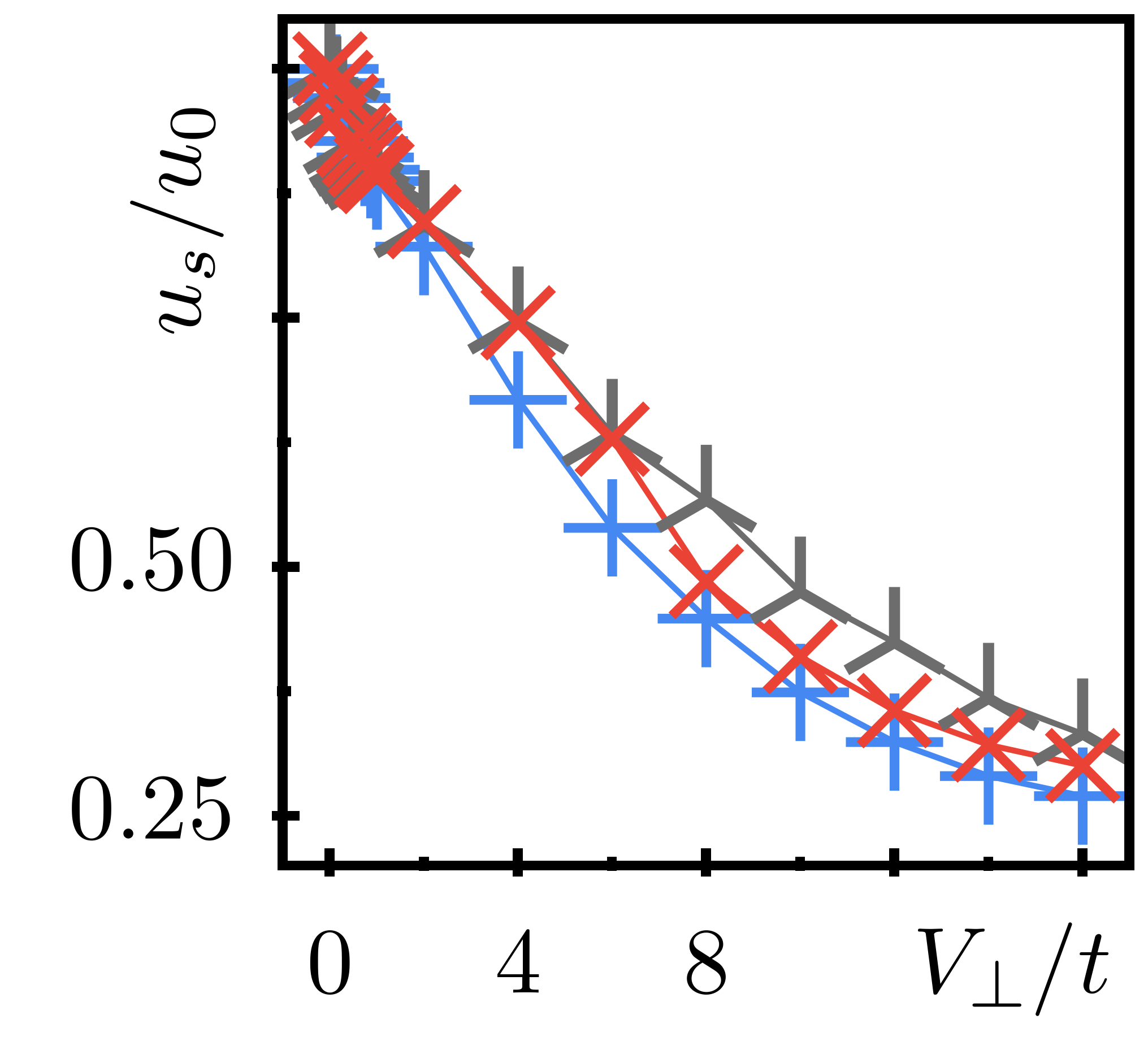}\llap{\parbox[b]{5cm}{\color{black}(b)\\\rule{0ex}{0.9cm}}}
    \includegraphics[width=.4944\columnwidth]{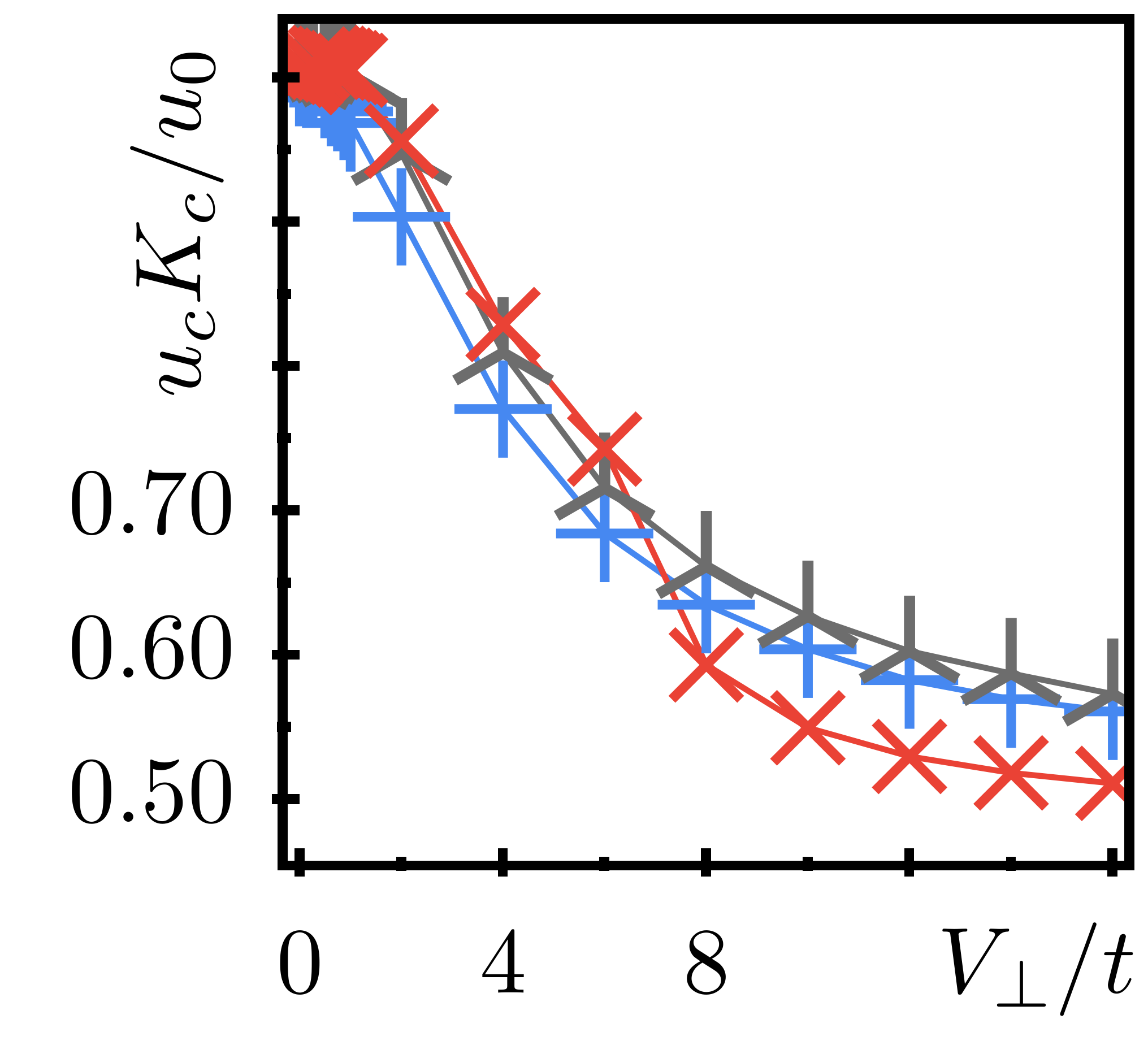}\llap{\parbox[b]{5cm}{\color{black}(c)\\\rule{0ex}{0.9cm}}}
    \includegraphics[width=.4944\columnwidth]{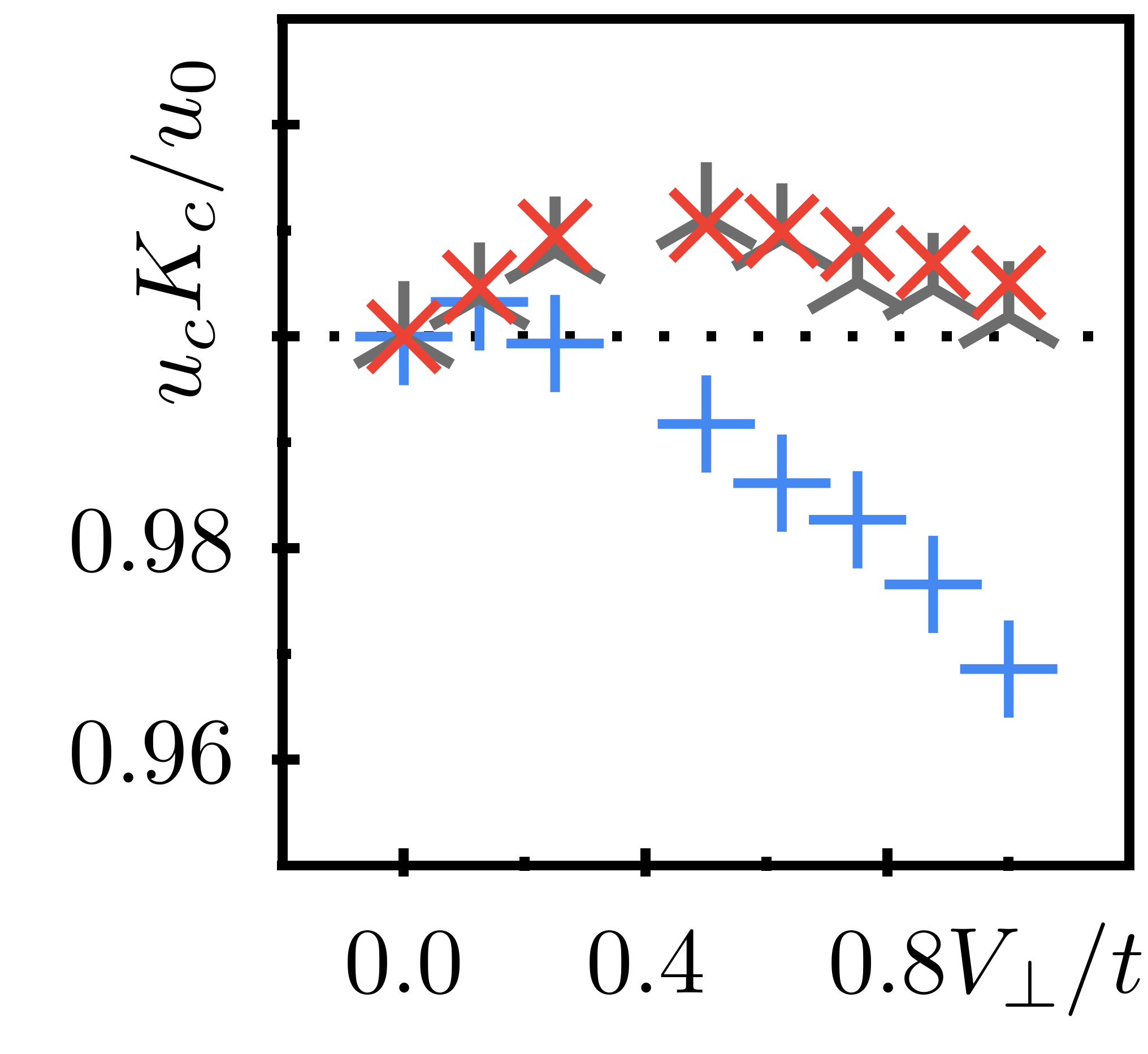}\llap{\parbox[b]{5cm}{\color{black}(d)\\\rule{0ex}{0.9cm}}}
    \caption{Extracted charge (a) and spin (b) velocities from the propagation of a hole. The colors/symbols represent the integer resonance in red/$\times$, the vortex/Luttinger liquid phase in gray/$\Yup$ and the Meissner phase in blue/+. We normalize the velocities to the bare value at $V_\perp=0$. In general, we observe interaction dependent velocities $u_c$ which are different in the Meissner and in the vortex phase, but do not distinguish between vortex and $h$-dominated phase. In contrast, we note a significant splitting of the red and gray data lines in $u_s$ after the critical value $V_\perp/t\gtrapprox 6$, signaling a phase transition.
    (c) Concerning the validity of Eq.~\eqref{eq:drude}, we see that it applies for small interactions only ($V_\perp < t$). If the spin fields are pinned from relevant interactions, $u_c K_c$ is suddenly renormalized, visible in the sudden bending of the red curve around $V_\perp=6t$. (d) Detail of the charge superfluid stiffness for small interactions. $u_cK_c$ is approximately constant in the vortex and $\nu=1$ states, whereas it decreases for weak interactions in the Meissner phase.}
    \label{fig:dynamics_velocity}
\end{figure}

By fitting the propagation of the wavefront in $n_{c/s}$ (see Fig.~\ref{fig:dynamics_density_propagation}), we obtain the estimates of $u_{c/s}$ presented in Fig.~\ref{fig:dynamics_velocity}.
In the charge velocities, we do not expect to find a difference between vortex and integer fillings, and indeed the simulations overlap.
On the contrary, we find two different curves distinguishing between non-Meissner and Meissner phase.
This behavior is expected as the pinning of the $\MF_s$ field in the Meissner phase results in a renormalization of the effective charge velocity.

An analog reasoning holds for the comparison of the spin velocities in the vortex phase and $\nu=1$ resonance [gray and red lines in Fig.~\ref{fig:dynamics_velocity}(b)]: $\mathcal{O}_h$ is irrelevant for small values of the rung repulsion $V_\perp$ such that the two $u_s$ velocities coincide for small interactions. However, for $\nu=1$ [red line in Fig.~\ref{fig:dynamics_velocity}(b)], $u_s$ displays a clear reduction beyond a critical value of $V_\perp \approx 6t$ at which it separates with the corresponding velocity in the vortex phase.
We interpret this as a renormalization effect and it is a further evidence of a transition between the vortex/Luttinger phase and the helical $h$-phase for $V_\perp\sim 6t$.

The estimates of $u_q$ and $K_q$ allow us to evaluate the superfluid stiffness in response to an infinitesimal variation of the vector potential.
Due to the Galilean invariance of the interaction $V_\perp$ and the small values of $\Omega$, the bare values of the superfluid stiffness are roughly constant in both charge and spin sectors:
\begin{align}
    u_q(l=0) K_q(l=0) \approx u_0 = 2 a t\sin(k_0 a).
    \label{eq:drude}
\end{align}
The spin stiffness is expected to be heavily renormalized by the interactions (which primarily act in the spin sector), and we focus in the following on the extraction of the charge stiffness only.
Panel (c) of Fig.~\ref{fig:dynamics_velocity} refers to the quasi-particle propagation in the charge sector for an interacting system. For weak interactions, $V_\perp \lesssim t$ , the data display an approximately constant charge stiffness for the vortex and resonant phases, consistently with a Luttinger liquid phase [see panel (d)]. On the contrary, in the Meissner phase the stiffness is renormalized and reduced also for weak interactions. By increasing the interaction, we see that the vortex and resonant states behave differently starting from  $V_\perp/t\gtrapprox6$. This is a further confirmation that a gap in the spin sector is opened for larger interactions, consistently with the onset of the $h$-phase.

\section{Conclusions and Perspectives} \label{sec:concl}

Ultracold atoms hopping in ladder geometries and subject to artificial magnetic fluxes are known to generate rich phase diagrams. For commensurate values of the ratio between the number of fluxes and the number of atoms, helical states with a net chiral current may appear. The simplest and most evident example of these helical states are the non-interacting Meissner phase for bosons and the helical state at flux $\chi=2k_F$ for fermions. It is known, however, that additional strongly correlated helical states originate for suitable values of the filling factor $\nu$ and suitable interactions \cite{Petrescu2015,Mazza2017,Taddia2017,Petrescu2017,Haller2018,CalvaneseStrinati2019,Rosson2019,Santos2019,Yang2020}.

In this work, we argued that a two-leg ladder of hardcore bosons at $\nu=1$ is characterized by one of these strongly correlated helical phases of matter, which we called the $h$-phase. We studied its signatures in terms of correlation functions, fluctuations and dynamical evolution.

Our DMRG simulations show that the strongly correlated $\nu=1$ helical $h$-phase can be accessed in systems with contact interactions only, similarly to the bosonic Laughlin-like state at filling factor $\nu=1/2$ \cite{Mazza2017}. With respect to the pretopological Laughlin-like state, however, the chiral current and gap signatures we observe are considerably stronger for a broad range of parameters when comparing systems with the same particle density, repulsive interaction and interleg tunneling.
 This is particularly relevant for experiments based on bosonic ladders like the Rb gases studied in the experiments in Refs. \cite{spielman2015,Genkina2019}.

Comparing our findings with the analogous fermionic systems, we also observe that the strongly-correlated phases of bosons can be reached through interactions with a shorter range than their fermionic counterpart, as common for several fractional quantum Hall states. This intuitively explains also why the signals we detect for bosons at filling $\nu=1$ are considerably larger than their fermionic counterpart (a pretopological $K=8$ fractional quantum Hall state) at filling $\nu=1/2$ \cite{Haller2018}. In this respect, we find that bosonic systems are more suitable for the experimental characterization of these helical and strongly correlated phases of matter (see the summary in Table \ref{tab:sum}).

The onset of the bosonic $h$-phase is caused by a particular mechanism that allows the interaction $\mathcal{O}_h$ acting on the spin sector of the theory to become relevant despite the presence of repulsive interactions only, which normally suppress it. This happens for bosons at the integer filling factor $\nu=1$ and it is analogous to the pairing mechanism determining the appearance of the pretopological $K=8$ phase in fermionic systems at $\nu=1/2$ \cite{Haller2018}. We stress that engineering (strong) pairing mechanisms analogous to the interaction $\mathcal{O}_h$ is recently at the focus of several proposals, due to its potential for the development of topological and strongly-correlated paired phases of matter \cite{vishwanath2020,Oreg2020}. The combination we adopt of artificial gauge potentials, equivalent to a spin-orbit coupling, and a commensurate particle density, can be adopted also in this framework.

\begin{table}[t]
\setlength{\tabcolsep}{0.5em}
\centering
\begin{tabular}{|c||c|c|c|}
\hline
{\textbf{Interactions}} & \makecell{Laughlin\\ $\nu=1/3$} & \makecell{Laughlin \\ $\nu=1/2$} & $h$-phase \\
\hline
\hline
Fermions & NNN \cite{Mazza2017} & -- & NN \cite{Haller2018} \\
\hline
HC Bosons & -- & Contact \cite{Mazza2017} & \makecell{\textbf{Contact} \\ (this work)} \\
\hline
\end{tabular}
\caption{The required interactions for the appearance of several pretopological helical states in ultracold atom ladder systems are compared. Bosonic pretopological states can be obtained from simple contact interactions, whereas their fermionic counterparts require nearest-neighbor [NN] or next-nearest-neighbor [NNN] repulsions.}
\label{tab:sum}
\end{table}

\begin{acknowledgements}
The authors wish to thank R. Citro and P. Schmoll for inspiring and enlightening discussions.
A.H. is thankful for the financial support from the Max Planck Graduate Center.
A.H. and M.R. acknowledge support from the Deutsche Forschungsgesellschaft (DFG) through the grant OSCAR 277810020 (RI 2345/2-1). M. B. is supported by the Villum Foundation (Research Grant No. 25310). Y. P. acknowledges financial support from the China Scholarships Council (grant No. 201906040093). The MPS simulations were run on the Mogon cluster of the Johannes Gutenberg-Universit\"at (made available by the CSM and AHRP), with a code based on a flexible Abelian Symmetric Tensor Networks Library, developed in collaboration with the group of S. Montangero at the University of Ulm (now moved to Padua). We   acknowledge   partial   support   from   the   Deutsche Forschungsgemeinschaft (DFG) – project grant 277101999 – within the CRC network TR 183 (subproject B01).

\end{acknowledgements}


\appendix

\section{Bosonization of the model and derivation of the renormalization group equations} \label{app:bosonization}

The field theoretical description of the hard-core boson ladder model has been developed based on bosonization techniques.
In particular, the analysis of the behavior at the resonance $\chi=2k_0$ requires a careful approach since there are two commonly used approximations that cannot be applied in this case.
The first is related to the use of several harmonics to map the creation and annihilation operators of the lattice model into the low-energy description in the continuum: the resonance at $\chi=2k_0$ appears evident only when taking into account higher harmonics, as in the case of the second-incommensurability effects studied in \cite{Orignac2016,Orignac2017,Citro2018}.
This is well-known in the analysis of the one-dimensional limits of fractional quantum Hall states (see, for example, \cite{Cornfeld2015,Petrescu2015,Mazza2017,Taddia2017,Petrescu2017,Haller2018,CalvaneseStrinati2019}), where higher harmonics can also be interpreted as multi-particle interaction processes appearing in higher orders of perturbation theory \cite{Kane2002,Teo2014,Oreg2014}.
The second important characteristics is that, for bosons at $\nu=1$, there are two of these multi-particle processes that resonate and compose $\mathcal{O}_g$ in Eq.~\eqref{og}.
These have exactly the same scaling behavior under the RG flow and hinder the possibility of adopting a simple semiclassical analysis.
This situation is analogous to the case of fermionic ladder models at $\nu=1/2$ \cite{Haller2018} and requires a second-order renormalization group analysis to be examined.
In this context the role of $\mathcal{O}_h$ in Eq.~\eqref{eq:LL_model} emerges and plays a crucial role,
since it mixes and competes with the mentioned $g$-terms.

In the following we will first present some of the details related to the derivation of the effective Hamiltonian~\eqref{eq:LL_model}, then we will discuss the main steps to derive the RG equation in (\ref{eq:RG_flow}) and how we deal with them numerically.

\subsection{Bosonization of the hard-core boson model}

We mentioned that the analysis of the resonant states appearing at specific values of the filling factor $\nu$ requires a bosonization of the lattice operators in terms of the series expansion of vertex operators in Eq.~\eqref{eq:bosonization_identity} (see, for example, \cite{Cazalilla2004}).
This series expansion relies on the non-universal coefficients $\beta_p$ which are difficult to evaluate in non-integrable models like ours.
Here we first obtain an estimate of the most relevant of these non-universal parameters in our hard-core bosonic model, $\beta_{\pm 2}$, via a Jordan-Wigner transformation and the standard bosonization of fermions.
This is rigorous when considering separate chains, i.e., in the limit $\Omega \to 0$ and $V_\perp\to 0$.
Then we briefly discuss our choice of setting them to such value, $\beta_{\pm 2} = 1/2$, irrespective of Hamiltonian parameters.

Hard-core boson operators $b^{(\dag)}$ can be related to one-dimensional fermionic operators $c^{(\dag)}$ via the Jordan-Wigner transformation:
\begin{equation}
b_{x,y}^{(\dag)} = (-1)^{\sum_{j<x}c^\dag_{j,y}c_{j,y}}c_{x,y}^{(\dag)}\, . \label{JW1}
\end{equation}
Since the total density of particles, $n_0=N_{\rm tot}/L$, coincides for both bosons and fermions, the Fermi momentum is given by $k_0=\pi n_0/2$.
The standard bosonization of the fermionic operators (see, for example, \cite{Cazalilla2004,giamarchibook}) reads:
\begin{equation}
c_{x,y} = \sqrt{\frac{k_0}{2\pi^2}} \left[\re^{-ik_0x+i\left(\varphi_y(x)+\theta_y(x)\right)} + \re^{ik_0x+i\left(\varphi_y(x)-\theta_y(x)\right)}\right] \, ,
\label{psi}
\end{equation}
where $\theta_y$ and $\varphi_y$ are two pairs of dual fields obeying commutation relations ($\Theta$ being the Heaviside function):
\begin{equation} \label{comm}
\left[\DF_{y'}(x'),\MF_{y\vphantom'}(x)\right] = \ri\pi\delta_{yy'}\Theta(x'-x)\, ,
\end{equation}
and we did not introduce any Klein factor, due to the bosonic nature of our particles which imposes operators in the two different legs to commute.

The density of each pseudo-spin species can be approximated at first order in the harmonics expansion as:
\begin{equation} \label{bos_dens}
n(x,y) \approx \frac{1}{\pi}\left[k_0 - \partial_x \theta_y(x)\right]\,.
\end{equation}
The symmetrized form of the Jordan-Wigner string of Eq.~\eqref{JW1} reads then:
\begin{multline} \label{JW3}
(-1)^{\sum_{j<x} n(j,y)} = \frac{\re^{i\pi\sum_{j<x} n(j,y)}+ \re^{-i\pi\sum_{j<x} n(j,y)}}{2}\\
 \to \frac{\re^{i\left(k_0x -\theta_y(x)\right)}+ \re^{-i\left(k_0x -\theta_y(x)\right)}}{2}\,.
\end{multline}
Substituting the expressions \eqref{psi} and \eqref{JW3} into Eq. \eqref{JW1}, we obtain:
\begin{widetext}
\begin{equation} \label{b1}
b_{x,y} \approx \sqrt{\frac{k_0}{2\pi^2}}\left[\re^{i\varphi_y(x)} + \frac{\re^{-2ik_0x}}{2} \re^{i\left(\varphi_y(x) + 2\theta_y(x)\right)}+ \frac{\re^{2ik_0x}}{2} \re^{i\left(\varphi_y(x) - 2\theta_y(x)\right)}\right]\,.
\end{equation}
\end{widetext}
By direct comparison with Eq.~\eqref{eq:bosonization_identity}, we finally find $\beta_2 = \beta_{-2} =1/2$ for separate chains.

In the limit $\Omega\to 0$, space-inversion  symmetry implies indeed that $\beta_p=\beta_{-p}$. When considering $\Omega > 0$, however, it relates the $\beta_p$ and $\beta_{-p}$ coefficients of the bosons of different legs. Since we focus in a regime with small values of $\Omega/t$, we assume hereafter that these coefficients are the same for both the bosonic species, such that $\beta_2=\beta_{-2}$.
Even in the case of more general coefficients, however, the analysis we present in the following is not qualitatively affected.

Concerning a precise evaluation of these coefficients, the estimate of the $\beta_p$ parameters for general values of the interactions, the transverse hopping $\Omega$ and the density of the system remains an open problem.
Analytical solutions and numerical approximation are known only for (several) integrable models (see, for example, \cite{Giamarchi2011,Imambekov2012}).

Our model, in particular, can be mapped into a fermionic Hubbard model with interaction $V_\perp > 0$ in the limit $\Omega\to 0$ (see also a similar analysis in \cite{Cornfeld2015}).
For the fermionic Hubbard model in the limit $V_\perp \to \infty$, the density modulations with momentum $2k_0$ are suppressed \cite{giamarchibook}, such that we expect $\beta_{\pm 2}$ to be equally suppressed for large $V_\perp$.
As we will discuss below, setting the value of $\beta_2$ might have important implications in the determination of the phase diagram through the RG equations~\eqref{eq:RG_flow}, due to the dependence of the bare values in \eqref{barehg} on $\beta_2$.
However, since a precise estimate of $\beta_{2}$ for large interactions is difficult to derive and beyond the scope of our work, we decide to approximate it with the constant $1/2$ irrespectively of the Hamiltonian parameters. This provides results compatible with the numerical simulations and we checked that the phase diagram in not affected by small variations of this parameter.

At this point, we are set to derive the $g-$ and $h$-terms of the effective model in Eq.~\eqref{eq:LL_model}. This is most conveniently done by using the charge $c$ and spin $s$ sectors of the model, as introduced in Sec. \ref{sec:bosonization}:
\begin{equation}
\theta_{c/s} = \frac{\theta_\Up \pm \theta_\Dn}{\sqrt{2}}\,,\quad  \varphi_{c/s} = \frac{\varphi_\Up \pm \varphi_\Dn}{\sqrt{2}}\, .
\end{equation}

We begin our analysis from a gauge-equivalent formulation of the non-interacting Hamiltonian of Eq.~\eqref{eq:kinetics}:
\begin{multline}\label{Ham_space}
 \MH_0 = -t\sum_{x,y}\left[b^\dag_{x,y}b_{x+1,y} + \hc\right] \\
     -\Omega\sum_{x}\left[\re^{\ri\chi x}b^\dag_{x,\Up}b_{x,\Dn} + \hc\right]
    \, .
\end{multline}
Although this gauge choice breaks the translational invariance, it turns out to be more convenient to make resonant terms evident.
By plugging in Eq.~\eqref{eq:bosonization_identity}, the interleg hopping term reads indeed:
\begin{multline} \label{H_omega}
    \mathcal{H}_\Omega \to -\int \rd x\, \Omega e^{i\chi x}\frac{k_0}{2\pi^2} \left[\re^{-i\sqrt{2}\varphi_s}  + {\rm F.O.} \right.\\
    \left.+\beta_2 \re^{-i 2 k_0 x} \left( \re^{-i\sqrt{2}\left(\varphi_s + \theta_s -\theta_c \right)}
    + \re^{-i\sqrt{2}\left(\varphi_s - \theta_s -\theta_c \right)}\right) \right] +{\rm H.c.}
\end{multline}
where we did not explicitly write additional fast oscillating terms (F.O.) which are not resonant for $\chi=2k_0$, i.e., $\nu=1$.
A direct comparison with Eq.~\eqref{eq:LL_model} yield the initial value of the coupling constant $g=\beta_2 k_0/2\pi^2$ in the RG flow, reported in Eq.~\eqref{barehg}.

Coming to the interacting part of the model, we need to supplement Eq.~\eqref{bos_dens} with additional higher-order oscillating terms \cite{Cazalilla2004}. The first of such terms can  be estimated from a point-splitting procedure, i.e., by evaluating $b^\dag\left(x-\delta x/2\right)b(x+\delta x/2)$ with $\delta x = \pi k_0^{-1}$.
In this way, we obtain the correction :
\begin{equation}
    n(x,y) \approx \frac{1}{\pi}\left[k_0 - \partial_x \theta_y(x)\right] \left[1 + \beta_2 \sin\left(2k_0x - 2\theta_y(x)\right)\right]\,.
\end{equation}
The interspecies interaction becomes then:
\begin{multline} \label{bos_int}
    \mathcal{H}_\perp \to \int \rd x \, \frac{V_\perp}{2\pi^2} \left[(\partial_x\theta_c)^2 - (\partial_x\theta_s)^2\right] +\\
    \int \rd x \, V_\perp \frac{\beta^2_2 k_0^2}{2\pi^2} \cos \left(2\sqrt{2}\theta_s(x)\right) \, ,
\end{multline}
where we neglected  fast oscillating addends, boundary terms (such as $\partial_x \theta_c$) and less relevant ones (such as $(\partial_x \theta_c) \cos (2 \sqrt{2} \theta_s)$).
The initial value $h= V_\perp \beta^2_2 k_0^2 / 2\pi^2$ for the RG flow, reported in Eq.~\eqref{barehg}, is then readily identified.
The first line of Eq.~\eqref{bos_int} is responsible for giving the initial condition for the Luttinger parameters $K_q$ of Eq.~\eqref{eq:initial_conditions}.

\subsection{Renormalization group equations} \label{app:RG}

The second-order renormalization group analysis of the Hamiltonian~\eqref{eq:LL_model} follows the analogous fermionic case at filling $\nu=1/2$ \cite{Haller2018}. We apply, in particular, the Wilsonian RG in momentum space at second order.
Here we derive the corresponding flow equations (\ref{eq:RG_flow}) in detail for the resonant regime, i.e., $\chi=2k_0$.

It is crucial to anticipate here that we consider the interaction terms $h$ and $g$ in Eq.~\eqref{eq:LL_model} as a perturbation of the gapless Luttinger liquid.
Therefore, we expect our results to be valid when the bare values of $g$ and $h$ are much smaller than 1.
In particular, by assuming $\Omega=0.05$ and $N/L = 48/64$ as in the numerical simulations, we obtain that the general features of the phase diagram presented in Fig.~\ref{fig:RG_flow} (b) are stable until $V_{\perp}<5t$ (cf. Fig. \ref{fig:RG_flow} (a)).
Beyond this threshold non-physical features emerge which are not compatible with the numerical simulations, thus indicating that a second-order perturbation approach fails beyond this limit.

For each of the bosonic fields we distinguish fast and slow modes, separated by an effective cutoff in momentum space that we label $\tilde{\Lambda}$.
Furthermore, we introduce an ultraviolet momentum cutoff $\Lambda>\tilde{\Lambda}$.
The fast oscillating modes are characterized by $\tilde{\Lambda}<k<\Lambda$ and we are interested in the limit $\Lambda/\tilde{\Lambda} = 1+\rd l$, with $\rd l$ infinitesimal.
The bosonic fields ($q = c, s$) can thus be decomposed into:
\begin{align}
& \varphi_q(x,t)= \varphi_{{\sf s},q}(x,t)+\varphi_{{\sf f},q}(x,t)\,,\\
&  \theta_q(x,t)= \theta_{{\sf s},q}(x,t)+\theta_{{\sf f},q}(x,t)\,.
\end{align}

Moreover, we resort to Euclidean space, i.e., we use coordinates $z=(x, \tau \equiv \ri t)$ and we denote $\rd^2 z = \rd x \, \rd\tau$, such that the following duality relations hold:
\begin{equation}
    \partial_\tau \theta_j = iu_jK_j\partial_x\varphi_j \,,\quad \partial_\tau \varphi_j = i\frac{u_j}{K_j}\partial_x\theta_j\,.
\end{equation}

We separate the action of the resonant model into a Gaussian and an interacting part, including both the $g-$ and $h-$terms of Eq.~\eqref{eq:LL_model}, and we consider the latter as a perturbation:
\bwt
\begin{equation}
 S = S_0 + S_I \equiv \frac{1}{2\pi} \int \rd^2 z \left[ \sum_{q=s,c} \frac{K_q}{u_q} (\partial_\tau \varphi_q)^2 + K_q u_q (\partial_x \varphi_q)^2 \right]
+ \int \rd^2x \left[g\left(\mathcal{O}_g + \mathcal{O}_g^\dag\right) + h\cos\left(2\sqrt 2\theta_s(x)\right)\right]\, .
\end{equation}
Our aim is to obtain an effective action for the slow modes only, by integrating out the fast degrees of freedom:
\begin{equation} \label{seff}
 S_{\rm eff}(\tilde{\Lambda})=S_0(\varphi_{\sf s}) - \ln \left\langle e^{-S_I(\varphi_{\sf s}+\varphi_{\sf f})} \right\rangle_{\sf f}
 \approx S_0(\varphi_{\sf s}) + \underbrace{\left\langle S_I(\varphi_{\sf s}+\varphi_{\sf f})\right\rangle_{\sf f}}_{\mathcal{A}}
-\frac{1}{2} \left(\underbrace{\left\langle S_I^2(\varphi_{\sf s}+\varphi_{\sf f}) \right\rangle_{\sf f}}_{\mathcal{B}} - \underbrace{\left\langle S_I(\varphi_{\sf s}+\varphi_{\sf f}) \right\rangle^2_{\sf f}}_{\mathcal{A}^2} \right) + \ldots\,,
\end{equation}
\ewt
where the expectation values are taken on the (fast) Gaussian action only, and we identified the effective action at the second order of perturbation theory.
In the following we make extensive use of the following well-known key property of Gaussian integrals ($\mathcal{G}$):
\begin{align} \label{eq:gaussian_int}
\ave{\re^{\ri \sum_k \alpha_k \phi_k}}_\mathcal{G}
	 = \re^{- \frac{1}{2} \sum_{k,k'} \alpha_k \alpha_{k'} \ave{\phi_k \phi_{k'}}_\mathcal{G}} \, ,
\end{align}
with $\phi$ the fields over which the integration is carried out.

The Gaussian correlation functions for the fields in first order of $\ln(\Lambda/\tilde\Lambda)$ read:
\begin{align}\label{eq:gaussian_corr_phi}
&\ave{\varphi_{{\sf f},q}(z_1) \varphi_{{\sf f},q}(z_2)}_{\sf f} =  \int_{\tilde{\Lambda}}^{\Lambda} \frac{dk}{2} \frac{J_0(kr)}{K_qk} = \frac{C_q(r)}{2K_q} \ln\frac{\Lambda}{\tilde{\Lambda}}\,,\\
\label{eq:gaussian_corr_theta}
&\ave{\theta_{{\sf f},q}(z_1) \theta_{{\sf f},q}(z_2)}_{\sf f} = \frac{C_q(r)K_q}{2} \ln\frac{\Lambda}{\tilde{\Lambda}}\,.
\end{align}
Here the logarithm captures the scaling behavior, and $C_q(r)$ is a function of $r=\sqrt{u_q^2(\tau_1-\tau_2)^2+(x_1-x_2)^2}$,
such that $C_q(0) = 1$.
In the following we will consider $C_q(r)$ to be suitably short-ranged; in the case of a sharp cutoff, $C_q(r) = J_0(\Lambda r)$ and the Bessel function $J_0$ does not satisfactorily fulfill this assumption, but $C_q(r)$ can be made sufficiently short-ranged with more refined cutoffs.
In particular, such an optimization can be achieved by a deformation of the integration $\int_{\tilde\Lambda}^\Lambda\rd p\rightarrow\int_0^\infty\rd p f_n(p,\Lambda)$ with $f_n = \Lambda^n/(p^n+\Lambda^n)$, in which the sharp cutoff is realized for $n\rightarrow\infty$.
An analytic result can be obtained in case of $n=2$, i.e. $C_q(r)=\Lambda r K_1(\Lambda r)$ with the modified Bessel function $K_1$ of the second kind.
This function has an exponentially decaying asymptotic form $zK_1(z)\approx\sqrt{z\pi/2}\re^{-z}$ which is thus sufficiently short-ranged.
To be more precise, the numerical values of $zK_1(z)$ above machine precision are confined to the interval $0\leq z<13\pi$, and moreover $K_1(2\pi)\approx0.0062$ is already pretty small for short distances of $r\sim2\pi/\Lambda$.
For this reason, we will later truncate the space-time integrations of a product containing the function $C_q$ to the interval $(0,\alpha)$ in which the cutoff parameter $\alpha \sim 2\pi/\Lambda = a$ is on the order of the lattice spacing.

For the sake of convenience, let us rewrite the interacting part of the action as:
\begin{equation}
S_I = \sum_\nu \int \rd^2z \, \left( \frac{h}{2} \mathcal{O}_h^\nu + g \sum_\mu  \mathcal{O}_{g,\mu}^\nu \right) \, ,
\end{equation}
where we introduced the shorthand notation:
\begin{equation}
\mathcal{O}_h^\nu =\re^{\ri \nu 2 \sqrt{2} \theta_s}
\quad \mbox{and} \quad
\mathcal{O}_{g,\mu}^\nu =\re^{\ri \nu \sqrt{2} \left(\theta_c + \varphi_s + \mu \theta_s \right)} \, .
\end{equation}
Thus we obtain for the first-order contribution:
\bwt
\begin{align}
\ave{\mathcal{O}_h^\nu}_{\sf f}  & = \,
	   \re^{\ri \nu 2 \sqrt{2} \theta_s} \, \re^{- 4 \ave{\theta_s^2}_{\sf f}} 
	   = \, \mathcal{O}_h^\nu \, \left(\frac{\Lambda}{\tilde{\Lambda}}\right)^{-2 K_s}
\label{eq:Oh_ave} \\
\ave{\mathcal{O}_{g,\mu}^\nu}_{\sf f}  & = \,
	\re^{\ri \nu \sqrt{2} \left(\theta_c + \varphi_s + \mu \theta_s \right)}  \re^{- \left(\ave{\theta_c^2}_{\sf f} + \ave{\varphi_s^2}_{\sf f} + \ave{\theta_s^2}_{\sf f} \right)}
	= \, \mathcal{O}_{g,\mu}^\nu \, \left(\frac{\Lambda}{\tilde{\Lambda}}\right)^{-\frac{1}{2} \left(K_c + \frac{1}{K_s} +K_s\right)}
\label{eq:Og_ave}
\end{align}
\ewt
where the operators on the right hand sides of the equations are intended to be constructed with slow fields only, and we dropped extra-labels for simplicity.
Together with the factor $(\Lambda/\tilde{\Lambda})^2$ from the rescaling of the integration domains for the new action after a RG (infinitesimal) step, and keeping in mind that $\Lambda/\tilde{\Lambda} = 1+\rd l$, Eqs.~\eqref{eq:Oh_ave}-\eqref{eq:Og_ave} return $S_I$ with the usual first-order dependence of the couplings $g$ and $h$ on the scaling dimensions, $D_h = 2 K_s$ and $D_g =  \left(K_c+K_s+K_s^{-1}\right)/2$, in the RG Eqs.~\eqref{eq:RG_flow}).

The many addends of the second-order contribution  could be compactly written as:
\bwt
\begin{equation} \label{eq:pert_second}
\mathcal{B}- \mathcal{A}^2 =
	\sum_{\nu, \nu'} \left(
	\frac{h^2}{4} \ave{\mathcal{O}_h^\nu \mathcal{O}_h^{\nu'}}_{\sf f \diamond} +
	\frac{g h}{2} \sum_\mu \left( \ave{\mathcal{O}_{g,\mu}^\nu \mathcal{O}_h^{\nu'}}_{\sf f \diamond} + \ave{\mathcal{O}_h^{\nu'} \mathcal{O}_{g,\mu}^\nu}_{\sf f \diamond} \right)
	+ g^2 \sum_{\mu,\mu'}  \ave{\mathcal{O}_{g,\mu}^\nu \mathcal{O}_{g,\mu'}^{\nu'}}_{\sf f \diamond}
	\right) \, ,
\end{equation}
\ewt
where we introduced the connected part of the quadratic expectation values, i.e., $\ave{\#_1 \#_2}_\diamond = \ave{\#_1 \#_2} - \ave{\#_1} \ave{\#_2}$.
It is straightforward to obtain which RG correction each addend gives rise to (hereafter, we drop the integration symbol over the Euclidean space, for the sake of space):
\bwt
\begin{align}
\ave{\mathcal{O}_h^\nu \mathcal{O}_h^{\nu'}}_{\sf f \diamond}  \longrightarrow & \,
	   \re^{\ri 2 \sqrt{2} \left(\nu \theta_{s,1} + \nu' \theta_{s,2} \right)} \,
	   		\re^{- 8 \ave{\theta_s^2}_{\sf f}}
				\left[ \re^{- 8 \nu \nu' \ave{\theta_{s,1} \theta_{s,2}}_{\sf f}} - 1 \right]
	\approx  -4 \nu \nu' K_s \rd l \,
		\re^{\ri 2 \sqrt{2} \left(\nu \theta_{s,1} + \nu' \theta_{s,2} \right)}
			C_s(z_{12}) \, ,
\label{eq:Oh_Oh_ave} \\
\ave{\mathcal{O}_{g,\mu}^\nu \mathcal{O}_h^{\nu'}}_{\sf f \diamond}  \longrightarrow & \,
	   \re^{\ri \sqrt{2} \left(\nu \theta_{c,1} - \nu \varphi_{s,1} +\nu \mu \theta_{s,1} + 2 \nu' \theta_{s,2}\right)} \,
	   		\re^{- \left(\ave{\theta_c^2}_{\sf f} + \ave{\varphi_s^2}_{\sf f} + 5 \ave{\theta_s^2}_{\sf f}\right)} \,
				\left[ \re^{- 4 \mu \nu \nu' \ave{\theta_{s,1} \theta_{s,2}}_{\sf f}} - 1 \right]
	\nonumber \\
	\approx &  -2 \mu \nu \nu' K_s \rd l \,
		\re^{\ri \sqrt{2} \left(\nu \theta_{c,1} - \nu \varphi_{s,1} +\nu \mu \theta_{s,1} + 2 \nu' \theta_{s,2}\right)}
			C_s(z_{12}) \, ,
\label{eq:Og_Oh_ave} \\
\ave{\mathcal{O}_h^{\nu'} \mathcal{O}_{g,\mu}^\nu}_{\sf f \diamond}   \longrightarrow & \, \ave{\mathcal{O}_{g,\mu}^\nu \mathcal{O}_h^{\nu'}}_{\sf f \diamond} \, \left( z_1 \leftrightarrow z_2\right) \, ,
\label{eq:Oh_Og_ave} \\
\ave{\mathcal{O}_{g,\mu}^\nu \mathcal{O}_{g,\mu'}^{\nu'}}_{\sf f \diamond}  \longrightarrow & \,
	   - \re^{\ri \sqrt{2} \left(\nu \theta_{c,1} + \nu' \theta_{c,2} - \nu \varphi_{s,1} - \nu' \varphi_{s,2} + \nu \mu \theta_{s,1} + \nu' \mu' \theta_{s,2}\right)} \,
	\nonumber \\ &
	   \hspace{2cm}	 \cdot \,	\re^{- 2 \left(\ave{\theta_c^2}_{\sf f} + \ave{\varphi_s^2}_{\sf f} + \ave{\theta_s^2}_{\sf f}\right)} \,
				\left[ \re^{- 2 \nu \nu' \left( \ave{\theta_{c,1} \theta_{c,2}}_{\sf f} + \ave{\varphi_{s,1} \varphi_{s,2}}_{\sf f} + \mu \mu' \ave{\theta_{s,1} \theta_{s,2}}_{\sf f} \right)} - 1 \right]
	\nonumber \\
	&\approx  \nu \nu' \rd l  \,
		\re^{\ri \sqrt{2} \left(\nu \theta_{c,1} + \nu' \theta_{c,2} - \nu \varphi_{s,1} - \nu' \varphi_{s,2} + \nu \mu \theta_{s,1} + \nu' \mu' \theta_{s,2}\right)}
			\left[ K_c C_c(z_{12})  + \left( \tfrac{1}{K_s} + \mu \mu' K_s\right) C_s(z_{12})\right] \, ,
\label{eq:Og_Og_ave}
\end{align}
\ewt
where we indicated the dependence on the Euclidean coordinates $z_1$ and $z_2$ by simple labels, and we took into account the additional Baker-Campbell-Hausdorff phase factor arising in Eq.~\eqref{eq:Og_Og_ave}.

We can now split the integrals into the integral of the center of mass coordinate $(z_1+z_2)/2$, and the relative coordinate $z_{12}=(z_1-z_2)$, and recall the assumed short-range character of the $C_q$ functions - see the previous discussion, just below Eqs.~\eqref{eq:gaussian_corr_phi}-\eqref{eq:gaussian_corr_theta}.
This constrains the two points to be in a neighbourhood of size $\delta z \simeq (\alpha, \alpha/u_q$) from each other, where $\alpha$ is of the order of the lattice spacing $a$.  The effect is two-fold:
first, this allows us to Taylor-expand the field operators in Eqs.~\eqref{eq:Oh_Oh_ave}-\eqref{eq:Og_Og_ave};
second, a factor $\alpha^2/u_q$ is generated by the integration.
%
Henceforth we group the addends of Eq.~\eqref{eq:pert_second} according to their RG behaviour:

\bwt
\begin{itemize}
\item The terms with $\nu'=-\nu$ in Eq.~\eqref{eq:Oh_Oh_ave} contribute to the quadratic part of the action:
\begin{align}
	& +  \rd l \, 4 h^2 \, \alpha^4 \, \tfrac{K_s}{u_s} \, \left[ (\partial_x \theta_{s})^2 +  \tfrac{1}{u_s^2}  (\partial_\tau \theta_{s})^2 \right]  
	= - \rd l \,  4 h^2 \, \alpha^4 \, \tfrac{K_s^3}{u_s^2}  \, \left[ \tfrac{1}{u_s} (\partial_\tau \varphi_{s})^2 +  u_s  (\partial_x \varphi_{s})^2 \right]  \, ,
\end{align}
as well as those with $\nu'=-\nu$ and $\mu'=+\mu$ in Eq.~\eqref{eq:Og_Og_ave}: 
\begin{align}
	- \rd l \, 2 g^2 \, \alpha^4 \, \left\{
		\left[\left(\partial_x \theta_{c} - \partial_x \varphi_{s}\right)^2 + \left(\partial_x \theta_{s} \right)^2\right]
			\overbrace{\left[ \tfrac{K_c}{u_c}  + \tfrac{1}{u_s} \left( \tfrac{1}{K_s} + K_s\right) \right]}^{p_1}
	+ 	\left[\left(\partial_\tau \theta_{c} - \partial_\tau \varphi_{s}\right)^2 + \left(\partial_\tau \theta_{s} \right)^2\right]
			\overbrace{\left[ \tfrac{K_c}{u_c^3}  + \tfrac{1}{u_s^3} \left( \tfrac{1}{K_s} + K_s\right) \right]}^{p_2}
	\right\}
\nonumber \\
	\approx + \rd l \, 2 g^2 \, \alpha^4 \, \left\{
		   \tfrac{1}{u_c} (\partial_\tau \varphi_{c})^2 \  K_c^2 \tfrac{1}{u_c} p_1
		+ u_c (\partial_x \varphi_{c})^2 \  K_c^2 u_c p_2
		+ \tfrac{1}{u_s} (\partial_\tau \varphi_{s})^2 \  \left[ \tfrac{K_s^2}{u_s} p_1- u_s p_2 \right]
		+ u_s (\partial_x \varphi_{s})^2 \  \left[ 	 K_s^2 u_s p_2 - \tfrac{ p_1}{u_s}\right]
		\right\} \, , \label{Kcorrections}
\end{align}
where we dropped two terms mixing $\varphi_s$ and $\theta_c$, thus the spin and change sectors.
Such terms break the possibility of rigorously splitting the system into these separate sectors and would require instead to introduce a unitary matrix, dependent on $l$, to diagonalize the Gaussian part of the action at each scale.
For the sake of simplicity, however, we neglect this mixing and we consider only the (standard) correction to the Luttinger parameters $K_s$ and $K_c$;

\item the terms with $\nu'=-\nu$ and $\mu'= -\mu$ in Eq.~\eqref{eq:Og_Og_ave} generate a correction to $\mathcal{O}_h$:
\begin{align}
	& + \rd l \, 2 g^2 \alpha^2 \,
		\cos\left( 2 \sqrt{2} \theta_{s} \right)
			\left[ K_c \tfrac{1}{u_c}  + \left( \tfrac{1}{K_s} - K_s\right) \tfrac{1}{u_s} \right] \, ;
\end{align}
\item finally, the terms with $\nu'= - \nu \mu$ in Eqs.~\eqref{eq:Og_Oh_ave}-\eqref{eq:Oh_Og_ave} give rise to a contribution proportional to $\mathcal{O}_g+\mathcal{O}_g^\dag$:
\begin{align}
	&  - \rd l \, 2 g h \alpha^2 K_s \tfrac{1}{u_s}
		\sum_\mu \cos \left(\sqrt{2} (\theta_{c} - \varphi_{s} - \mu \theta_{s})\right)  \, ;
\end{align}
\item all other terms are considerably more irrelevant.
\end{itemize}
\ewt

\begin{figure*}[ht]
    \centering
    \includegraphics[width=0.245\textwidth]{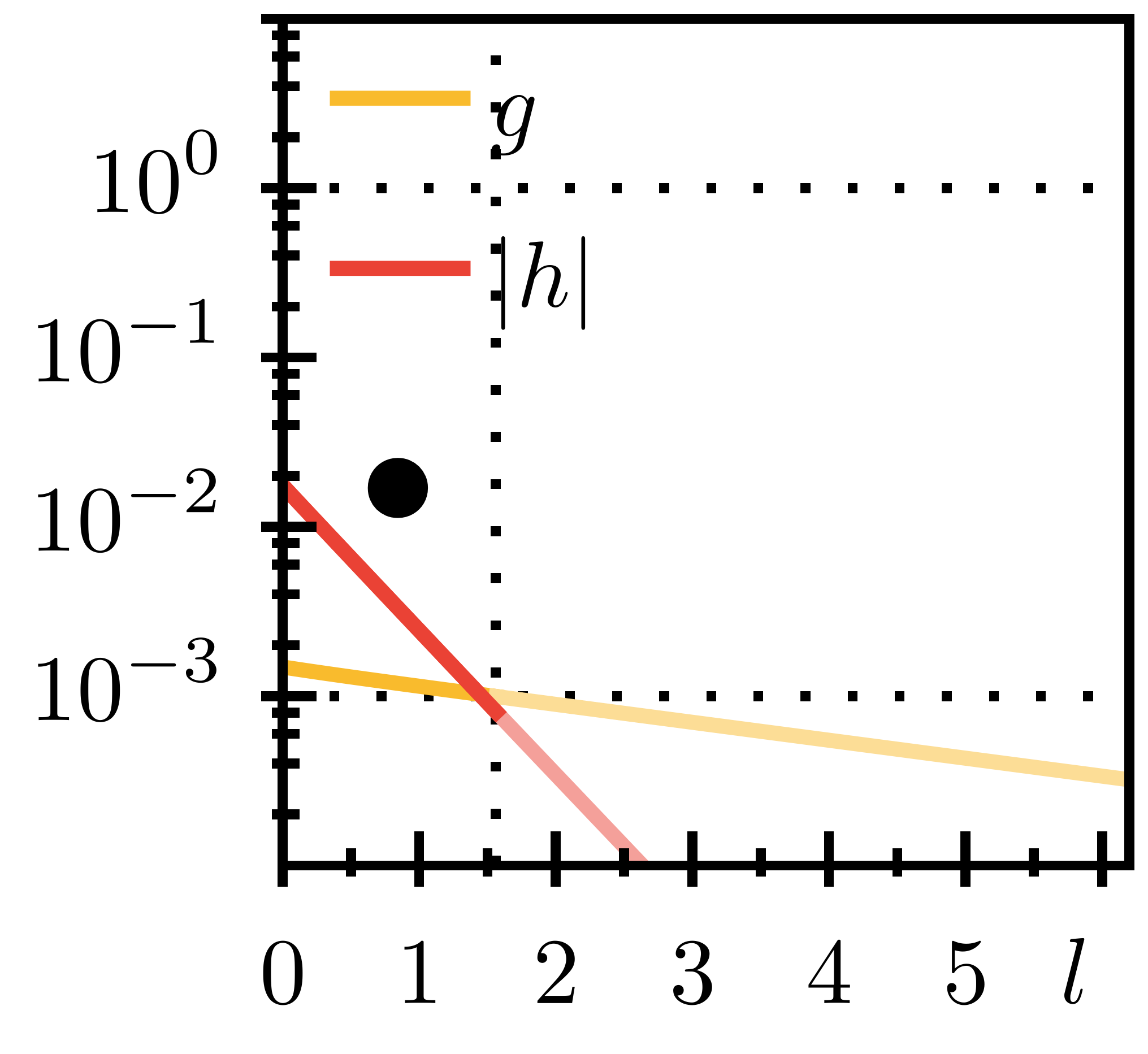}\llap{\parbox[b]{5.8cm}{(a)\\\rule{0ex}{0.9cm}}}\llap{\parbox[b]{5.cm}{$l^*$\\\rule{0ex}{0.3cm}}}\llap{\parbox[b]{3.cm}{$t^*$\\\rule{0ex}{3.2cm}}}
    \includegraphics[width=0.245\textwidth]{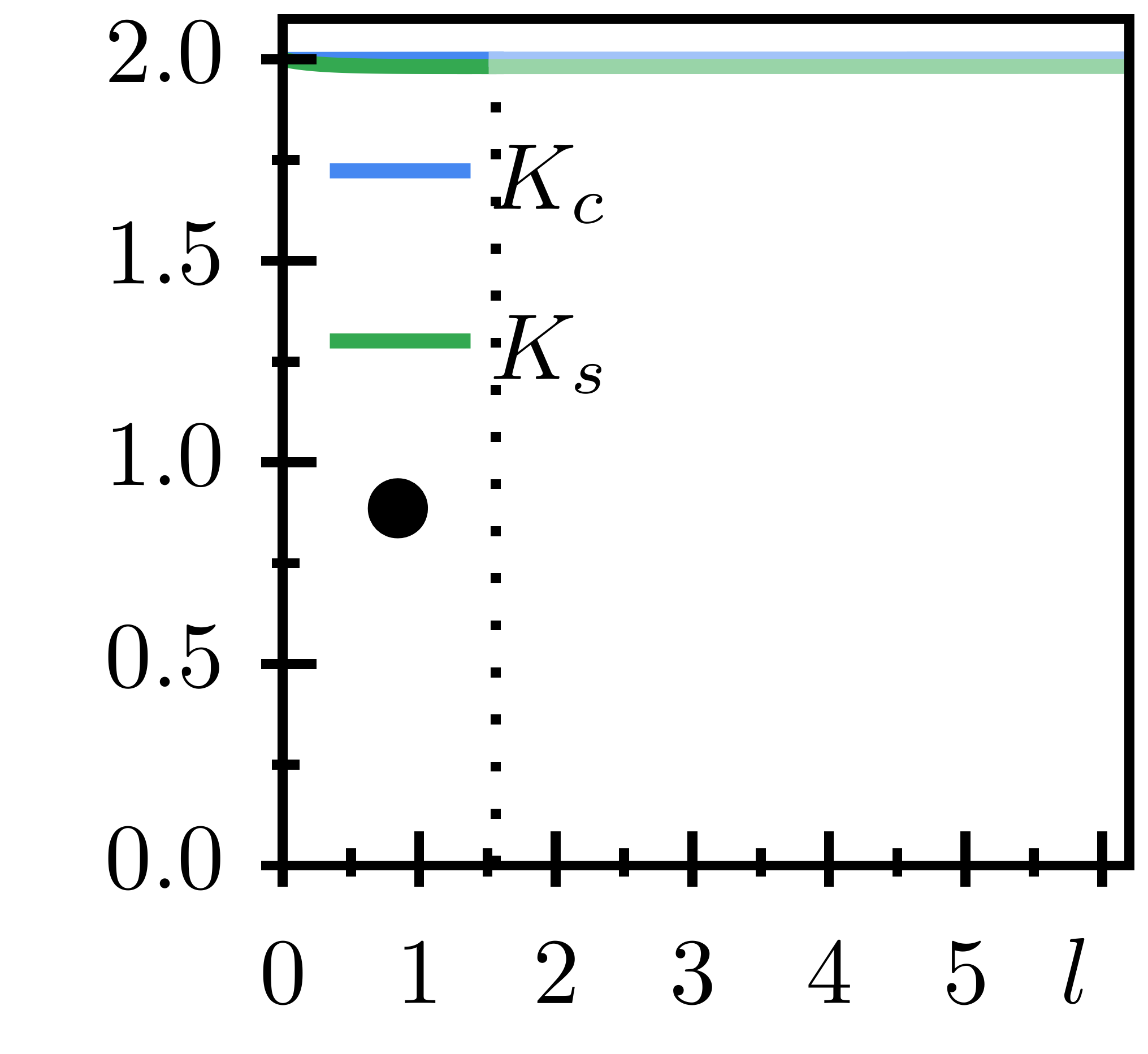}\llap{\parbox[b]{5.8cm}{(b)\\\rule{0ex}{0.9cm}}}\llap{\parbox[b]{5.cm}{$l^*$\\\rule{0ex}{0.3cm}}}
    \hfil
    \includegraphics{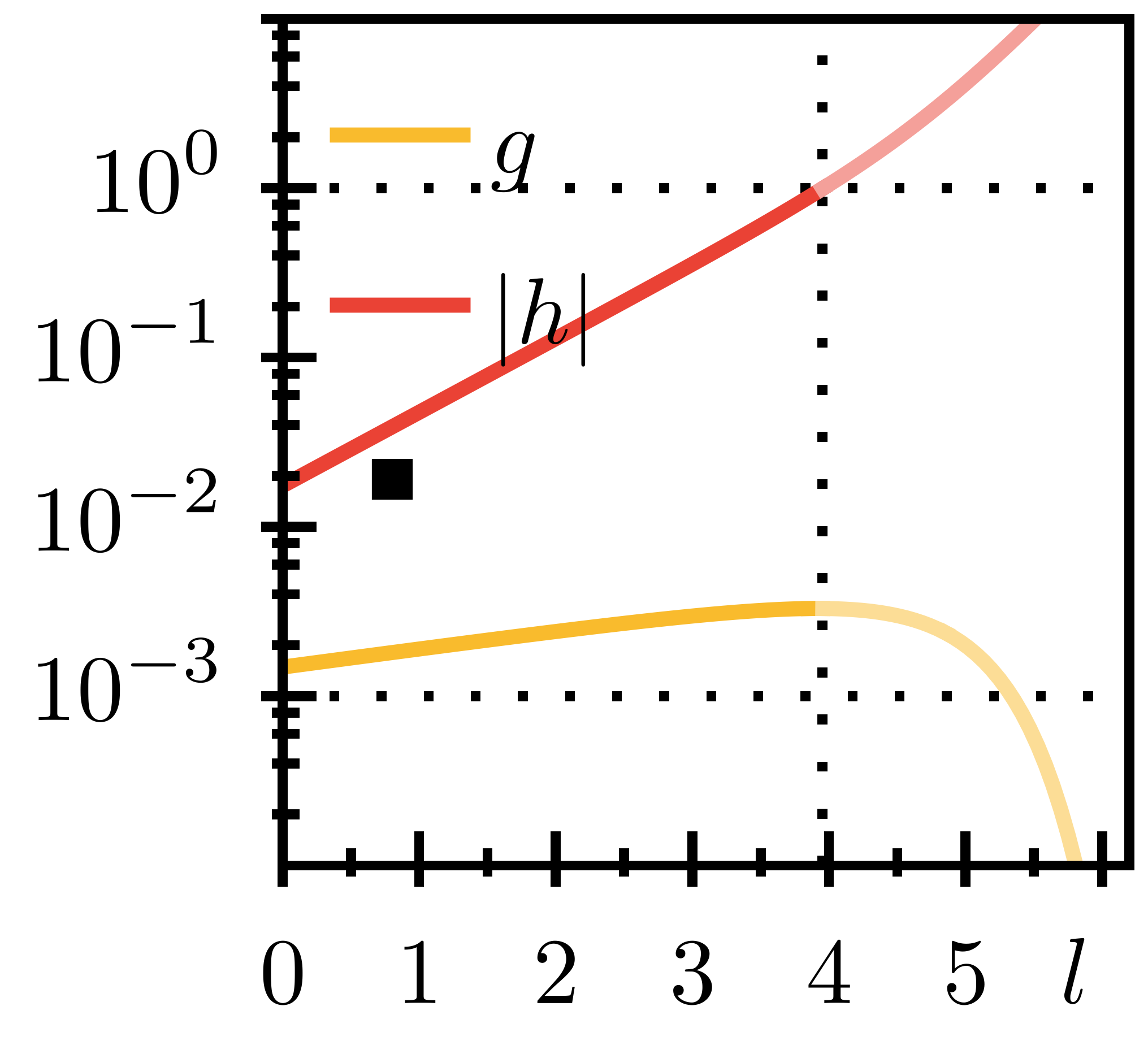}\llap{\parbox[b]{5.8cm}{(c)\\\rule{0ex}{0.9cm}}}
    \hfil
    \includegraphics{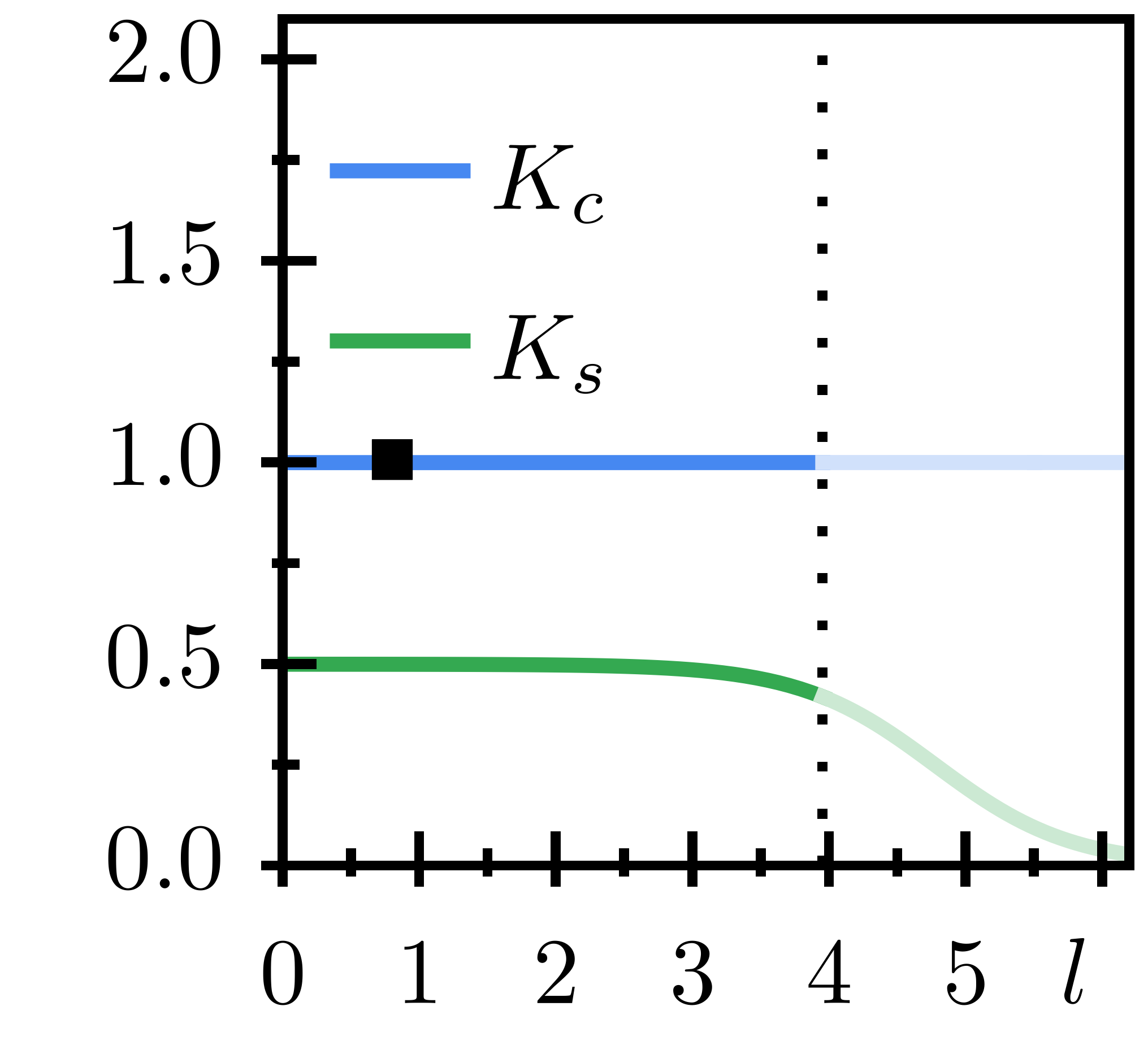}\llap{\parbox[b]{5.8cm}{(d)\\\rule{0ex}{0.9cm}}}
    \\
    \includegraphics{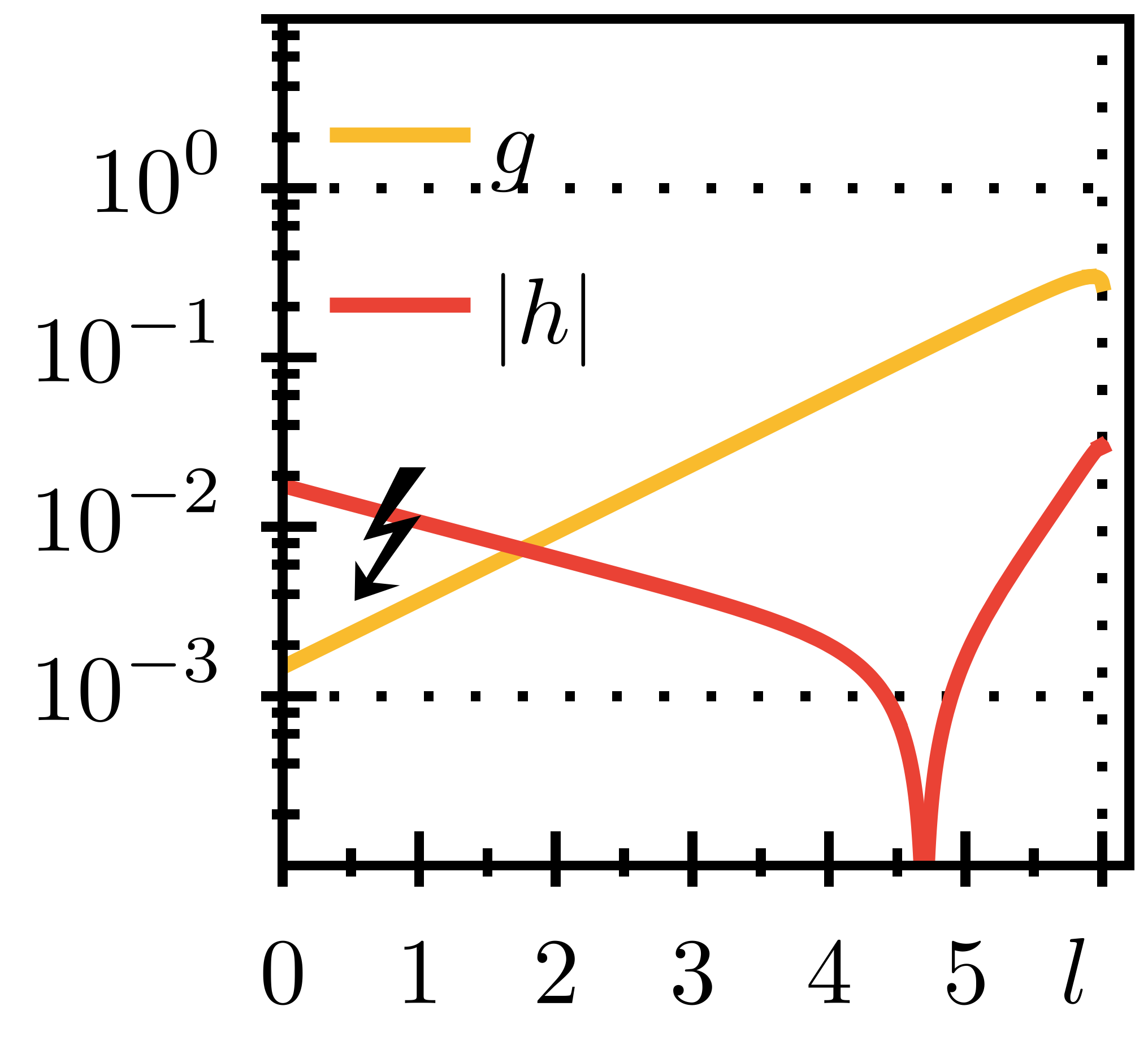}\llap{\parbox[b]{5.8cm}{(e)\\\rule{0ex}{0.9cm}}}
    \hfil
    \includegraphics{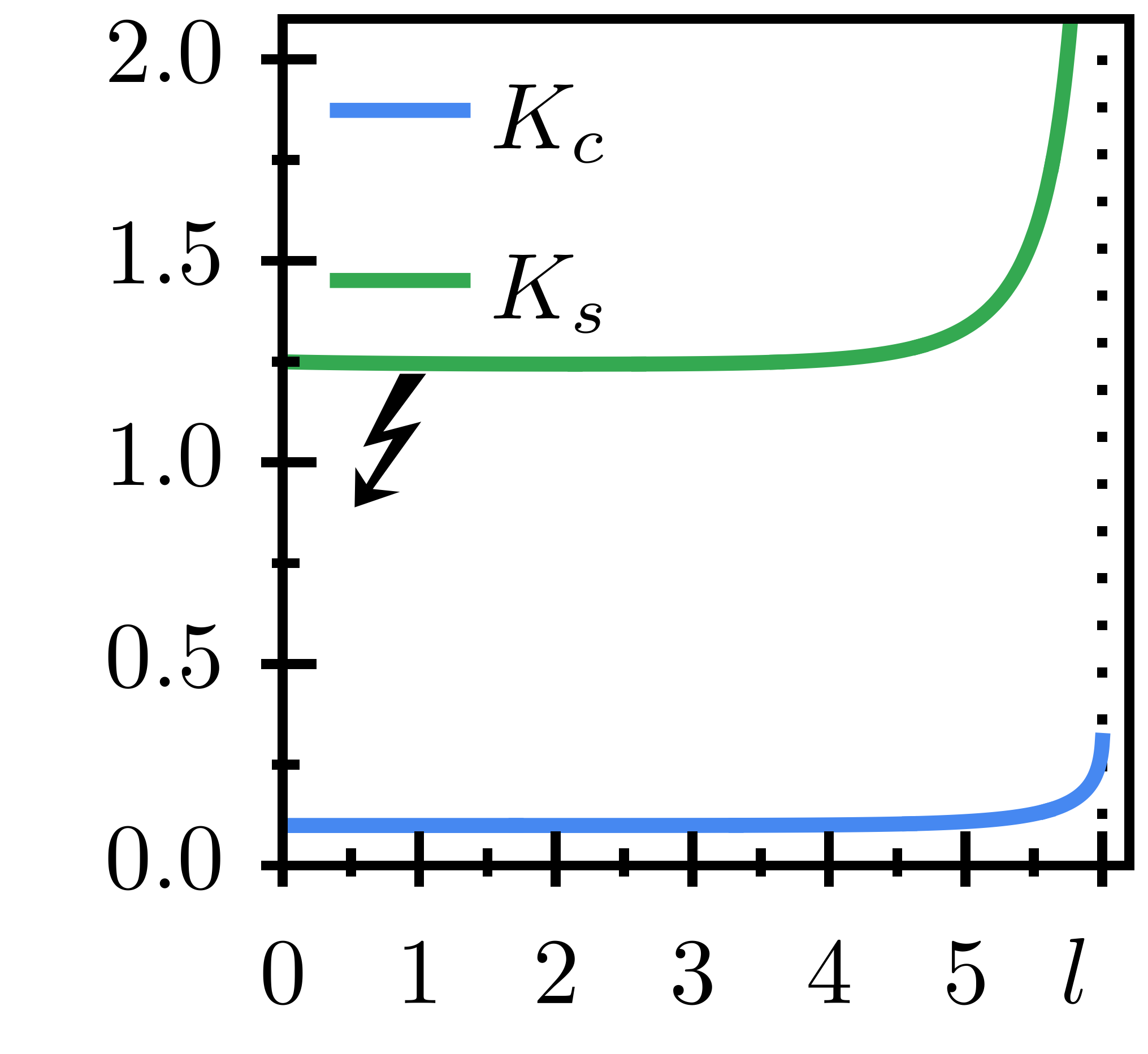}\llap{\parbox[b]{5.8cm}{(f)\\\rule{0ex}{0.9cm}}}
    \hfil
    \includegraphics{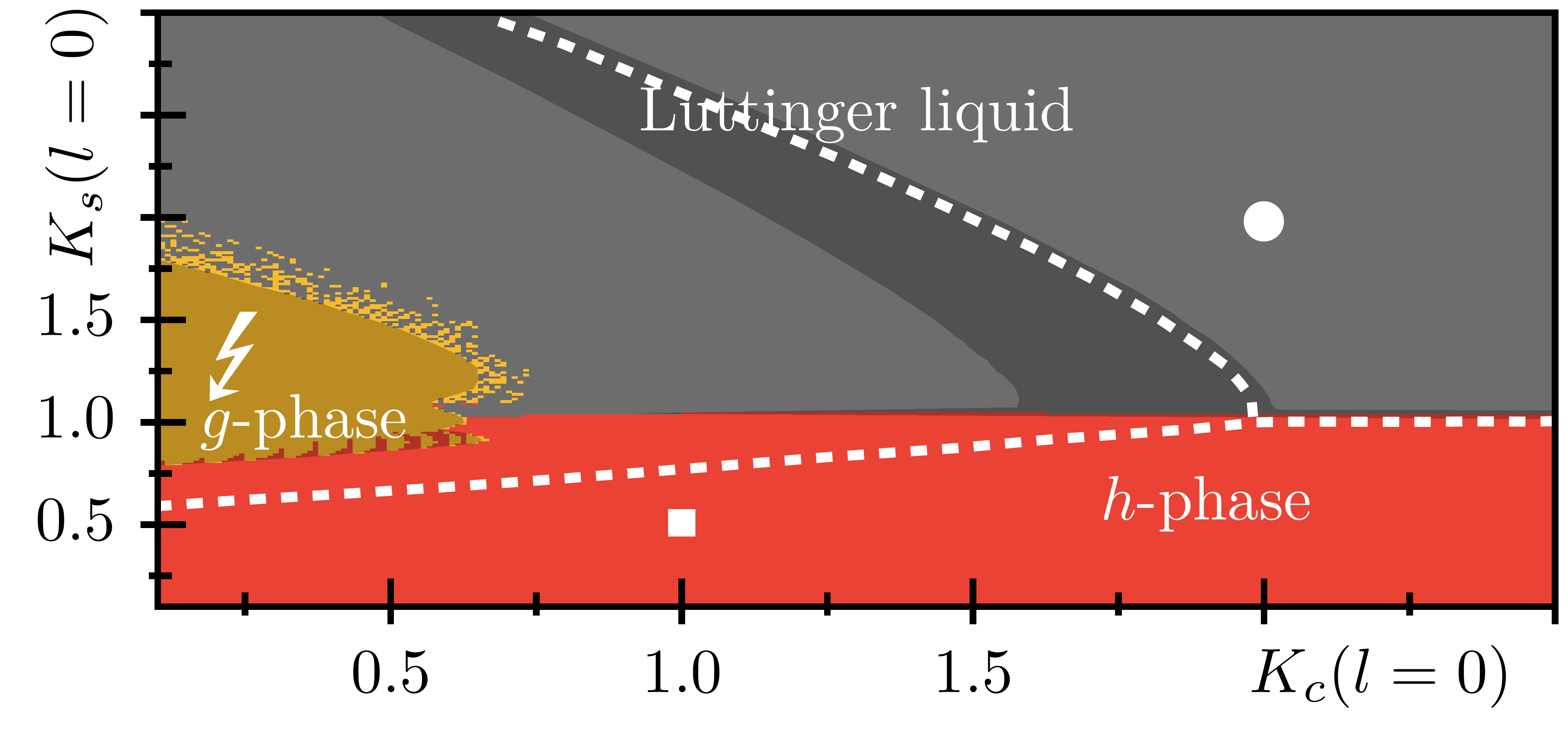}\llap{\parbox[b]{15cm}{\color{white}(g)\\\rule{0ex}{3.5cm}}}
    \caption{(a)-(f) Flow of the couplings $g$, $h$ and Luttinger parameters $K_c$, $K_s$ versus the flow parameter $l$ at fixed $V_\perp=t$ and $\Omega=0.05t$. The initial conditions of $K_c(l=0)$, $K_s(l=0)$ are indicated in the phase diagram presented in (g).
    We focus here on the second order RG refinements deep inside the phases predicted by the first order RG equations.
    The Luttinger liquid in the top right part of the phase diagram remains fairly standard, as indicated in panels (a) and (b):
    Initial values of the couplings flow exponentially towards zero, and the Luttinger liquid parameters remain constant.
    The $h$-dominated phase is slightly refined: we note the expected exponential growth of the coupling while the Luttinger liquid parameter $K_s$ flows towards zero.
    On the contrary, the $g$-dominated phase is drastically refined, as the Luttinger parameters show sudden divergences [see panels (e), (f), and an explanation in the text], causing the couplings to diverge accordingly.
    These divergences are followed by a non-monotonicity in the growth of the couplings which is better visualized by gradually moving the initial values of $K_{c/s}$ from the Luttinger towards the $g$-dominated phase, a path we present in Fig.~\ref{fig:RG_flow_details_2}.}
    \label{fig:RG_flow_details}
\end{figure*}

In our numerical solutions of the RG equations, we checked that small variations of the non-universal parameter $\alpha$ around the lattice spacing $a$ do not qualitatively affect the phase diagram in Fig. \ref{fig:RG_flow} (a), consistently with having a sufficiently sharp cutoff function $C_q$.

Finally, we point out that, differently from the standard sine-Gordon interactions, the terms in Eq. \eqref{Kcorrections} generate also a non-trivial flow of the velocities. In particular, there are second-order corrections to $u_c$ and $u_s$ proportional to $\rd l g^2 \left(u_c^2 -u_s^2\right)/u_c^2$. In the regime with a bare value $g \ll t$ and small interactions, the flow of the velocities is expected to have only negligible effects and we do not take in consideration their flow. To observe a quantitative difference between the bare and physical values of the velocity, we can compare the gray (vortex phase) and red ($\nu=1$ resonance) curves in Fig. \eqref{fig:dynamics_velocity}. In the vortex phase the system is described by a standard sine-Gordon model which fulfills Lorentz invariance (thus the velocities do not flow). Only a small difference in $u_s$ can be observed when comparing the red and gray curves, thus confirming that the flow of the velocities is practically negligible in our regime of interest.

\begin{figure*}[ht]
    \centering
    \includegraphics{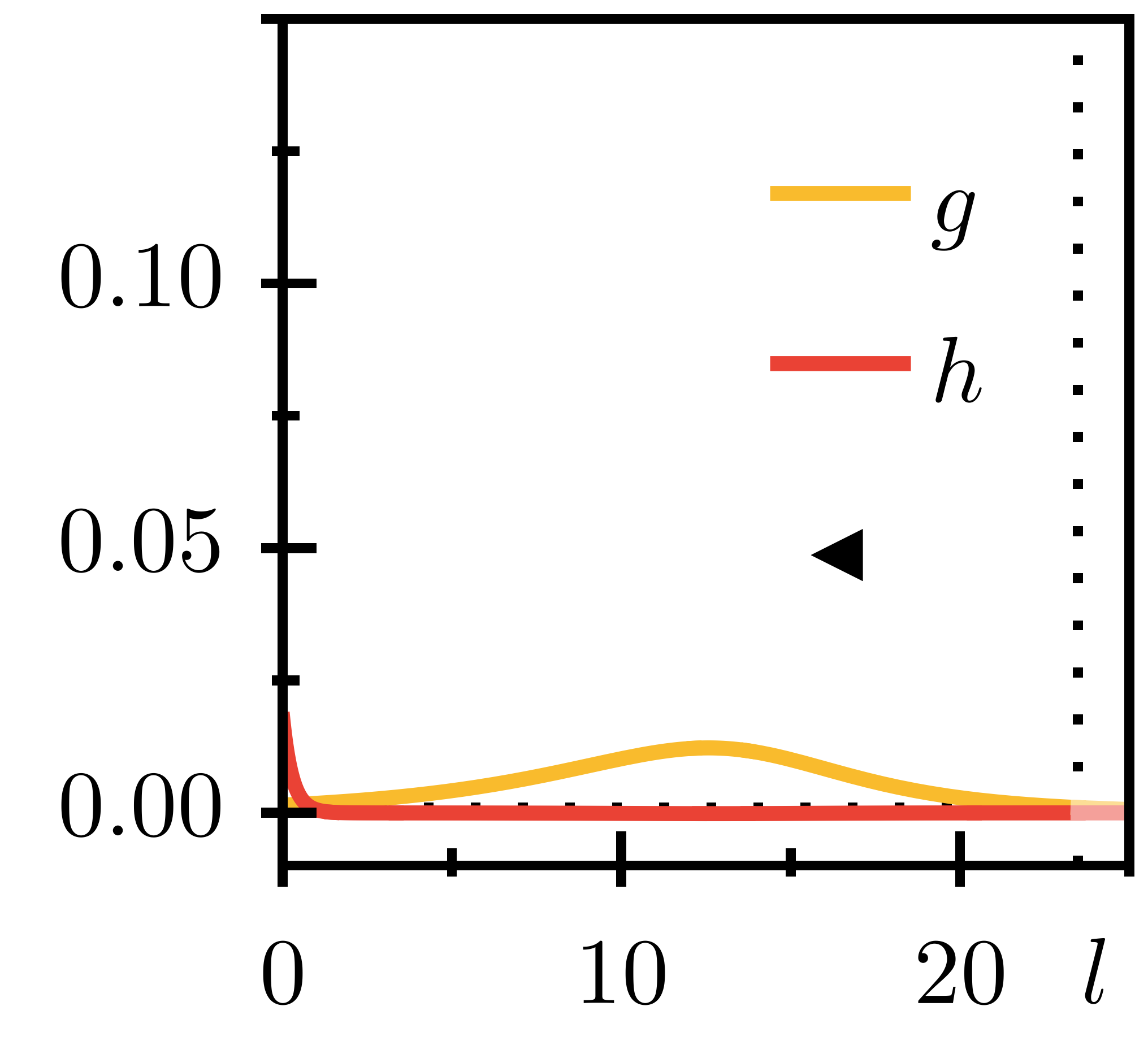}\llap{\parbox[b]{5.8cm}{(a)\\\rule{0ex}{3.35cm}}}
    \hfil
    \includegraphics{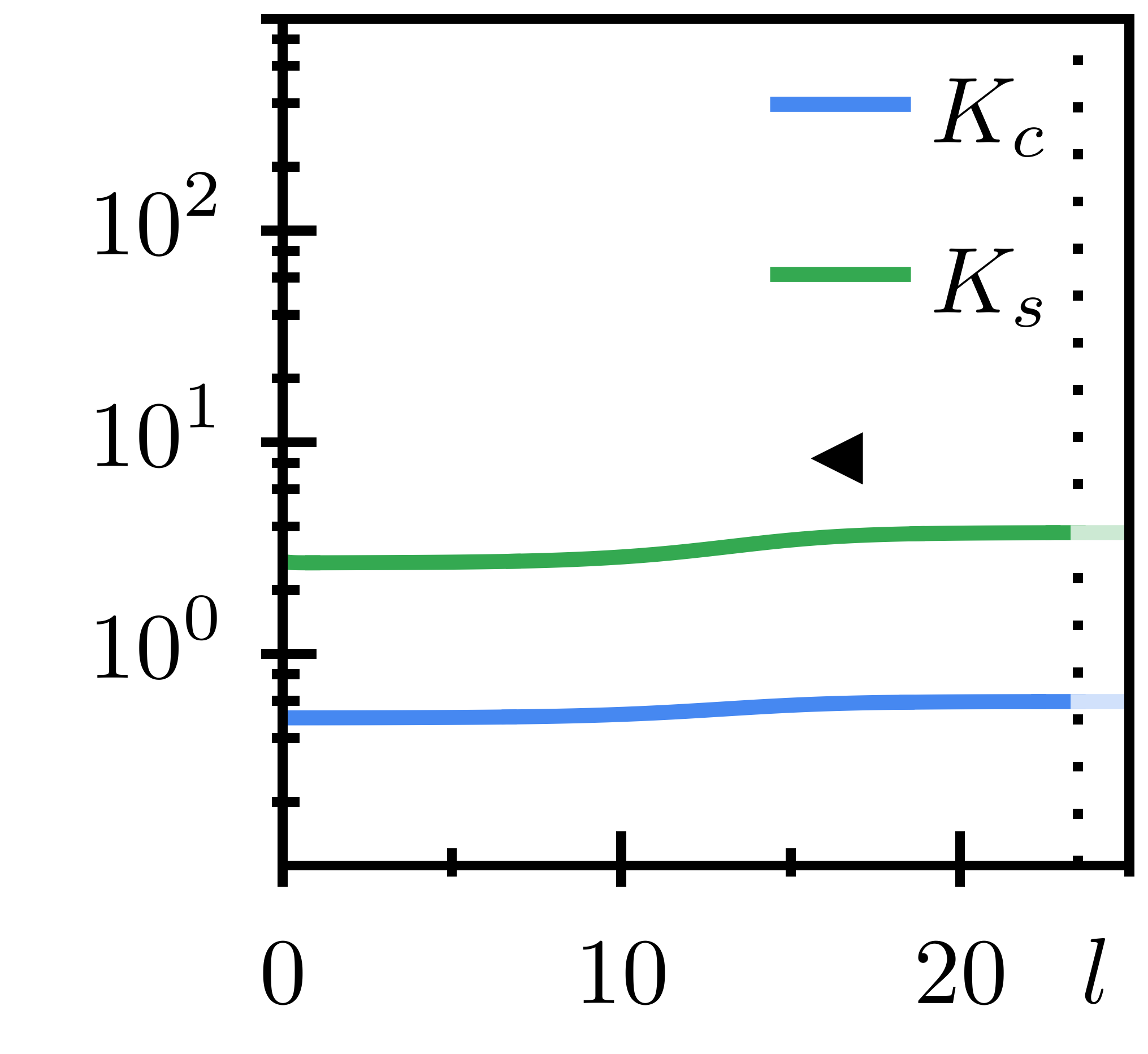}\llap{\parbox[b]{5.8cm}{(b)\\\rule{0ex}{3.35cm}}}
    \hfil
    \includegraphics{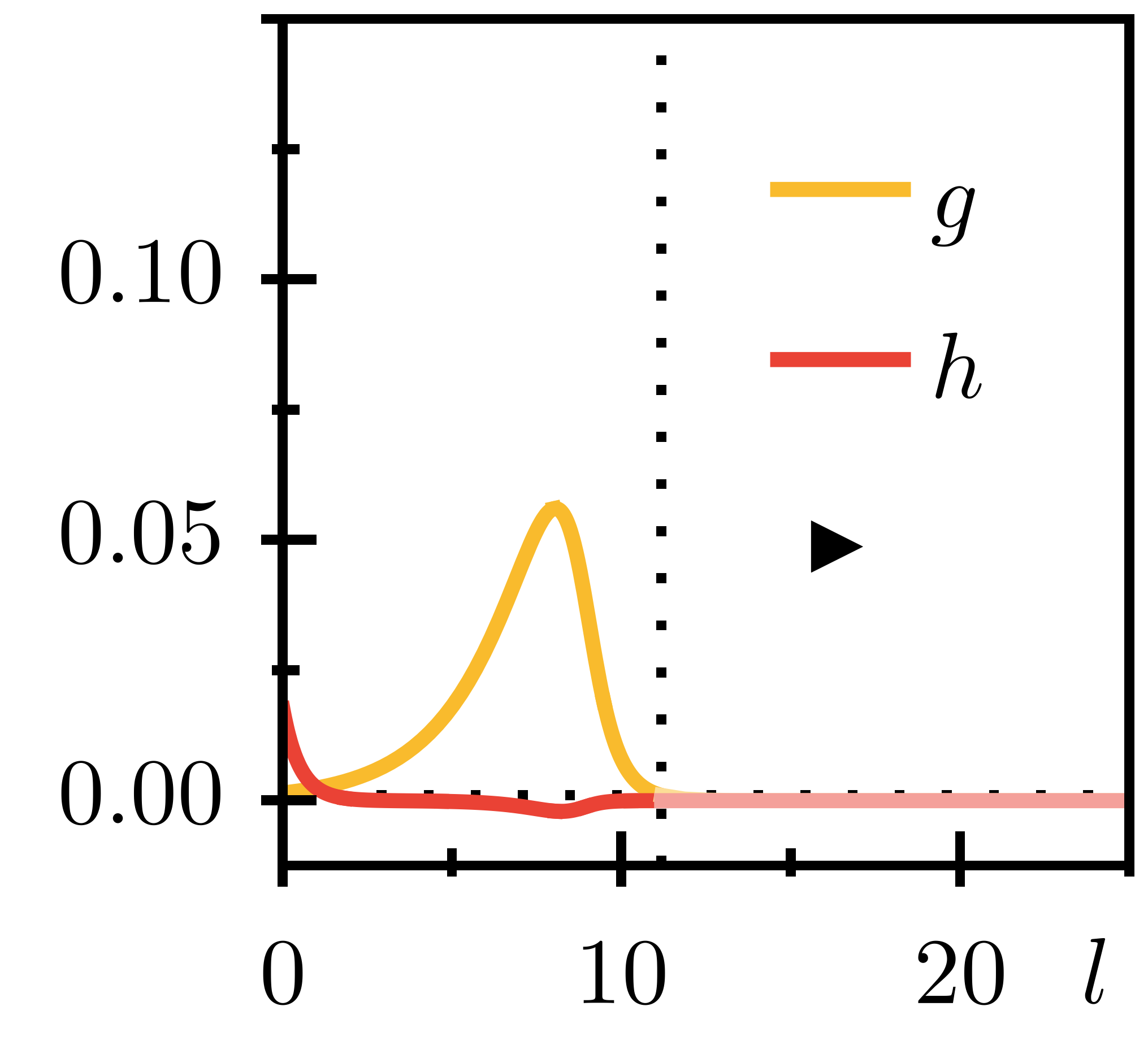}\llap{\parbox[b]{5.8cm}{(c)\\\rule  {0ex}{3.35cm}}}
    \hfil
    \includegraphics{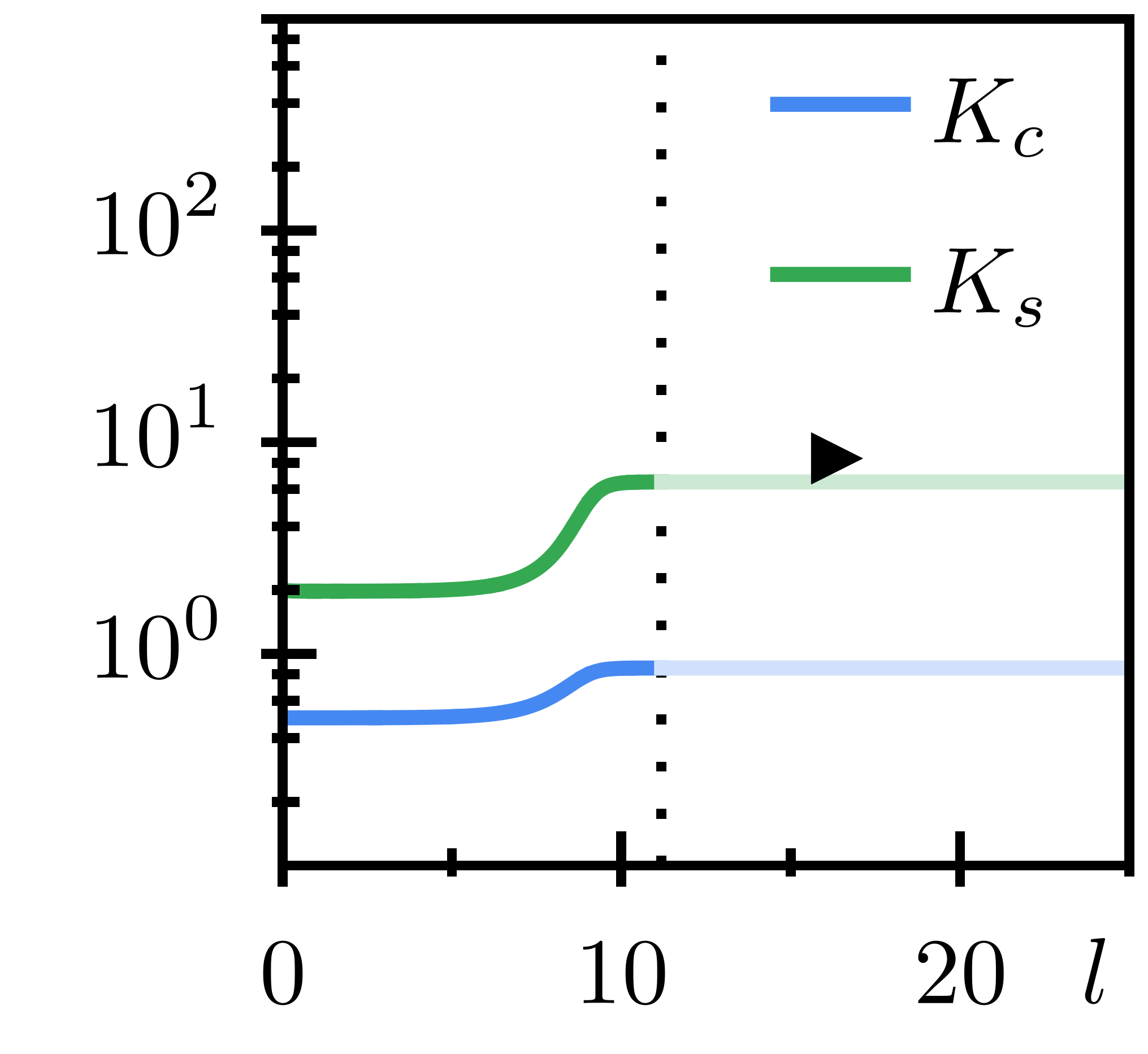}\llap{\parbox[b]{5.8cm}{(d)\\\rule{0ex}{3.35cm}}}
    \\
    \includegraphics{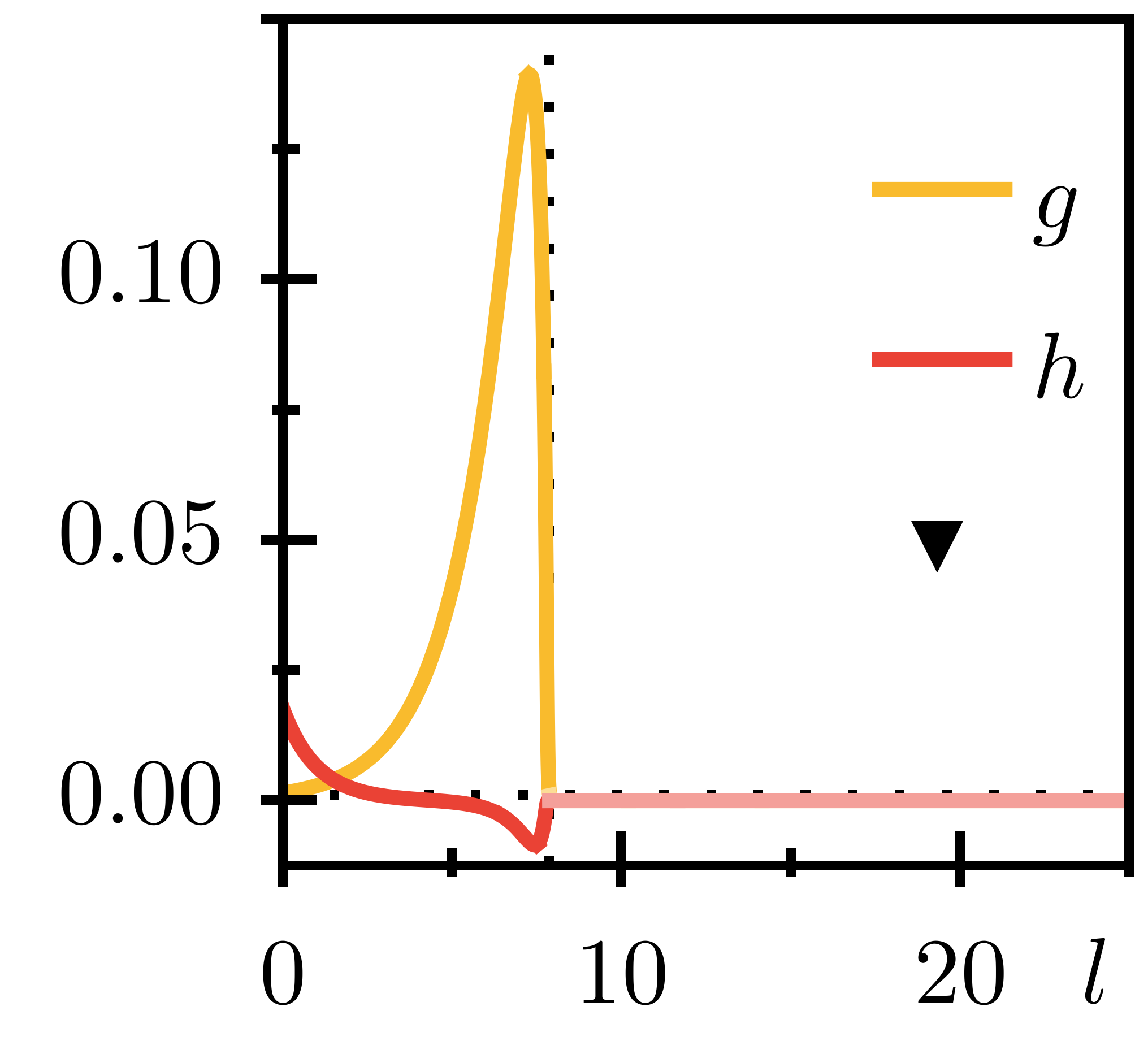}\llap{\parbox[b]{5.8cm}{(e)\\\rule{0ex}{3.35cm}}}
    \hfil
    \includegraphics{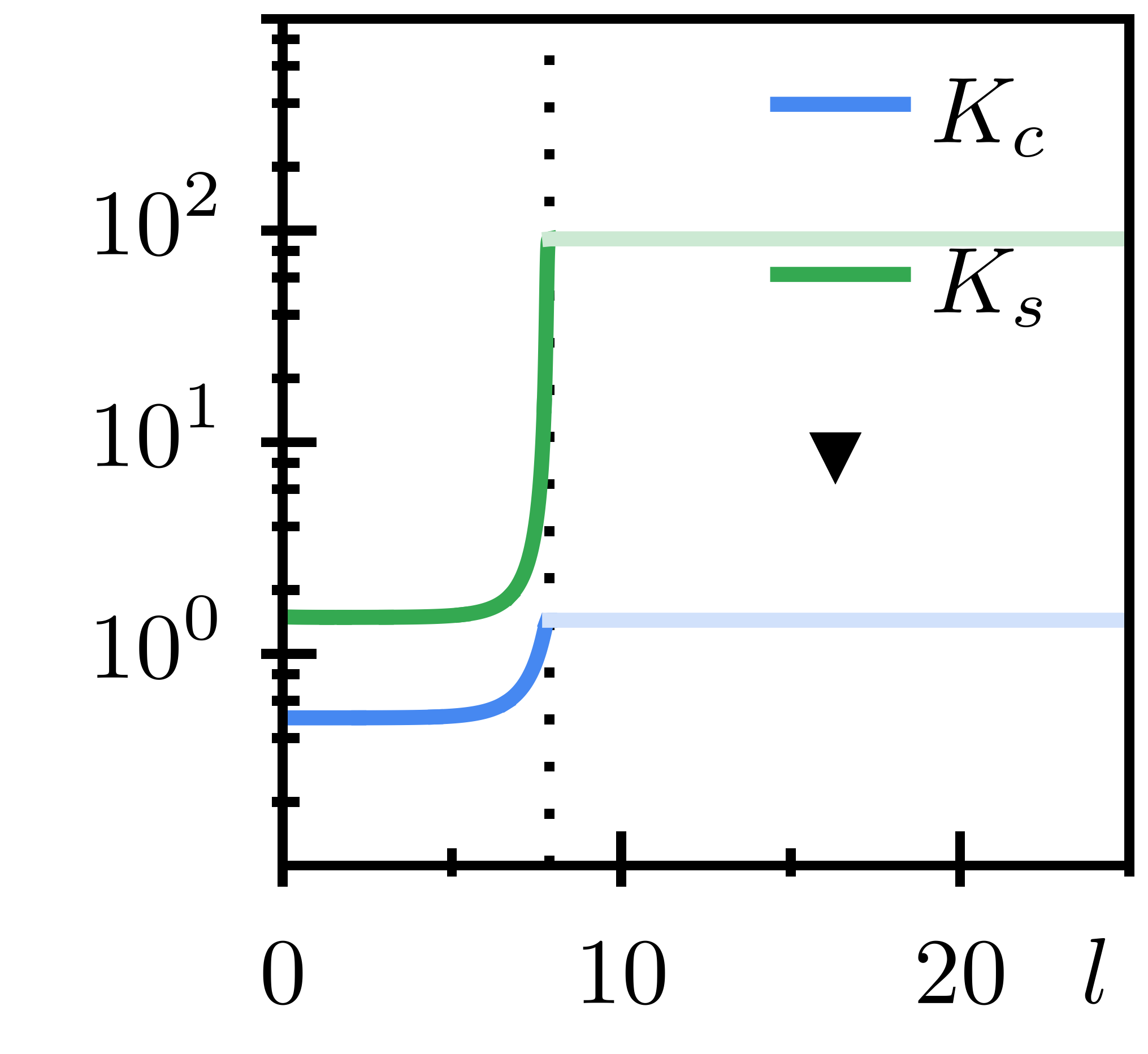}\llap{\parbox[b]{5.8cm}{(f)\\\rule{0ex}{3.35cm}}}
    \hfil
    \includegraphics{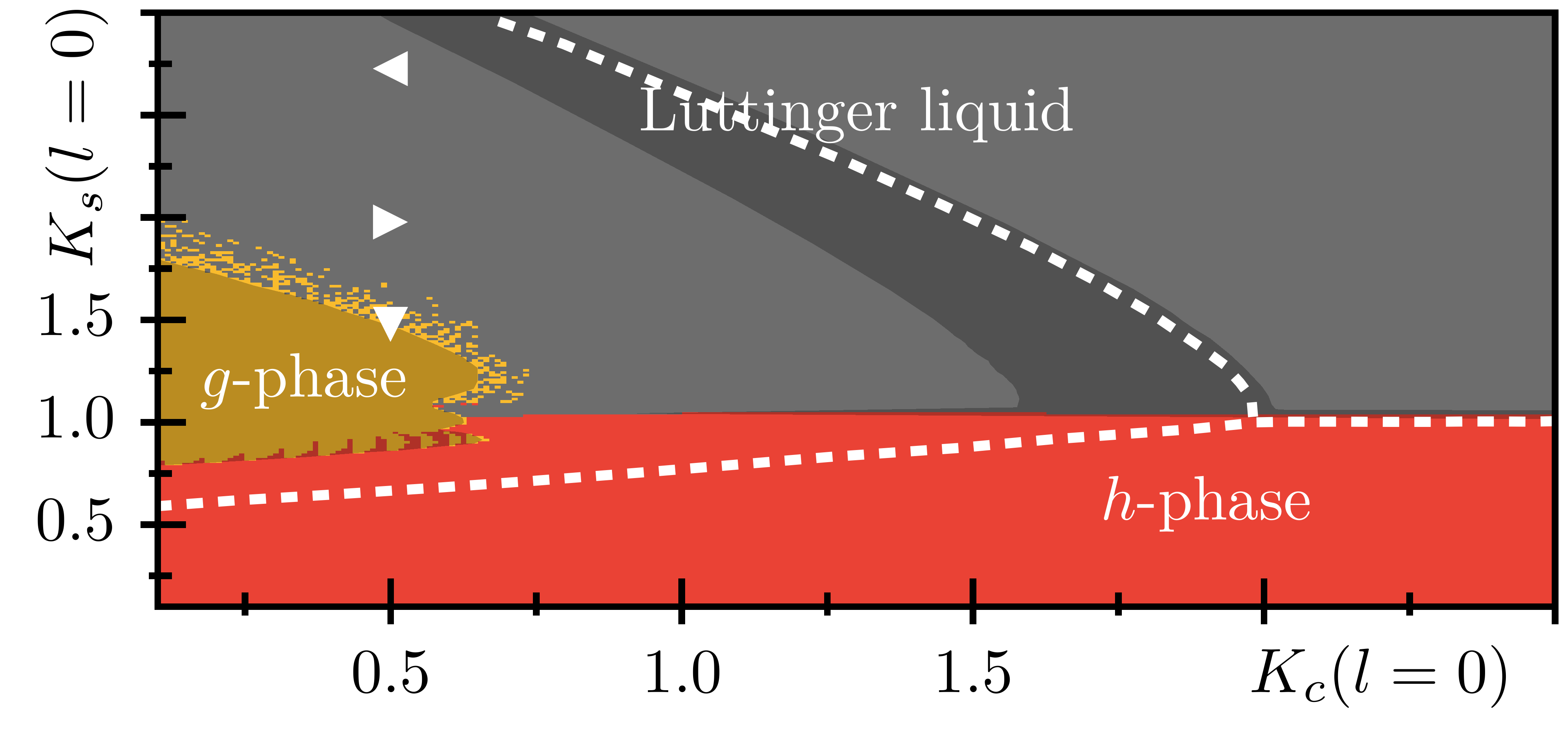}\llap{\parbox[b]{15cm}{\color{white}(g)\\\rule{0ex}{3.35cm}}}
    \caption{(a)-(f) Flow of the couplings $g$, $h$ and Luttinger parameters $K_c$, $K_s$ versus the flow parameter $l$ at fixed $V_\perp=t$ and $\Omega=0.05t$. The initial conditions of $K_c(l=0)$, $K_s(l=0)$ are indicated in the phase diagram presented in (g).
    By extending the first order RG approximation to second order, we note that the Luttinger liquid phase grows drastically larger, reaching far inside the region predicting the relevance of $g$ at first order.
    This Luttinger liquid phase is not trivial, as the Luttinger parameters flow to asymptotic values significantly different from the initial ones, followed by a rapid change of the coupling amplitudes.
    We observe thus a subtle non-monotonic behavior of the coupling constants, which smoothly extends towards the $g$-dominated region.
    Towards the $g$-dominated region the asymptotic value of $K_s$ approaches infinite. We report thus a sudden divergence of $K_s$ inside the $g$ phase (see Fig.~\ref{fig:RG_flow_details} (f)), demonstrating the breakdown of the second order RG approach as indicated by the shaded regions in the phase diagrams.}
    \label{fig:RG_flow_details_2}
\end{figure*}

\section{Numerical solutions of the renormalization group equations} \label{App:nrg}
In the following, we provide additional information about the numerical solutions of the renormalization group equations presented in Eqs. (\ref{eq:RG_flow}).

Fig.~\ref{fig:RG_flow} is obtained by explicitly solving the differential equations using an implict Runge-Kutta method, in particular by fixing the initial/boundary conditions of Eqs. (\ref{eq:RG_flow}).
In this four-dimensional phase space, we put on the axes the Luttinger parameters before the RG flow $K_q\equiv K_q(l=0)$, while fixing $g/h(l=0)$.
We thus manage to present two-dimensional cuts of the general phase diagram which display in which regions of these bare parameters the operators are relevant.
Representatives of the solutions of the RG equations for $\Omega=0.05t$ and $V_\perp=t$ are presented in Fig.~\ref{fig:RG_flow_details}.
The flow of the Luttinger parameters in the MPS explored region is in general slow, and the bare values roughly correspond to the renormalized ones at $l=l^*$ (see panels (c) and (d) presenting the flow in the $h$-phase).
This way we can conveniently combine the phase diagram and the path of our MPS simulations.
In particular, feeding the extracted Luttinger parameters from the MPS simulations as initial conditions, we confirm by estimating the energy gap $\Delta$ presented in Fig. 2 (b) that the RG predicts a flow inside the $h$ phase in the thermodynamic limit, after a critical value of $V_\perp\approx6t$.
It must be stressed that the estimate of the RG predicted energy gap are given in arbitrary units, and, being unable to resolve it with the MPS simulations, its quantitative size remains unknown.

By observing the different points of the phase diagram presented in Fig.~\ref{fig:RG_flow_details}, we observe a Luttinger liquid in which the couplings flow very fast towards the lower threshold $\tilde{\tau}$ and the Luttinger parameters remain constant [panels (a) and (b)].
In the $h$-dominated phase [panels (c) and (d)], we always find a fast exponential growth of the couplings, hitting the upper threshold at $l=l^*<6$.
On the contrary, the entire $g$-phase suffers from divergences due to a divergent $K_s$ [see Fig. \ref{fig:RG_flow_details} panels (e) and (f)].
These divergent solutions of the RG equations can be qualitatively understood by assuming that $K_s$ reaches a large value $K_s \gg 1,K_c$ during its flow: in this limit, we can approximate the dominant part of its RG equations with $\frac{{\rm d}K_s}{{\rm d}l}=a K_s^3$, which upon integration becomes $K_s^{-1} = \sqrt{c-3al}$, diverging at $l=c/3a$.

Interestingly, the second order RG equations result in a non-monotonic curvature of $g(l)$ and $h(l)$ in parts of the Luttinger liquid phase [see \ref{fig:RG_flow_details_2}].
In this cases, the Luttinger parameters reach asymptotic values such that the coupling constants $h$ and $g$ decay to zero after a non-monotonic flow.
This happens in particular in the portion of the Luttinger liquid phase where $g$ is predicted to be relevant according to a simple first order analysis.
By examining the flow in proximity of the $g$-dominated phase, we observe that $h$ flows to negative values before decaying towards zero.
Naturally, this non-monotonicity of the coupling constants connects smoothly to the reported divergences of the $g$-dominated phase in which the asymptotic value of $K_s$ appears to be infinite.
In conclusion, the divergence is not caused by numerical errors, but it is rather a unique feature of the second order RG equations in proximity of the $g$-dominated phase.
This also suggests that a more thorough study of the intricate $g$-phase must include third order corrections.

Concerning the velocities of the system, their value is considered to be approximately constant throughout the RG flow and has been estimated from the relation of Eq.~\eqref{eq:drude}, resulting in $v_q=v_0/K_q$. This relation, however, is supposed to be affected by large deviations for large values of the interaction $V_\perp$, as suggested by the behavior in Fig. \ref{fig:dynamics_velocity} (b) which shows $u_s$ asymptotically reaching a minimum value $\sim u_0/4$ for large interactions.
The failure of the approximation $v_s=v_0/K_s$ for large interactions concurs to increase the errors in the RG flow predictions for this strong interaction regime. In particular, we see that in the off-resonant case $g=0$, the RG flow erroneously predicts a new onset of the $h$ phase for $V_\perp>6t$ at large values of $K_s$. This is a combined effect of the limitations of the second-order perturbation theory and the approximation $v_s=v_0/K_s$, which lead to $\rd K_s/\rd l \propto -h^2K_s^5$ at the beginning of the flow. In this specific regime of large interactions, this decrease of the Luttinger parameter $K_s$ appears to be too fast and yields a non-physical reappearance of the $h$-phase for large $V_\perp$ which is not observed anywhere outside of the resonance in our numerical simulations. Based on the numerical solution of the RG equation, we conclude that our RG results are reliable only in the range $V_\perp<5t$ when addressing large values of the Luttinger parameter $K_s$.

\section{Further details about the correlation functions} \label{App:corr}

The calculation of the correlation functions of the $\theta$ and $\varphi$ fields for finite systems depends on the boundary conditions. In principle, we should separately consider the charge and spin sectors. For the charge sector indeed, no particle can tunnel in or out of the system from the boundaries, and the charge current must exactly vanish at the boundaries. Therefore $\partial_x \varphi_c =0$ for both $x=0$ and $x=L$. For the spin sector, instead, the total spin density is not conserved and, although the chiral current must vanish in average at the boundaries, the previous Neumann boundary conditions for $\varphi_s$ is too restrictive, as demonstrated by the non-vanishing rung current at the edges of the system (Fig. \ref{fig:correlations} (d)). Hence, one should consider more general kinds of boundary conditions \cite{Liguori1998,Mintchev2006,Bellazzini2008,Calabrese2012}. We expect, however, the deviation from the Neumann boundary conditions to be small (proportional to $\Omega/t$), and, also for the spin case, we will consider $\partial_x \varphi_s =0$ at the edges.
Based on the construction in \cite{Cazalilla2004} (see also \cite{Liguori1998,Mintchev2006}), we can define the charge and spin fields (at time $t=0$) in the form:
\begin{align}
\sqrt{K_q} \varphi_q &= \varphi_0 + \sideset{}{'}\sum_{k} \frac{1}{\sqrt{2|k|}} \left[e^{i\frac{k \pi x}{2L}} b_{q,k}  + e^{-i\frac{k \pi x}{2L}} b^\dag_{q,k}\right] ,\\
\frac{\theta_q}{\sqrt{K_q}}&= \theta_0 + \sideset{}{'}\sum_{k} \frac{{\rm sign} k}{\sqrt{2|k|}} \left[e^{i\frac{k \pi x}{2L}} b_{q,k}  + e^{-i\frac{k \pi x}{2L}} b^\dag_{q,k}\right] ;
\end{align}
where $\sideset{}{'}\sum$ labels the sum over the even integers $k$, with $k\neq 0$ from $-\infty$ to $+\infty$. Here $b_{q,k}$ labels two sets of bosonic annihilation and creation operators related to the spin and charge sectors respectively. In order to fulfill the Neumann boundary conditions, we must impose:
\begin{equation}
 b_{q,k} = b_{q,-k}\,.
\end{equation}

From the previous relations we derive:
\begin{align} \label{eq:elementary_expval}
\langle\DF_q(x)\DF_q(x')\rangle &= -\frac{K_q}2\ln\left[\frac{d(x-x'|2L)}{d(x+x'|2L)}\right],\\
\langle\MF_q(x)\MF_q(x')\rangle &= -\frac{1}{2K_q}\ln\left[d(x-x'|2L)d(x+x'|2L)\right],
\end{align}

Considering the case $x'\to 0$, and neglecting functions of $x'$ only, we obtain the behavior of Eq.~\eqref{fluct2} for the spin and charge fluctuations. Given the deviation from the Neumann boundary conditions of the spin sector, however, we must account for an additional weak space dependence represented by the function $f_s(\ell)$.

\bibliography{biblio}

\end{document}